\documentclass[fp,twocolumn]{jpsj3}

\title{Higher-Order Topological Insulator on a Martini Lattice and Its Square Root Descendant}
\author{Daiki Matsumoto$^1$, Tomonari Mizoguchi$^2$, and Yasuhiro Hatsugai$^2$}
\inst{
	$^1$Graduate School of Pure and Applied Sciences, University of Tsukuba, Tsukuba, Ibaraki 305-8571, Japan
$^2$Department of Physics, University of Tsukuba, Tsukuba, Ibaraki 305-8571, Japan}
\abst{
	The notion of square-root topological insulators has been recently generalized to higher-order topological insulators.
    In two-dimensional square-root higher-order topological insulators, the emergence of in-gap corner states is inherited from the squared Hamiltonian, which hosts higher-order topology.
    In this paper, we first propose that the martini lattice model serves as a concrete example of a higher-order topological insulator.
    We further propose a square-root higher-order topological insulator based on the martini lattice model.
    Specifically, we propose that the honeycomb model with two-site decoration, whose squared Hamiltonian consists of two martini lattice models, realizes square-root higher-order topological insulators. 
    We show, for both the martini lattice and the decorated honeycomb model, that in-gap corner states appear at finite energies and that non-trivial bulk $\mathbb{Z}_3$ topological invariant protects them.
}

\usepackage{txfonts}
\usepackage{bm}
\usepackage{braket}
\usepackage[dvipdfmx]{graphicx}
\usepackage{here}
\usepackage{float}
\usepackage{nidanfloat}
\usepackage{mathtools}
\usepackage{color}
\usepackage{cite}
\usepackage{amsmath,amssymb}
\usepackage{amsfonts}
\usepackage{mathrsfs}
\usepackage{cite}

\makeatletter
\newcommand\newblock{\hskip .11em\@plus.33em\@minus.07em}
\makeatother

\begin{document}
	\maketitle
    \section{Introduction}
    The topological phase of matter is one of the hot topics in condensed matter physics.
    Topological insulators(TIs) are the typical example of the topological phase.~\cite{Haldane1988,Kane2005,Kane2005_2,Bernevig2006,Hasan2010,Qi2011}. 
    They have robust gapless states at $(d-1)$-dimensional boundaries of samples due to non-trivial topological indices($d$ is the dimension of bulk).
    This relation between bulk topological indices and boundary states is known as bulk-boundary correspondence~\cite{Hatsugai1993,Hatsugai1993_2}, which is one of the most characteristic nature of the topological phase.
    
    Nowadays, various kinds of TIs have been actively explored.
    Higher-order topological insulators (HOTIs) are one of the examples of such novel topological phases~\cite{Benalcazar2017,Benalcazar2017_2,Hashimoto-Wu-Kimura2017,Schindler2018,Ezawa2018,Hayashi2018}.
    HOTIs in $d$ dimensions have $(d-n)$-dimensional topologically-protected boundary states (with $n \geq 2$) rather than $(d-1)$-dimensional topological boundary states as conventional TIs.
    For instance, a second-order TI in two dimensions has $0-$dimensional boundary states, i.e, the corner states\cite{Araki2019,Takane2019,Watanabe2020,Takahashi2021}.

    \begin{figure}[!b]
        \centering
        \includegraphics[width=1\hsize]{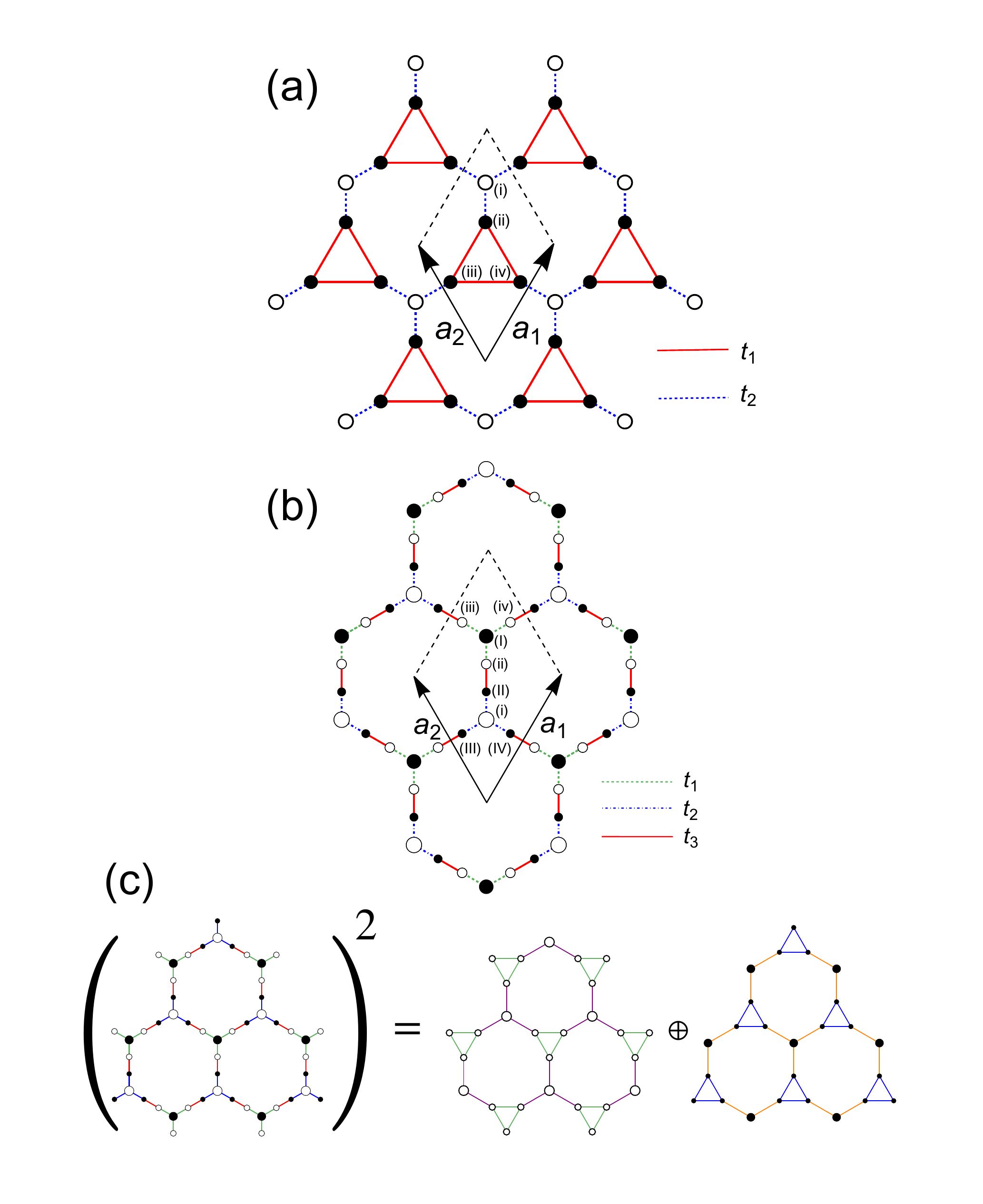}
        \label{fig:SR_IMG}
        \caption{(Color online)Schematic figures of (a) the martini lattice model, and (b) the honeycomb lattice model with two-site decorations. 
        Here $\bm{a}_1$ and $\bm{a}_2$ represent the lattice vectors: $\bm{a}_1 = \left( \frac{1}{2}, \frac{\sqrt{3}}{2}\right)$ and $\bm{a}_2 = \left( -\frac{1}{2}, \frac{\sqrt{3}}{2}\right)$. 
        (c) An illustration of the relation between the original Hamiltonian and its square. 
        The squared Hamiltonian of the decorated honeycomb lattice model is the direct sum of two martini lattice models with sublattice-dependent on-site potentials.}
    \end{figure}

    Square-root TIs are another example of a novel topological phase~\cite{Arkinstall2017}.
    Historically, the concept of square-root plays an essential role in various aspects of physics~\cite{Dirac1928,Kane2014,Attig2017,Attig2019,Naumis2021,Navarro-Labastida2022}.
    In the context of the topological phase, square-root TIs stand for the TIs that inherit their non-trivial topological nature from the square of the Hamiltonian, which is referred to as a parent Hamiltonian. Consequently, a Hamiltonian for a square-root TI is referred to as a child Hamiltonian. 
    Since its discovery, experimental realizations of square-root TIs have actively been pursued, mainly in artificial topological materials~\cite{Kremer2020,Song2020,Yan2020,Wu2021,Yan2021,Kang2021,Geng2021} 
    such as photonic crystals, phononic crystals, and electric circuits.
    On the theoretical side, the generation of square-root descendant  
    is applied to various topological phases~\cite{Mizoguchi2020_sq,Ezawa2020,Mizoguchi2021,Marques2021,Dias2021,Yoshida2021,Marques2021_2,Song2022,Cheng2022,Zhang2022,Roychowdhury2022,PhysRevB.106.L060305,PhysRevResearch.4.033109,10.21468/SciPostPhys.13.2.015}.
    One such example is a generalization to HOTIs, which is dubbed as square-root HOTIs~\cite{Mizoguchi2020_sq}.
    Since the square-root HOTIs are a novel phase and only a few examples are known so far, it is desirable to search for additional concrete examples to verify their ubiquity.
    
    In this paper, we first propose a tight-binding model which hosts a HOTI, that is, the martini lattice model~\cite{Miyahara2005,Scullard2006,McClarty2020} shown in Fig.~\ref{fig:SR_IMG}(a). 
    Then, based on this, we also propose a realization of square-root HOTI in the honeycomb lattice model with two-site decoration~\cite{Barreteau2017,Mizoguchi2021_FB}, which we simply call the decorated honeycomb lattice model in what follows, shown in Fig.~\ref{fig:SR_IMG}(b).
    The squared Hamiltonian of the decorated honeycomb model corresponds to the direct sum of two martini lattices as schematically shown in Fig.~\ref{fig:SR_IMG}(c).
    In other words, the martini lattice model is the parent of the decorated honeycomb lattice model.
    Thus, the square-root HOTI is realized in the decorated honeycomb lattice model.
    Remarkably, we find that changing the corner termination changes the nature of the topological corner modes, which originates from the fact that the corner termination specifies the corresponding parent Hamiltonian from which the corner modes are inherited.

    The rest of this paper is organized as follows.
    In Sect.~\ref{Sec:model}, we introduce the martini lattice model and the decorated honeycomb lattice model and clarify the key symmetries.
    Then, we point out their square-root relation with a special focus on the inheritance of the energy dispersion and Bloch wave functions.
    We also show the dispersion relations obtained by numerical calculation for each model to seek topological phase transition points.
    In Sect.~\ref{Sec:edge}, we elucidate the higher-order topology of the martini and the decorated honeycomb models, by demonstrating the existence of in-gap corner states in the finite system under the open boundary conditions.
    We also numerically calculate $\mathbb{Z}_3$ topological indices related to polarization for each model to confirm the bulk-corner correspondence.
    It is revealed that the decorated honeycomb model realizes the square-root HOTI inherited from the martini model. In Sect.~\ref{Sec:summary}, we present a summary of this paper.
    
    \section{Model}
    \label{Sec:model}
    \subsection{Martini model}
\begin{figure*}[!tb]
        \centering
        \begin{minipage}[c]{0.24\hsize}
            \includegraphics[width=1\hsize]{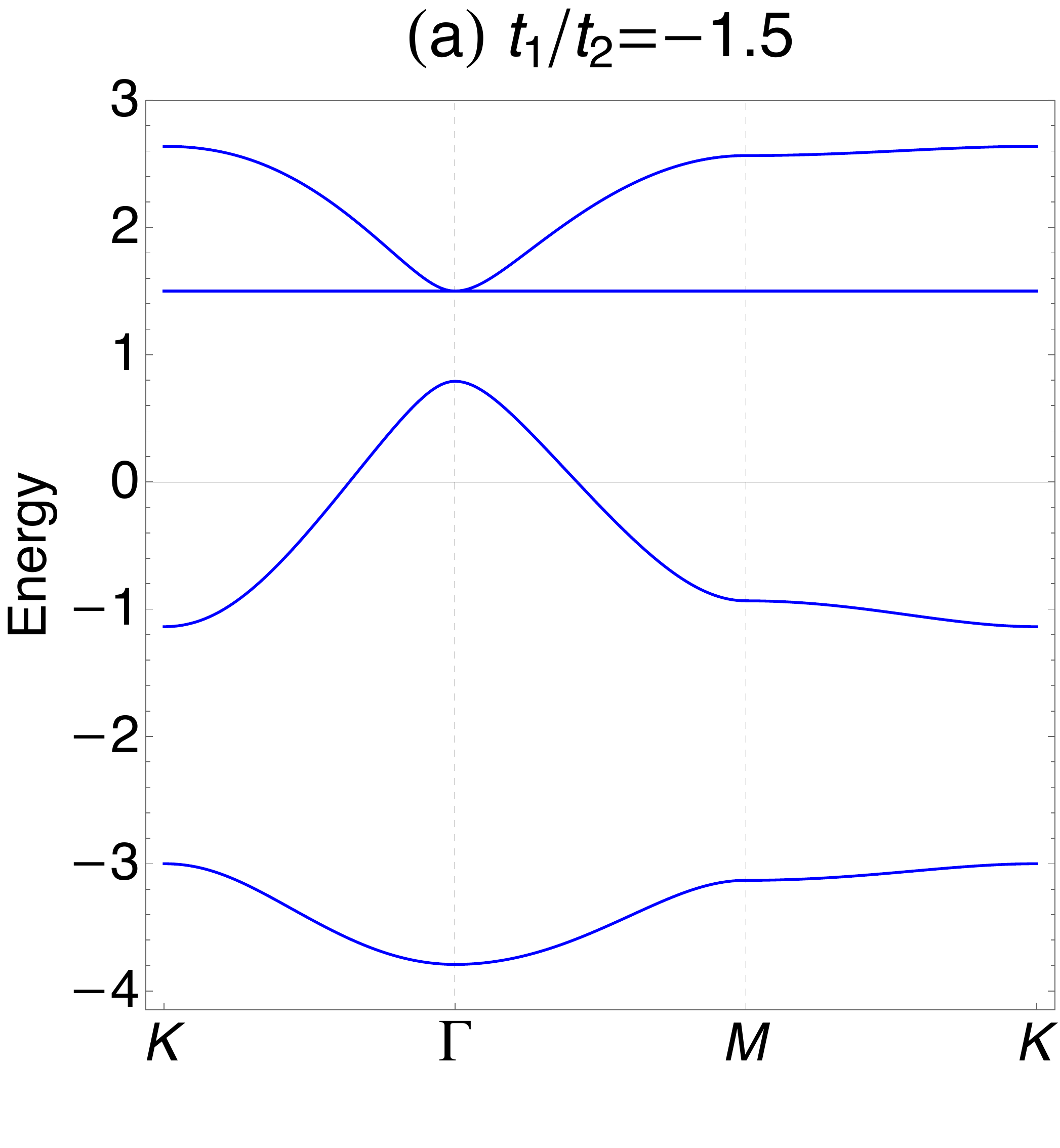}
        \end{minipage}
        \begin{minipage}[c]{0.24\hsize}
            \includegraphics[width=1\hsize]{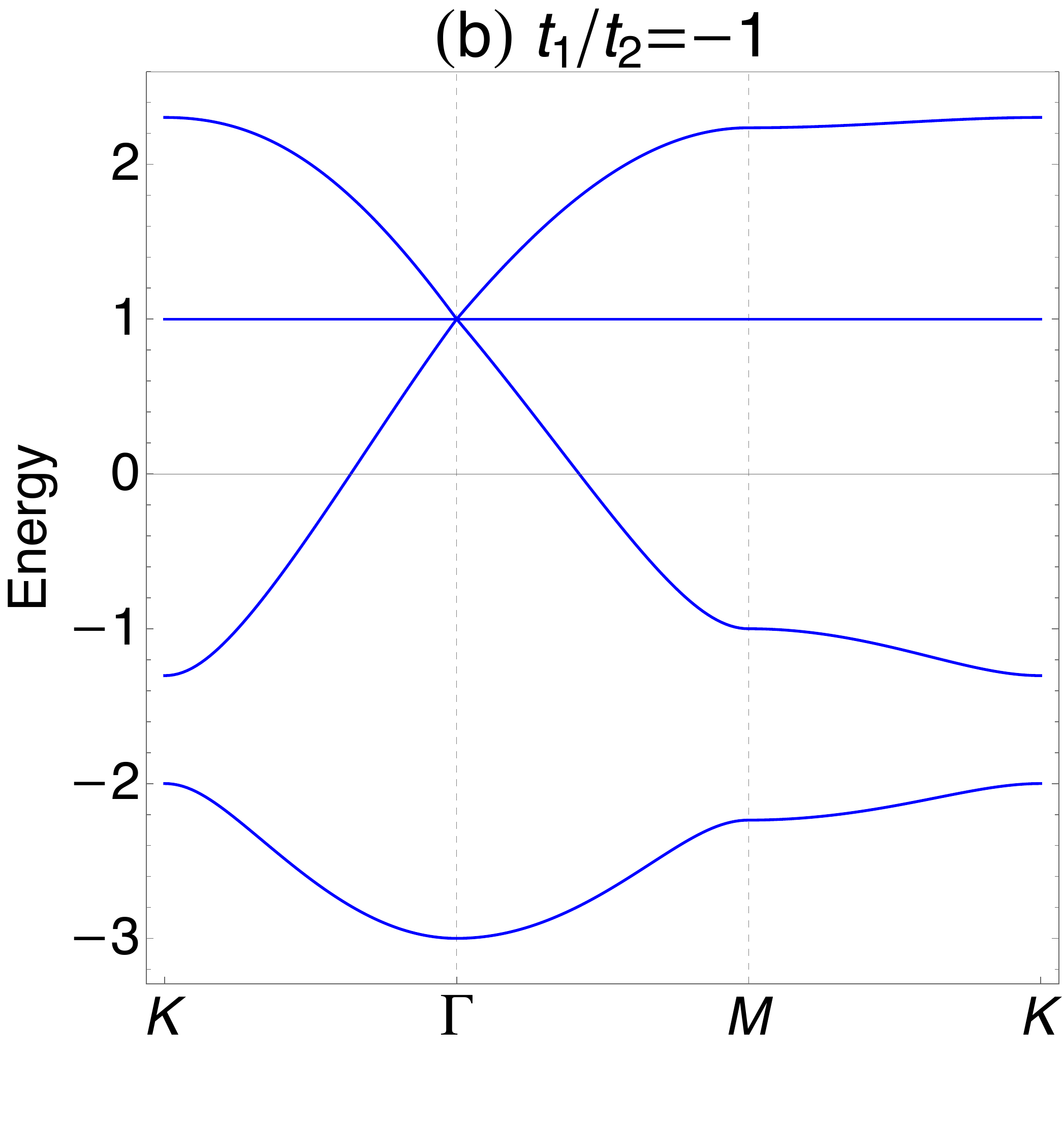}
        \end{minipage}
        \begin{minipage}[c]{0.24\hsize}
            \includegraphics[width=1\hsize]{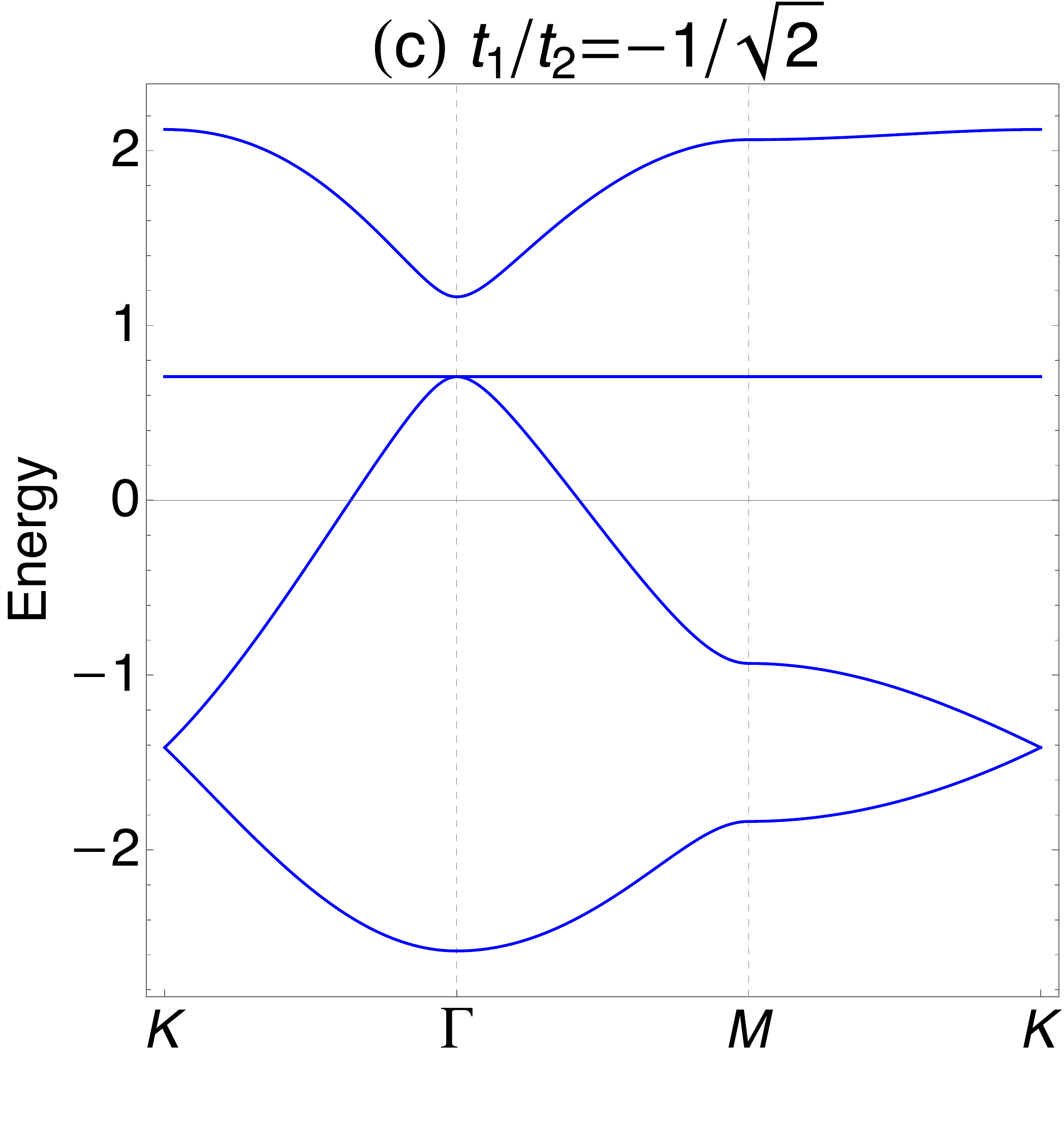}
        \end{minipage}
        \begin{minipage}[c]{0.24\hsize}
            \includegraphics[width=1\hsize]{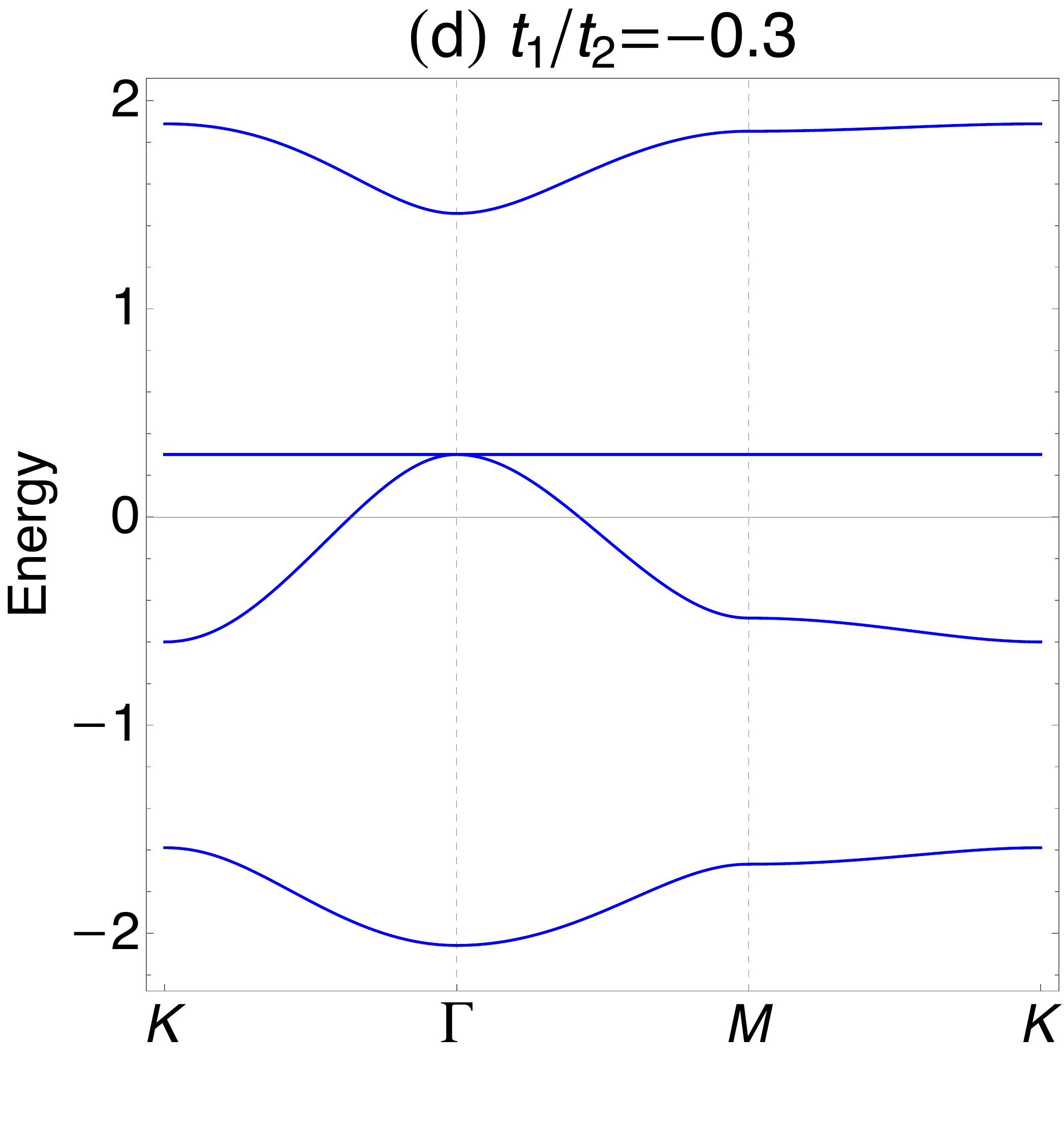}
        \end{minipage}
        \\
        \begin{minipage}[c]{0.24\hsize}
            \includegraphics[width=1\hsize]{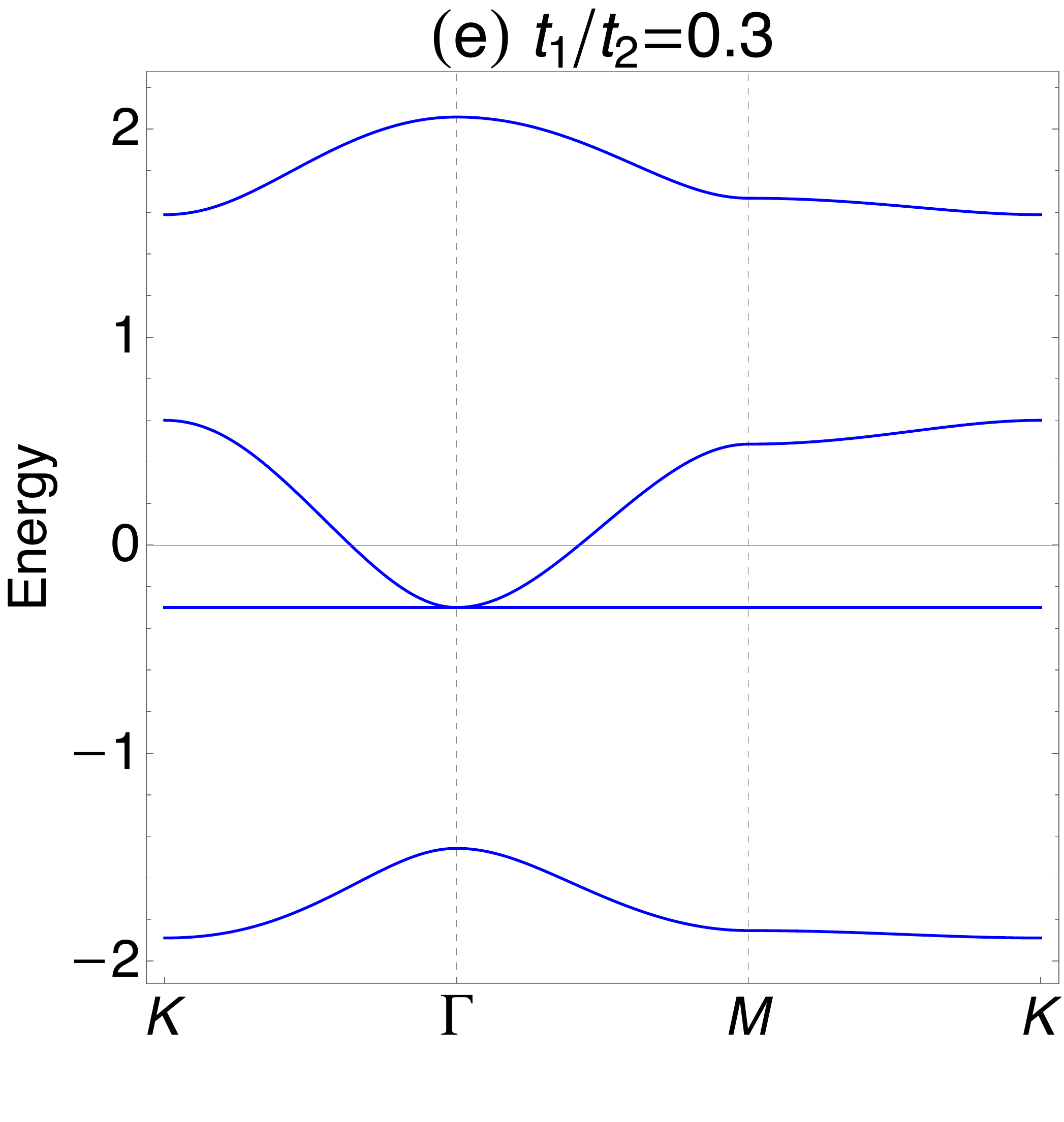}
        \end{minipage}
        \begin{minipage}[c]{0.24\hsize}
            \includegraphics[width=1\hsize]{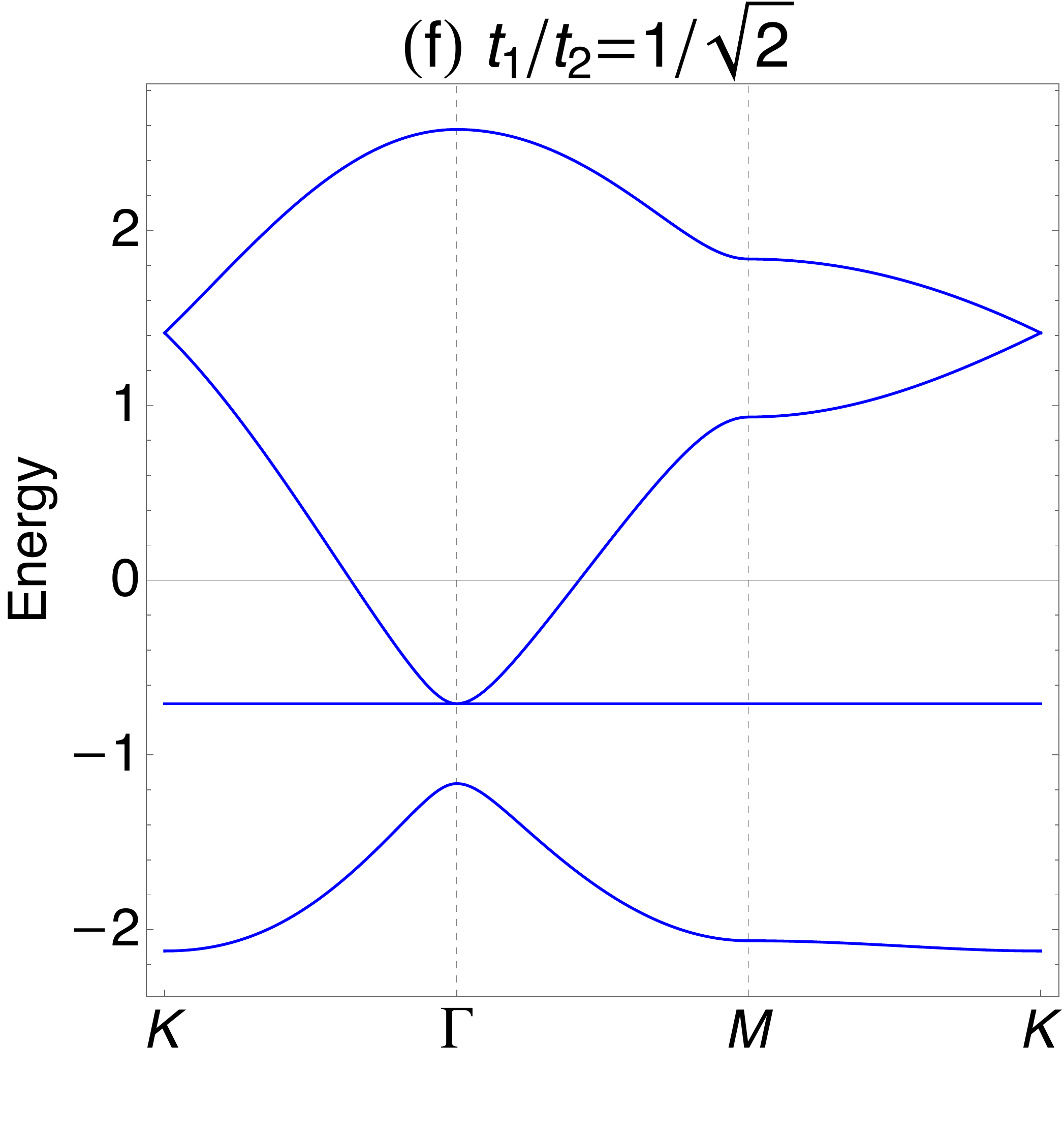}
        \end{minipage}
        \begin{minipage}[c]{0.24\hsize}
            \includegraphics[width=1\hsize]{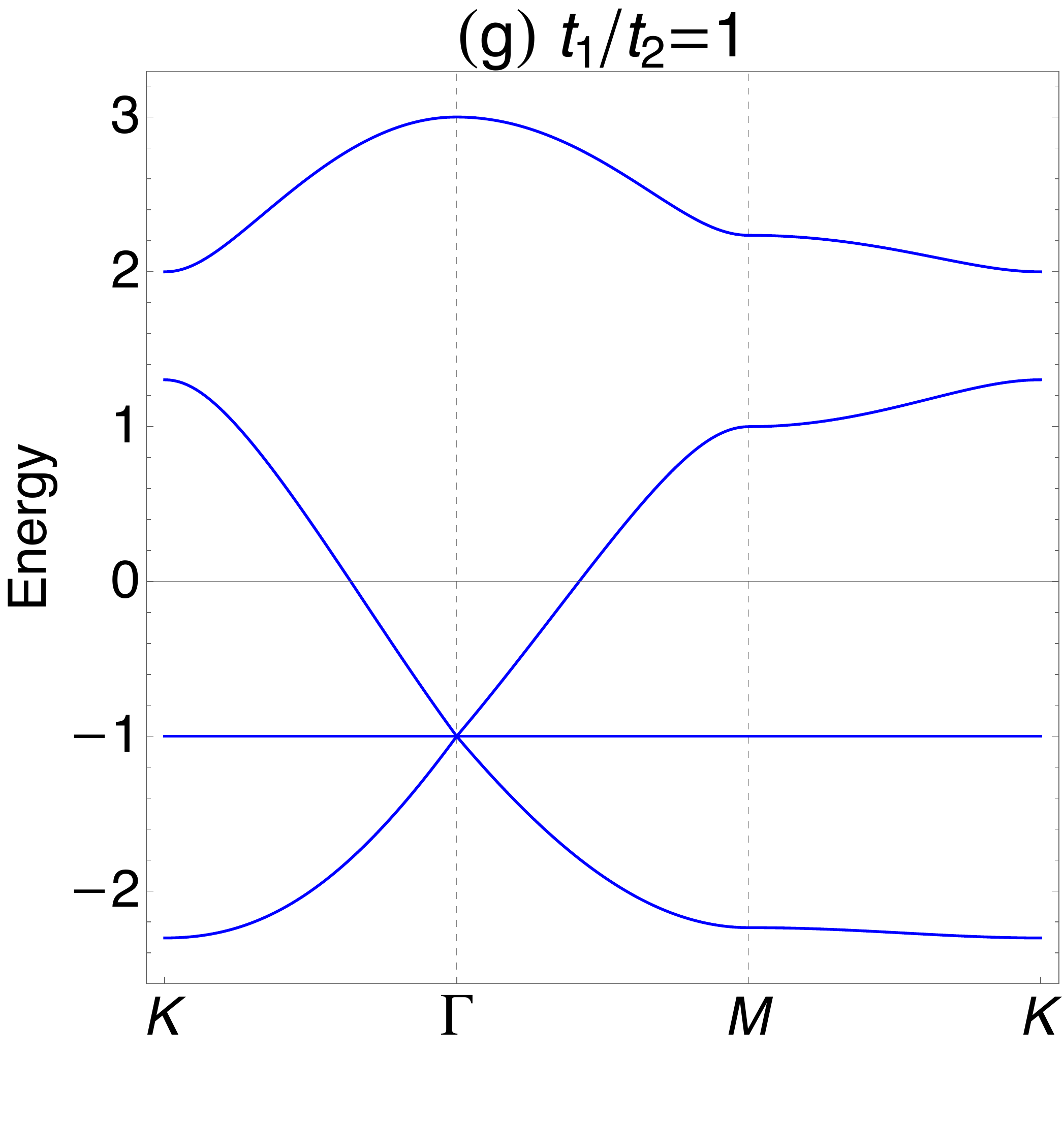}
        \end{minipage}
        \begin{minipage}[c]{0.24\hsize}
            \includegraphics[width=1\hsize]{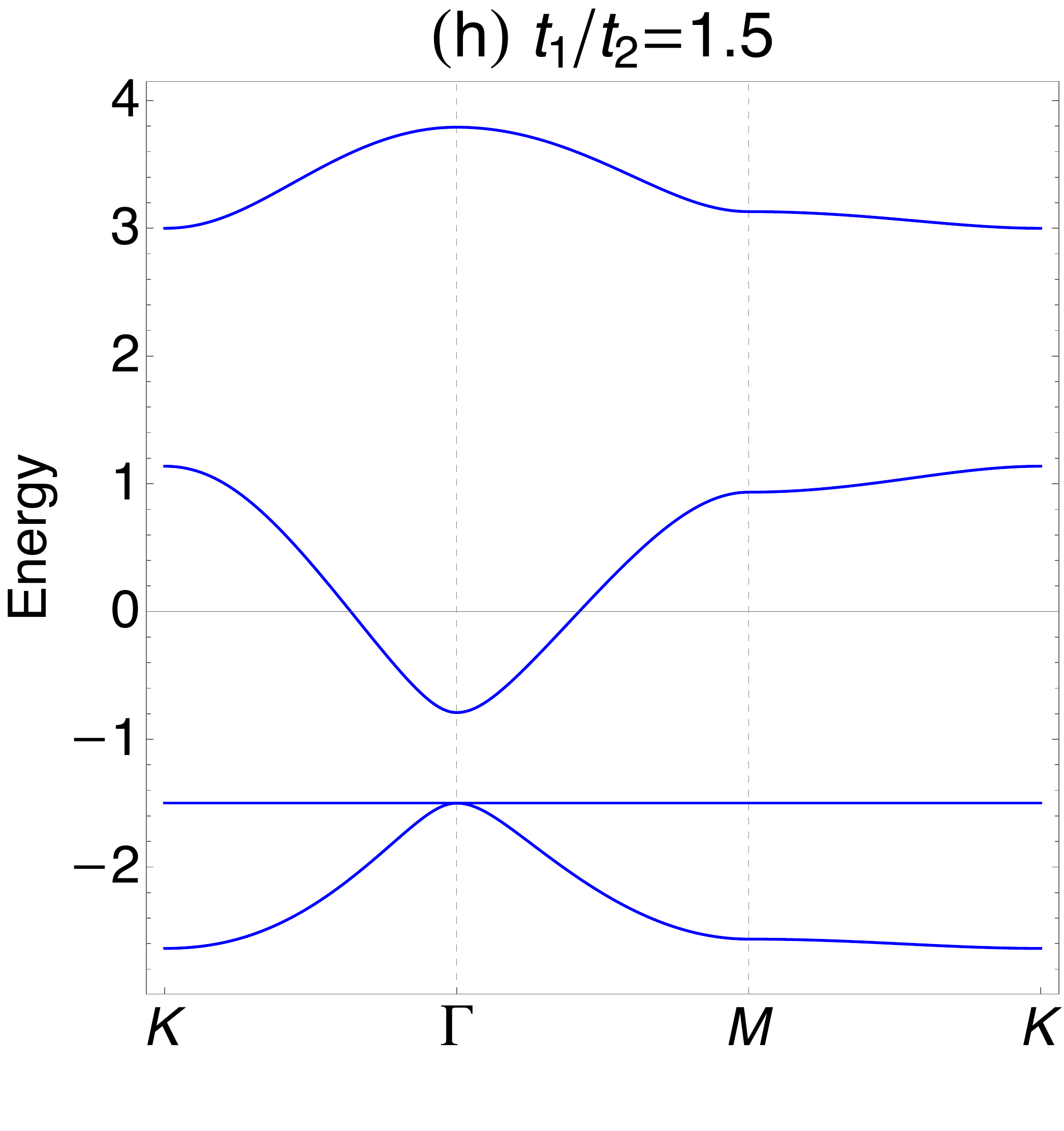}
        \end{minipage}
        \caption{(Color online)Band structure of the martini lattice model with $t_2=1$,$V=0$ and (a) $t_1/t_2=-1.5$, (b) $t_1/t_2=-1$, (c) $t_1/t_2=-1/\sqrt{2}$, 
        (d) $t_1/t_2=-0.3$, (e) $t_1/t_2=0.3$, (f) $t_1/t_2=1/\sqrt{2}$, (g) $t_1/t_2=1$, (h) $t_1/t_2=1.5$. $\Gamma=(0,0), {{\rm{K}}}=\left(\frac{4\pi}{3},0 \right)$ and 
        ${\rm{M}}=\left(\pi ,\frac{\pi}{\sqrt{3}} \right)$ are high-symmetry points in the Brillouin zone.}
        \label{fig:martini_band}
        \vspace{-15pt}
    \end{figure*}

    As a parent Hamiltonian of the following discussions, we consider a tight-binding Hamiltonian on a martini lattice.
    This lattice structure has been considered in several previous works~\cite{Kubo2006,Scullard2006,McClarty2020}.
    It has four sites per unit cell, as shown in Fig.~\ref{fig:SR_IMG}(a).
    The tight-binding Hamiltonian reads
    \begin{equation}
        H^{(\bm{M})}=\sum_{\bm{k}}\bm{C}_{\bm{k}}^{\dagger} H^{(\bm{M})}_{\bm{k}} \bm{C}_{\bm{k}}, 
    \end{equation}
    where $\bm{C}_{\bm{k}}=(C_{\bm{k},(\rm{i})},C_{\bm{k},(\rm{ii})},C_{\bm{k},(\rm{iii})},C_{\bm{k},(\rm{iv})})^T$ are the annihilation operators,
    and
    \begin{equation}
        H^{(\bm{M})}_{{\bm{k}}} = \left(
                \begin{array}{cccc}
                   V & t_2 & t_2e^{i\bm{k}\cdot\bm{a}_1} & t_2e^{i\bm{k}\cdot\bm{a}_2} \\
                   t_2 & 0 & t_1 & t_1 \\
                   t_2e^{-i\bm{k}\cdot\bm{a}_1}& t_1 & 0 & t_1 \\
                   t_2e^{-i\bm{k}\cdot\bm{a}_2} & t_1 & t_1 & 0
                \end{array}
                \right), \label{eq:martini}
    \end{equation}
    is the Hamiltonian matrix.
    For the definition of $\bm{a}_1$ and $\bm{a}_2$, see Fig.~\ref{fig:SR_IMG}(a).
    This model includes three parameters: nearest-neighbor hoppings with two different parameters, namely, $t_1$ for intra-triangles and $t_2$ for inter-triangles, and the difference of on-site potentials between the black and white sites, $V$.
    We note, for later discussions, that the Hamiltonian has $C_3$ symmetry centered at the center of the red triangle consisting of sublattice (i), (ii), and (iii) in a unit cell: $H^{(\bm{M})}_{\bm{k}}$ satisfies
    \begin{equation}
        \label{eq:c3_martini}
        H^{(\bm{M})}_{C_3 \bm{k}}=U^{(\bm{M})}_{\bm{k}} H^{(\bm{M})}_{\bm{k}}(U^{(\bm{M})}_{\bm{k}})^{\dagger},
    \end{equation}
    where $C_3 \bm{k}=(-\frac{1}{2}k_x+\frac{\sqrt{3}}{2}k_y,-\frac{\sqrt{3}}{2}k_x-\frac{1}{2}k_y )$ and
    \begin{equation} 
        U^{(\bm{M})}_{\bm{k}}=\left(
                \begin{array}{cccc}
                    e^{-i\bm{k}\cdot\bm{a}_2}&0&0&0 \\
                    0&0&0&1 \\
                    0&1&0&0 \\
                    0&0&1&0
                \end{array}
            \right).
    \end{equation}
    The Hamiltonian has also $C_3$ symmetry centered at sublattice(i):
    \begin{align}
        H^{(\bm{M})}_{C_3 \bm{k}} &= U^{(\bm{dM})}_{\bm{k}} H^{(\bm{M})}_{\bm{k}}(U^{(\bm{dM})}_{\bm{k}})^{\dagger},
    \end{align}
    where
    \begin{align}
        U^{(\bm{dM})}_{\bm{k}} &= e^{-i\bm{k}\cdot \bm{a}_2}(U^{(\bm{M})}_{\bm{k}})^{*}.
    \end{align}
    We depict the band structure for several sets of hopping parameters with fixing $V=0$ in Fig.~\ref{fig:martini_band}.
    We see that a flat band appears at $E=-t_1$ regardless of parameters, which originates from the fact that the martini lattice belongs to a class of lattices called partial line graphs\cite{Miyahara2005}.
    When $t_1/t_2 < 0$, the third band is a flat band [Fig.~\ref{fig:martini_band} (a)-(d)], whereas when $t_1/t_2 > 0$, the second band is a flat band [Fig.~\ref{fig:martini_band} (e)-(h)].  
    In general, we see two gaps in the band structure and we focus on these gaps when discussing the topological phases.
    The band touching between the first and the second bands occurs for $t_1/t_2=-1/\sqrt{2}$, and that between the third and the fourth bands occurs for $t_1/t_2 = 1/\sqrt{2}$; they occur at K point.
    Additionally, the triple band touching among the second, the third, and the fourth bands occur for $t_1/t_2=-1$, and that among the first, the second, and the third bands occur for $t_1/t_2=1$; they occur at $\Gamma$ point.
    \subsection{Decorated honeycomb model}
    Next, we introduce a tight-binding model on a decorated honeycomb lattice. 
    This lattice structure has been considered theoretically~\cite{Barreteau2017,Mizoguchi2021_FB}, and is also relevant to some
    solid-state materials~\cite{Baughman1987,Longuinhos2014,Li2015,Lee2020}.
    The Hamiltonian is
    \begin{equation}
        H^{(\bm{DH})}=\sum_{\bm{k}}\bm{C}_{\bm{k}}^{\dagger} H^{(\bm{DH})}_{\bm{k}} \bm{C}_{\bm{k}},
    \end{equation}
    where $\bm{C}_{\bm{k}}=(C_{\bm{k},\bullet (\rm{I})},C_{\bm{k},\bullet (\rm{II})},C_{\bm{k},\bullet (\rm{III})},C_{\bm{k},\bullet (\rm{IV})}, \\
    C_{\bm{k},\circ (\rm{i})},C_{\bm{k},\circ (\rm{ii})},C_{\bm{k},\circ(\rm{iii})},C_{\bm{k},\circ(\rm{iv})})^T$ are the annihilation operators, 
    and
    \begin{equation}
        H^{(\bm{DH})}_{\bm{k}} = \left(
                    \begin{array}{cc}
                       \mathcal{O}_{4,4}&\Phi_{\bm{k}}^{\dagger} \\
                       \Phi_{\bm{k}}&\mathcal{O}_{4,4}
                    \end{array}
                    \right), \label{eq:Ham_DH}
    \end{equation}
    is the Hamiltonian matrix. 
    In Eq.~(\ref{eq:Ham_DH}), we have introduced
    \begin{equation}
        \Phi_{\bm{k}} = \left(
            \begin{array}{cccc}
              0&t_2&t_2&t_2\\
              t_1&t_3&0&0 \\
              t_1&0&t_3e^{i\bm{k}\cdot\bm{a}_1}&0 \\
              t_1&0&0&t_3e^{i\bm{k}\cdot\bm{a}_2}
            \end{array}
          \right),
    \end{equation}
    and $n\times m$ zero matrix, $\mathcal{O}_{n,m}$.
    Note that this model includes only nearest-neighbor hoppings with three different parameters $t_1,t_2$ and $t_3$. 
    The vectors $\bm{a}_1$ and $\bm{a}_2$ are defined in the same way as for a martini lattice model.
    
    As for the symmetries of this Hamiltonian, we note that 
    the Hamiltonian has chiral symmetry:
    \begin{equation}
    \label{eq:chiral}
        \gamma H^{(\bm{DH})}_{\bm{k}}\gamma = -H^{(\bm{DH})}_{\bm{k}},
    \end{equation}
    where 
    \begin{equation}
        \gamma = \left(
                    \begin{array}{cc}
                       I_4&\mathcal{O}_{4,4}\\
                       \mathcal{O}_{4,4}&-I_4
                    \end{array}
                    \right).
    \end{equation}
    Here $I_n$ is a $n$ dimensional identity matrix.
    The Hamiltonian also has $C_3$ symmetry centered at sublattice $\circ$(i):
    \begin{equation}
    \label{Eq:c3-honey}
        H^{(\bm{DH})}_{C_3 \bm{k}}=U^{(\bm{DH})}_{\bm{k}} H^{(\bm{DH})}_{\bm{k}}(U^{(\bm{DH})}_{\bm{k}})^{\dagger},
    \end{equation}
    where
    \begin{equation}
        \label{eq:c3-operator_honey}
        U^{(\bm{DH})}_{\bm{k}} = \left(
                    \begin{array}{cc}
                        U^{(\bm{M})}_{\bm{k}}&\mathcal{O}_{4,4} \\
                       \mathcal{O}_{4,4}&U^{(\bm{dM})}_{\bm{k}}
                    \end{array}
                    \right).
    \end{equation}
    \begin{figure*}[!b]
        \centering
        \begin{minipage}{0.19\hsize}
            \includegraphics[width=1\hsize]{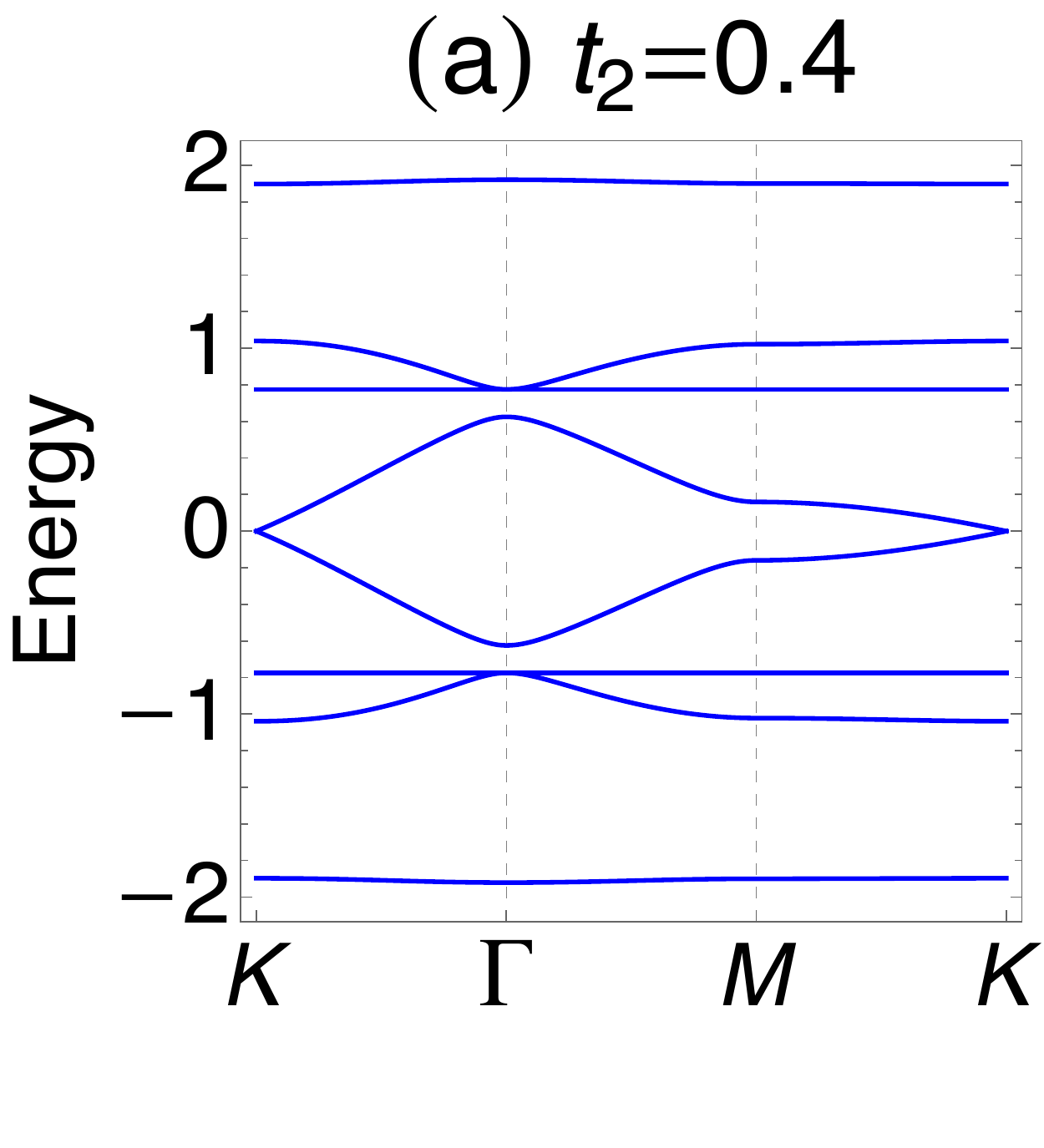}
        \end{minipage}
        \begin{minipage}{0.19\hsize}
            \includegraphics[width=1\hsize]{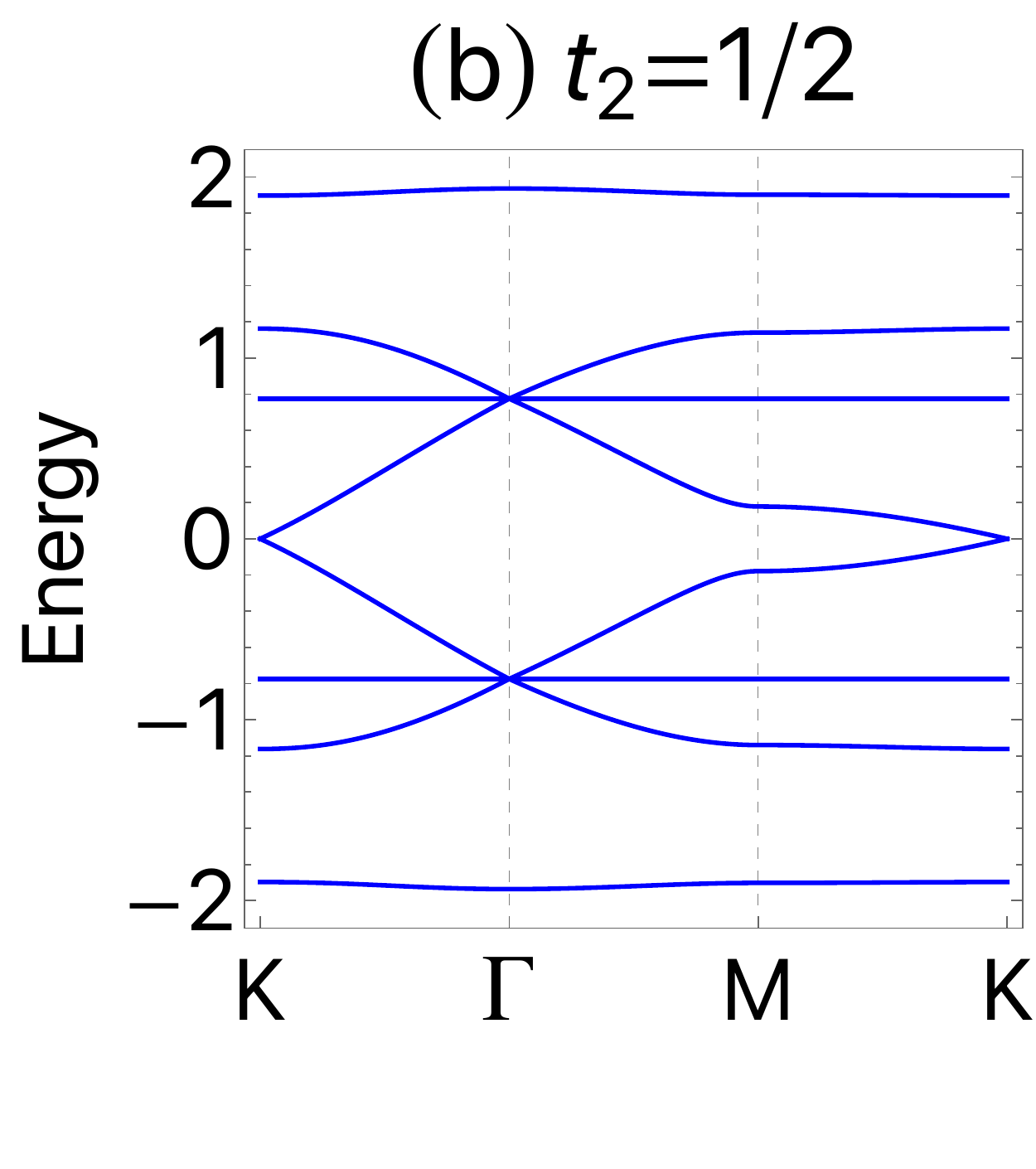}
        \end{minipage}
        \begin{minipage}{0.19\hsize}
            \includegraphics[width=1\hsize]{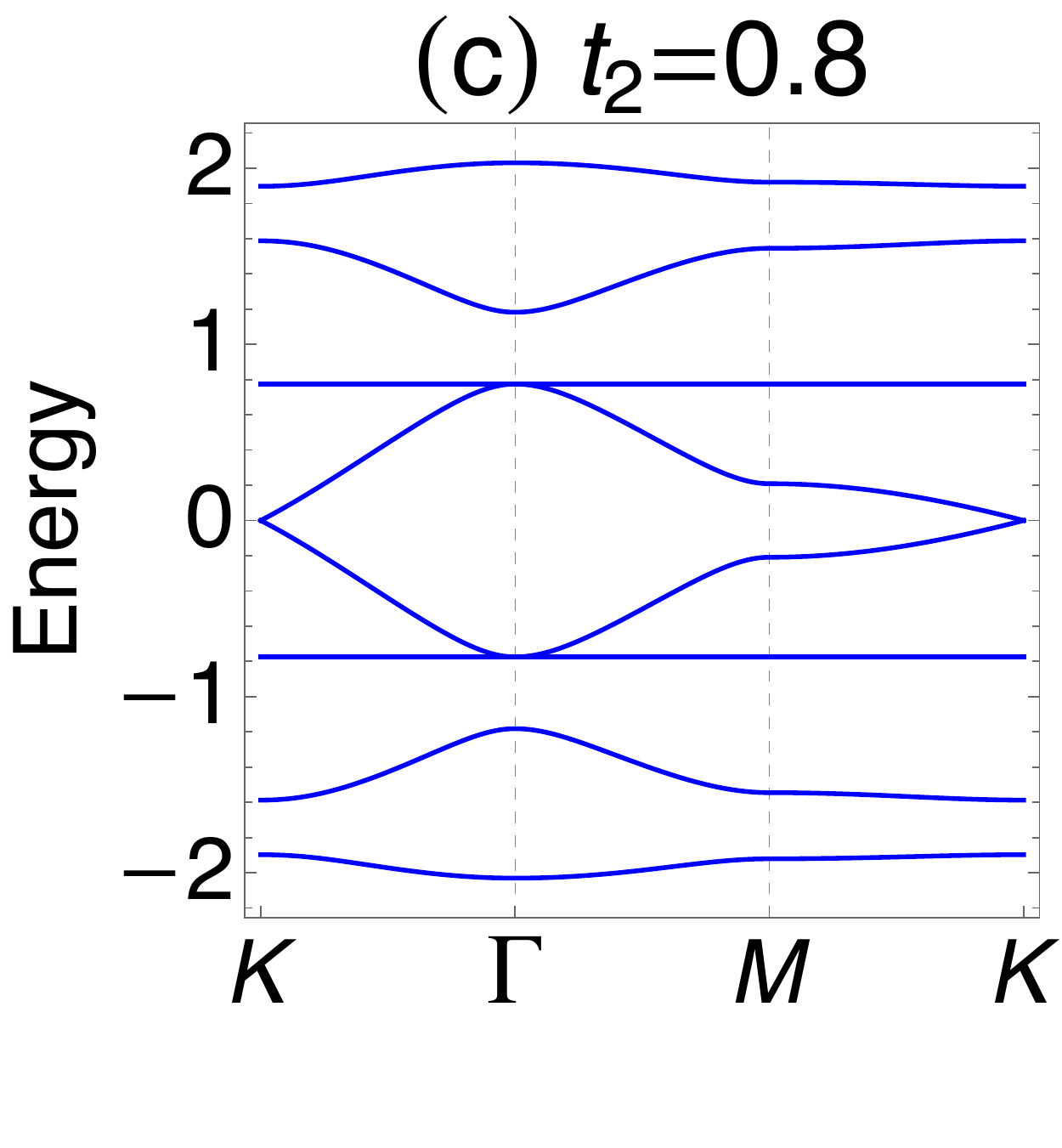}
        \end{minipage}
        \begin{minipage}{0.19\hsize}
            \includegraphics[width=1\hsize]{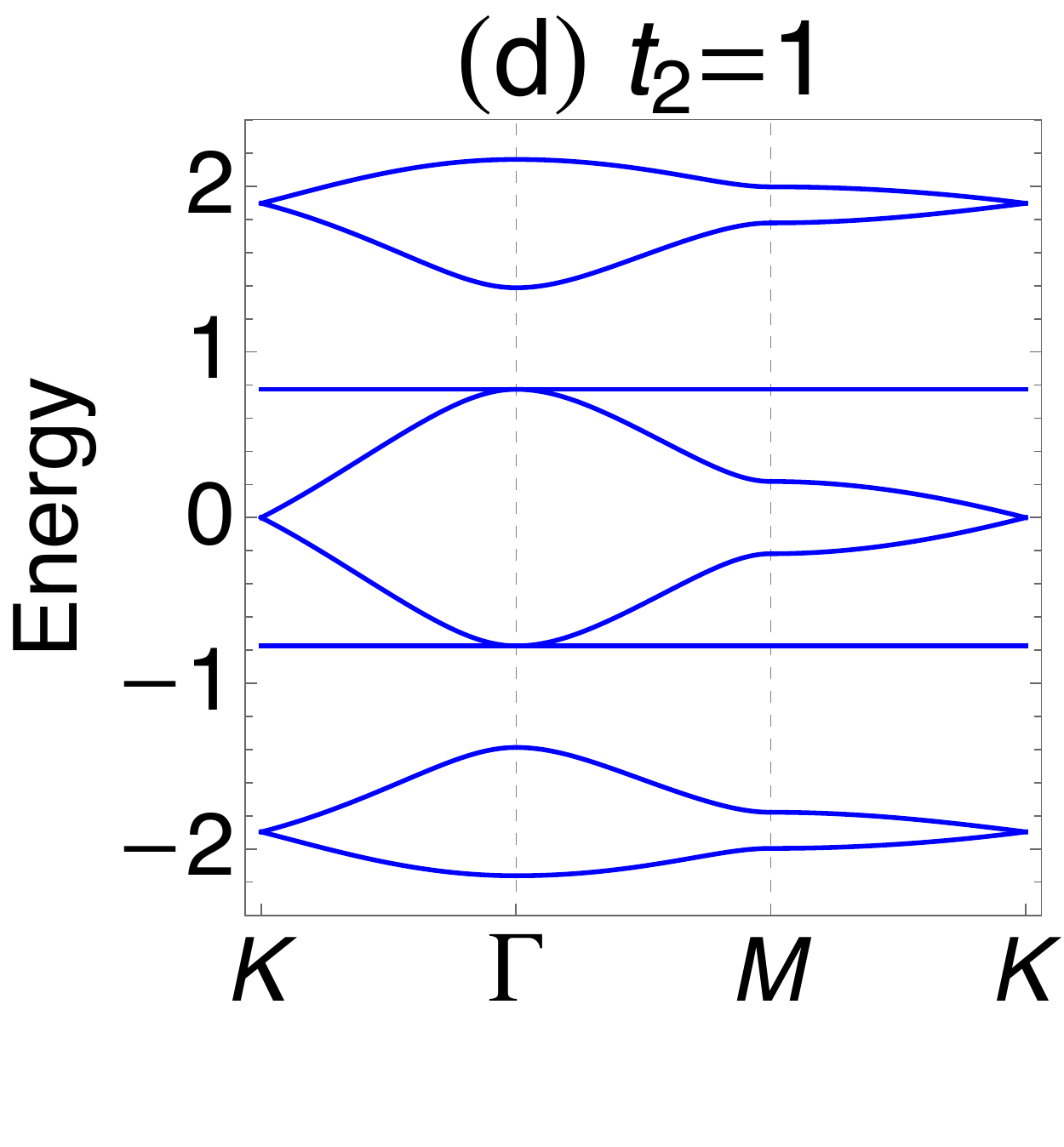}
        \end{minipage}
        \begin{minipage}{0.19\hsize}
            \includegraphics[width=1\hsize]{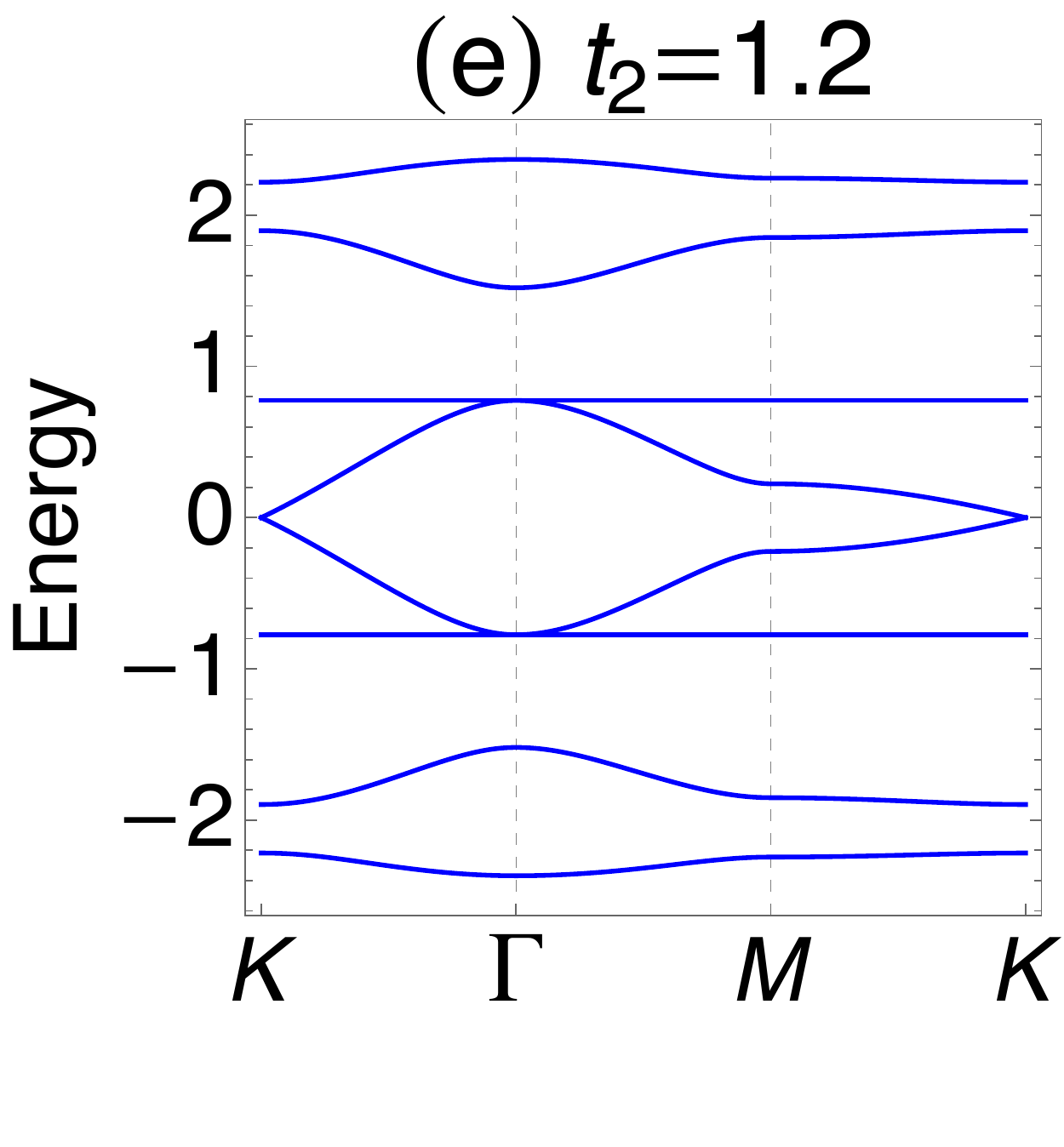}
        \end{minipage}
        \caption{(Color online)Band structures of decorated honeycomb lattice with $t_1=1, t_3=\sqrt{\frac{3}{5}}$ and (a) $t_2=0.4$, (b) $t_2=0.5$, (c) $t_2=0.8$, (d) $t_2=1$, (e) $t_2=1.2$. $\Gamma=(0,0), {{\rm{K}}}=\left(\frac{4\pi}{3},0 \right)$ and ${\rm{M}}=\left(\pi ,\frac{\pi}{\sqrt{3}} \right)$ are high-symmetry points in the Brillouin zone.}
        \label{fig:band_dhoney}
    \end{figure*}
    
    In Fig.~\ref{fig:band_dhoney}, we show the band structures of the decorated honeycomb model. 
    Here, and also in Sect.~\ref{Sec:edge},
    we fix the parameters $t_1$ and $t_3$ as $\left(t_1,t_3 \right) = \left(1,\sqrt{3/5} \right)$, and retain $t_2$ as the only variable parameter, for the purpose of concreteness.
    We see that two flat bands appear for all the panels of Fig.~\ref{fig:band_dhoney}. In fact, the flat bands appear at $E=\pm t_3$ regardless of the choice of parameters\cite{Mizoguchi2021_FB}.
    In Fig.~\ref{fig:band_dhoney}(b), we see that triple band touching among the second, the third, and the fourth bands
    and that among the fifth, the sixth, and the seventh bands
    occur at $\Gamma$ point with $t_2=1/2$.
    In fact, it is analytically shown 
    that the triple band touching at $\Gamma$ point occurs for $(1/t_1^2 + 1/t_2^2)t_3^2 = 3$~\cite{Mizoguchi2021_FB}.
    Besides $\Gamma$ point, the band touchings between the first and the second bands as well as the seventh and the eighth bands
    occur at K point for $|t_1|=|t_2|$ [Figs.~\ref{fig:band_dhoney}(d)]. 
    For $|t_1| \neq |t_2|$ they are gapped, and we focus on these gaps when discussing the square-root HOTIs, in the decorated honeycomb model.
    
    \subsection{Square-root relation}
    We now show the square-root relation between the decorated honeycomb lattice model and the martini lattice model.
    A simple understanding of this relation can be obtained from the real-space viewpoint.
    Since the Hamiltonian includes only NN hoppings which connect only white sites with black sites, a particle on a black (white) site can moves to the nearest black (white) sites or comes back to the original site by operating the Hamiltonian twice.
    The former paths network black (white) sites that form a martini (downward martini) lattice, while the latter corresponds to the on-site potentials.
    Turning to the square-root relation in the reciprocal space,
    i.e., that for the Bloch Hamiltonian, we see that the square of the decorated honeycomb lattice model becomes
    \begin{equation}
        \label{Eq:squaredHamiltonian}
        [ H^{(\bm{DH})}_{\bm{k}} ]^2 = \left(
            \begin{array}{cc}
                h_{\bm{k}}^{(\bm{M})} & \mathcal{O}_{4,4}\\
                \mathcal{O}_{4,4} & (h_{\bm{k}}^{(\bm{M})})^{\dagger}     
            \end{array}
        \right),
    \end{equation}
    where
    \begin{eqnarray}
        \label{eq:squaredHamiltonian_martini}
        h_{\bm{k}}^{(\bm{M})}&=&\Phi_{\bm{k}}^{\dagger}\Phi_{\bm{k}} \\
        &=& \left(
            \begin{array}{cccc}
              3t_1^2&t_1t_3&t_1t_3e^{i\bm{k}\cdot \bm{a}_1}&t_1t_3e^{i\bm{k}\cdot \bm{a}_2}\\
              t_1t_3&t_2^2+t_3^2&t_2^2&t_2^2 \\
              t_1t_3e^{-i\bm{k}\cdot \bm{a}_1}&t_2^2&t_2^2 + t_3^2&t_2^2 \\
              t_1t_3e^{-i\bm{k}\cdot \bm{a}_2}&t_2^2&t_2^2&t_2^2 + t_3^2
            \end{array}
          \right), \nonumber
    \end{eqnarray}
    is equivalent to the Hamiltonian of a martini lattice of Eq.~(\ref{eq:martini}),
    with 
    the nearest-neighbor hoppings being $t_1 t_3$ and $t_2^2$, 
    and the on-site potential being $3t_1^2 -\left(t_2^2+ t_3^2 \right)$,
    up to the constant energy shift $t_2^2+ t_3^2 $.
    Further, 
    for the lower-right part of Eq. (\ref{Eq:squaredHamiltonian}), we have
    \begin{eqnarray}
        \label{eq:squaredHamiltonian_dmartini}
        (h_{\bm{k}}^{(\bm{M})})^{\dagger}&=&\Phi_{\bm{k}}\Phi_{\bm{k}}^{\dagger} \\
        &=& \left(
            \begin{array}{cccc}
              3t_2^2&t_2t_3&t_2t_3e^{-i\bm{k}\cdot \bm{a}_1}&t_2t_3e^{-i\bm{k}\cdot \bm{a}_2}\\
              t_2t_3&t_1^2+t_3^2&t_1^2&t_1^2 \\
              t_2t_3e^{i\bm{k}\cdot \bm{a}_1}&t_1^2&t_1^2 + t_3^2&t_1^2 \\
              t_2t_3e^{i\bm{k}\cdot \bm{a}_2}&t_1^2&t_1^2&t_1^2 + t_3^2
            \end{array}
          \right).\nonumber
    \end{eqnarray}
    Importantly, $(h_{\bm{k}}^{(\bm{M})})^{\dagger}$ is also the Hamiltonian of the martini lattice with the upside-down orientation to that of $h_{\bm{k}}^{(\bm{M})}$ [see Fig.~\ref{fig:SR_IMG}(c)].
    The corresponding tight-binding parameters are given as follows:
    the nearest-neighbor hoppings are $t_2 t_3$ and $t_1^2$,
    and the on-site potential is $3t_2^2 -\left(t_1^2+ t_3^2 \right)$,
    and the constant energy shift is $t_1^2+ t_3^2$.
    We call the lattice for $(h_{\bm{k}}^{(\bm{M})})^{\dagger}$ as a downward martini lattice and write $(h^{(\bm{M})}_{\bm{k}})^{\dagger} $ as $h^{(\bm{dM})}_{\bm{k}}$ henceforth.
    Note that the rotation operator $U^{(\bm{dM})}_{\bm{k}}$ in Eq.~(\ref{eq:c3-operator_honey}) acts on the subspace of the downward martini lattice.
    Based on this relation,  we elucidate that the energy dispersion and the eigenstates of the decorated honeycomb lattice can be expressed by using those
    of squared Hamiltonian~\cite{Mizoguchi2020_sq}.
    Let $\ket{u_{\bm{k},i}^{(\bm{M})}}$ be the normalized eigenstate of $h_{\bm{k}}^{(\bm{M})}$ with the eigenenergy $E^{(\bm{M})}_{{\bm{k}},i}$($i=1,2,3,4$).
    Note that we label the eigenstate such that $E^{({\bm{M}})}_{{\bm{k}},1} \leq \cdots \leq E^{({\bm{M}})}_{{\bm{k}},4}$ holds.
    Namely, $\ket{u_{\bm{k},i}^{(\bm{M})}}$ satisfies
    \begin{equation}
        \label{eq:squared_martini}
        h_{\bm{k}}^{(\bm{M})} \ket{u_{\bm{k},i}^{(\bm{M})}} = \Phi_{\bm{k}}^{\dagger}\Phi_{\bm{k}} \ket{u_{\bm{k},i}^{(\bm{M})}} = E^{(\bm{M})}_{{\bm{k}},i} \ket{u_{\bm{k},i}^{(\bm{M})}}.
    \end{equation}
    Then, multiplying $\Phi_{\bm{k}}$ on Eq.~(\ref{eq:squared_martini}) from left, we get
    \begin{eqnarray}
        \Phi_{\bm{k}} \Phi_{\bm{k}}^{\dagger} (\Phi_{\bm{k}} \ket{u_{\bm{k},i}^{(\bm{M})}}) &=& h^{(\bm{dM})}_{\bm{k}}\Phi_{\bm{k}} \ket{u_{\bm{k},i}^{(\bm{M})}} \nonumber \\
        &=&  E^{(\bm{M})}_{{\bm{k}},i} \Phi_{\bm{k}} \ket{u_{\bm{k},i}^{(\bm{M})}}.
    \end{eqnarray}
    This relation means that 
    $h_{\bm{k}}^{(\bm{dM})}$ has the same eigenenergies as $h_{\bm{k}}^{(\bm{M})}$,
    and that the corresponding eigenstate is
    \begin{equation}
        \label{eq:martini-dmartini}
        \ket{u_{\bm{k},i}^{(\bm{dM})}} = \Phi_{\bm{k}} \ket{u_{\bm{k},i}^{(\bm{M})}}.
    \end{equation}
    
    Using these relations, we are ready to construct the eigenstates of $H^{({\bm{DH}})}_{{\bm{k}}}$. 
    To be concrete, for the non-zero energy state, the positive energy eigenstate of $H^{({\bm{DH}})}_{{\bm{k}}}$ is given as
    \begin{equation}\label{Eq:eigstate_honey}
        \ket{u_{\bm{k},i+4}} = \frac{1}{N_{\bm{k},i}} \left(
        \begin{array}{c}
             \sqrt{E_{\bm{k},i}^{(\bm{M})}} \ket{u_{\bm{k},i}^{(\bm{M})}}   \\
             \ket{u_{\bm{k},i}^{(\bm{dM})}} 
        \end{array}
        \right),
    \end{equation}
    with the eigenenergy
    \begin{equation}
        \varepsilon_{\bm{k},i+4} = \sqrt{E^{(\bm{M})}_{\bm{k},i}}.
    \end{equation}
    Note that $N_{\bm{k},i}=\sqrt{2E_{\bm{k},i}^{(\bm{M})}}$ is the normalization constant. 
    One can explicitly check this as
    \begin{eqnarray}
        H^{(\bm{DH})}_{\bm{k}}\ket{u_{\bm{k},i+4}} &=&\frac{1}{N_{\bm{k},i}} \left(
            \begin{array}{c}
                 \Phi^{\dagger}_{\bm{k}} \ket{u_{\bm{k},i}^{(\bm{dM})}}   \\
                 \sqrt{E_{\bm{k},i}^{(\bm{M})}}  \Phi_{\bm{k}} \ket{u_{\bm{k},i}^{(\bm{M})}} 
            \end{array}
            \right) \nonumber \\
            &=& \frac{1}{N_{\bm{k},i}}\left(
            \begin{array}{c}
                \Phi^{\dagger}_{\bm{k}} \Phi_{\bm{k}} \ket{u_{\bm{k},i}^{(\bm{M})}}   \\
                \sqrt{E_{\bm{k},i}^{(\bm{M})}}  \Phi_{\bm{k}} \ket{u_{\bm{k},i}^{(\bm{M})}} 
            \end{array}
            \right) \nonumber \\
            &=& \frac{1}{N_{\bm{k},i}}\left(
                \begin{array}{c}
                    h^{(\bm{M})}_{\bm{k}} \ket{u_{\bm{k},i}^{(\bm{M})}}   \\
                    \sqrt{E_{\bm{k},i}^{(\bm{M})}} \ket{u_{\bm{k},i}^{(\bm{dM})}} 
                \end{array}
                \right) \nonumber \\
            &=& \frac{1}{N_{\bm{k},i}}\left(
                \begin{array}{c}
                    E_{\bm{k},i}^{(\bm{M})} \ket{u_{\bm{k},i}^{(\bm{M})}}   \\
                    \sqrt{E_{\bm{k},i}^{(\bm{M})}} \ket{u_{\bm{k},i}^{(\bm{dM})}} 
                \end{array}
                \right) \nonumber \\
            &=&\sqrt{E_{\bm{k},i}^{(\bm{M})}}\ket{u_{\bm{k},i+4}},
    \end{eqnarray}
    where we use Eq.~(\ref{eq:martini-dmartini}) 
    on the second and third lines.
    We note that, due to the chiral symmetry of Eq.~(\ref{eq:chiral}), 
    the negative-energy eigenstate is given as 
    $\gamma \ket{u_{\bm{k},i+4}}$, and its eigenenergy is $-\sqrt{E_{\bm{k},i}^{(\bm{M})}}$.
    We also note that for the zero-energy state which occurs at K and $\rm{K}'$ points with $i=1$, the above construction of eigenstate can not be applied because $E^{({\bm{M}})}_{{\rm{K}},1}=0$. The zero-energy modes are doubly degenerate and they can be chosen as (for K point)
    \begin{align}
        \label{eq:zero-energy_eigenstate}
        \ket{u_{{\rm{K}},4}}=\left(
            \begin{array}{c}
                 \ket{u_{{\rm{K}},1}^{(\bm{M})}}   \\
                 \bm{0}
            \end{array}
            \right),
            \ket{u_{{\rm{K}},5}}=\left(
            \begin{array}{c}
                 \bm{0}   \\
                 \ket{u_{{\rm{K}},1}^{(\bm{dM})}} 
            \end{array}
            \right).
    \end{align}
    Here $\bm{0}$ is a four-component zero vector.
    \section{Topological Corner State}
    \label{Sec:edge}
    In this section, we analyze the higher-order topology of the martini and the decorated honeycomb lattice model, putting emphasis on the corner-shape dependence as well as the consequences of the square-root relation.
    \subsection{Martini lattice}
\begin{figure*}[!tb]
        \centering
        \begin{minipage}[c]{0.35\hsize}
            \includegraphics[width=\hsize]{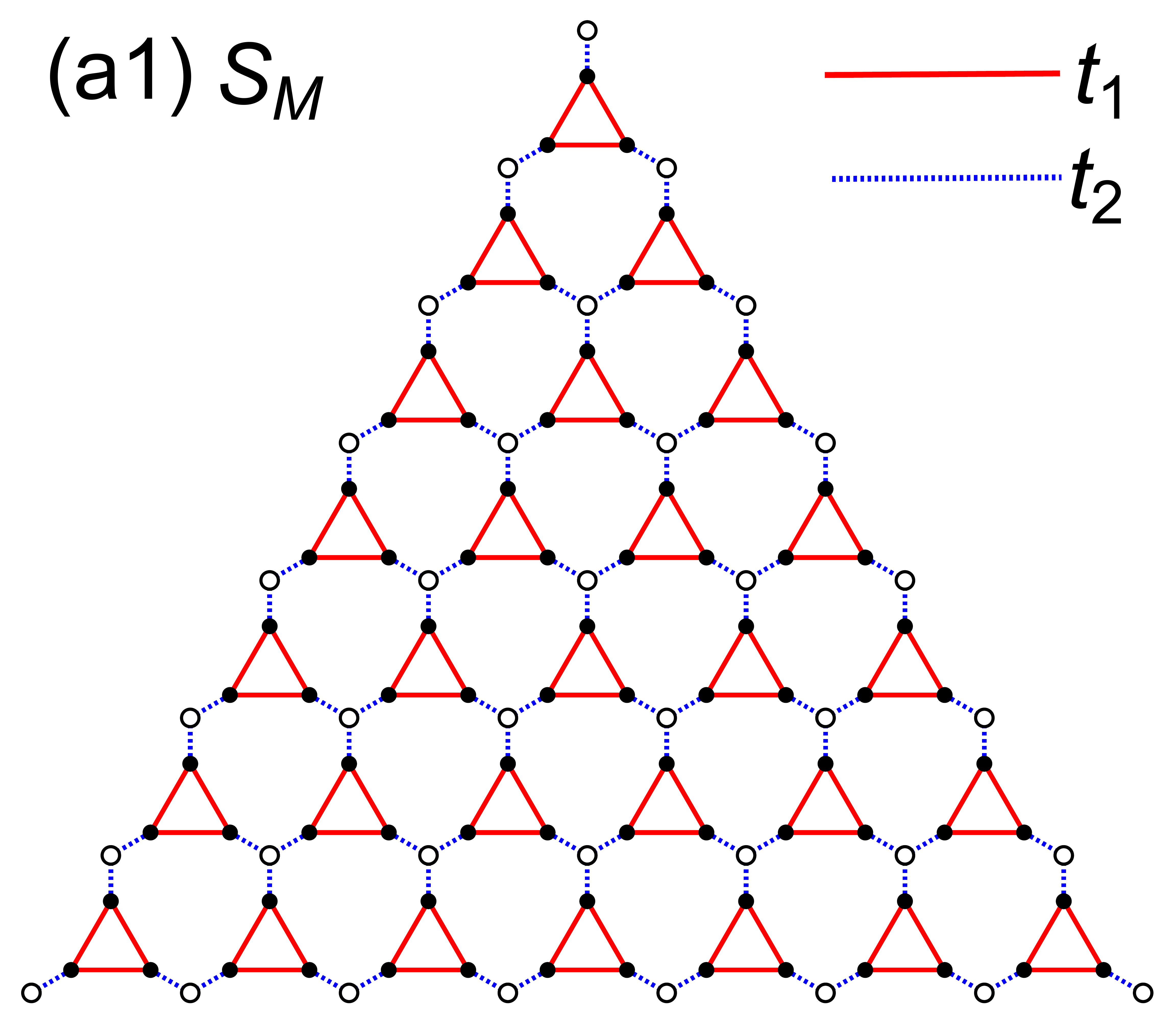}
        \end{minipage}
        \begin{minipage}[c]{0.35\hsize}
            \includegraphics[width=\hsize]{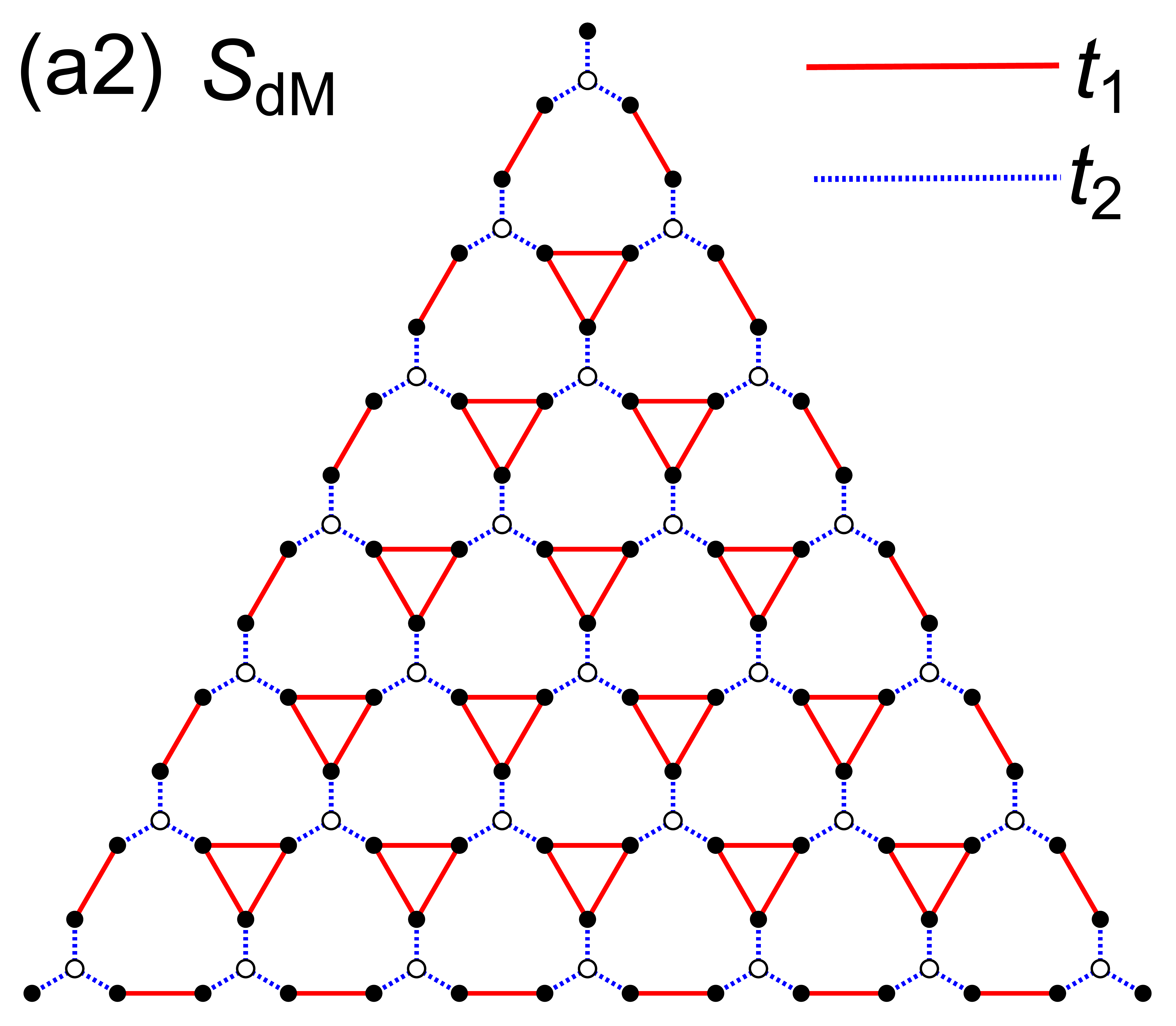}
        \end{minipage}
        \begin{minipage}[c]{0.35\hsize}
            \hspace{-15pt}
            \includegraphics[width=\hsize]{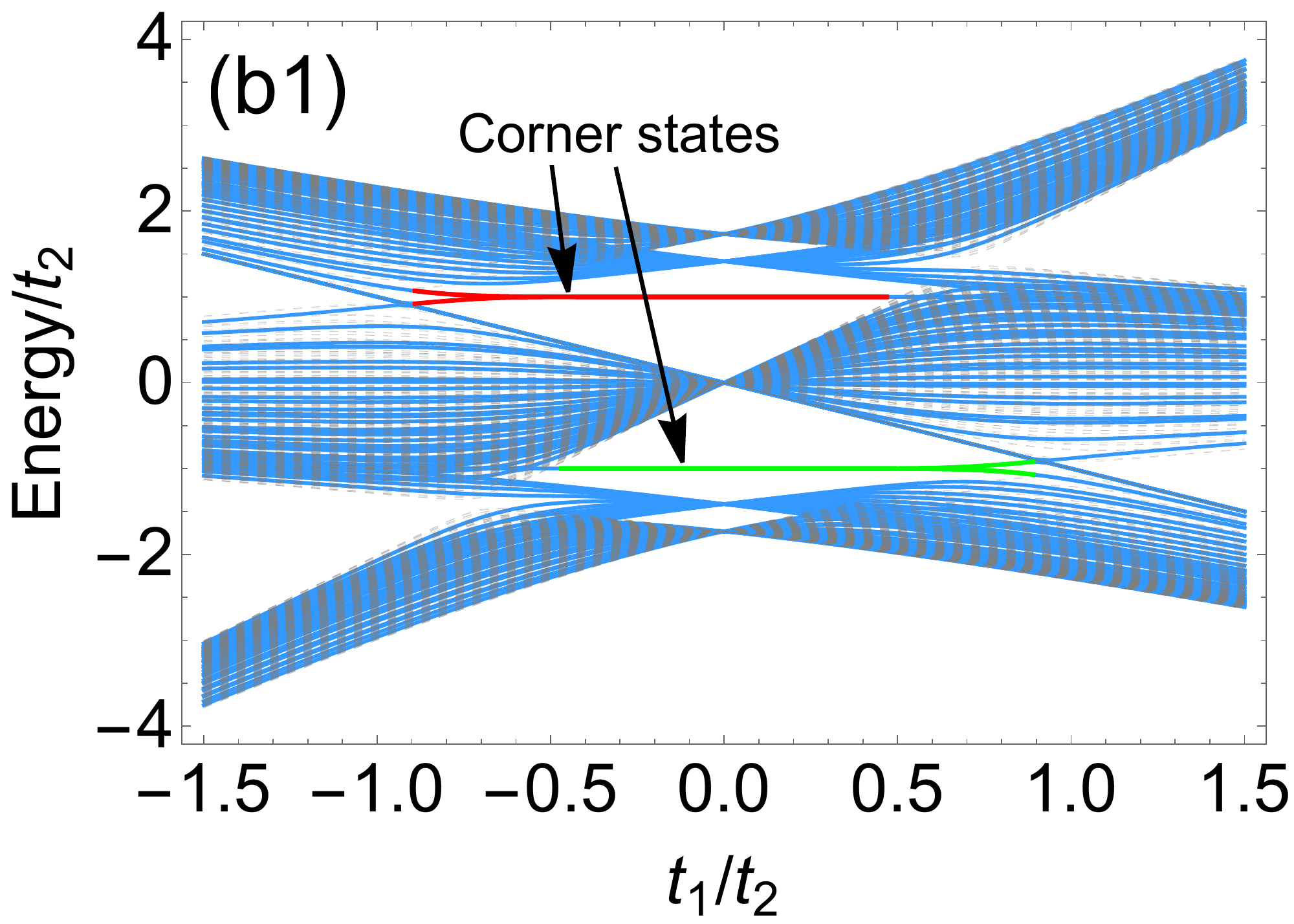}
        \end{minipage}
        \begin{minipage}[c]{0.35\hsize}
            \hspace{-15pt}
            \includegraphics[width=\hsize]{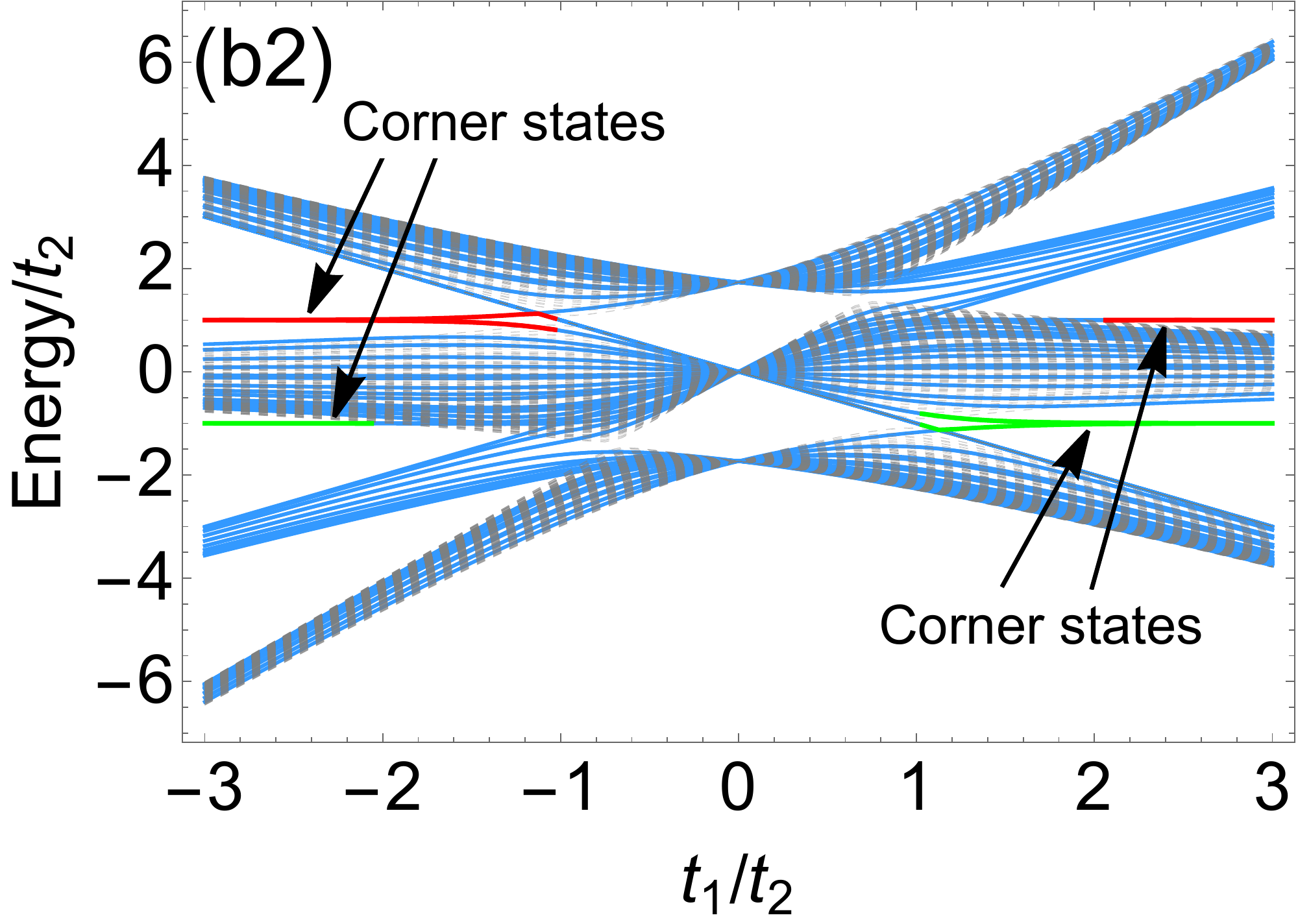}
        \end{minipage}
        \\
        \begin{minipage}[c]{0.35\hsize}
            \hspace{-15pt}
            \includegraphics[width=\hsize]{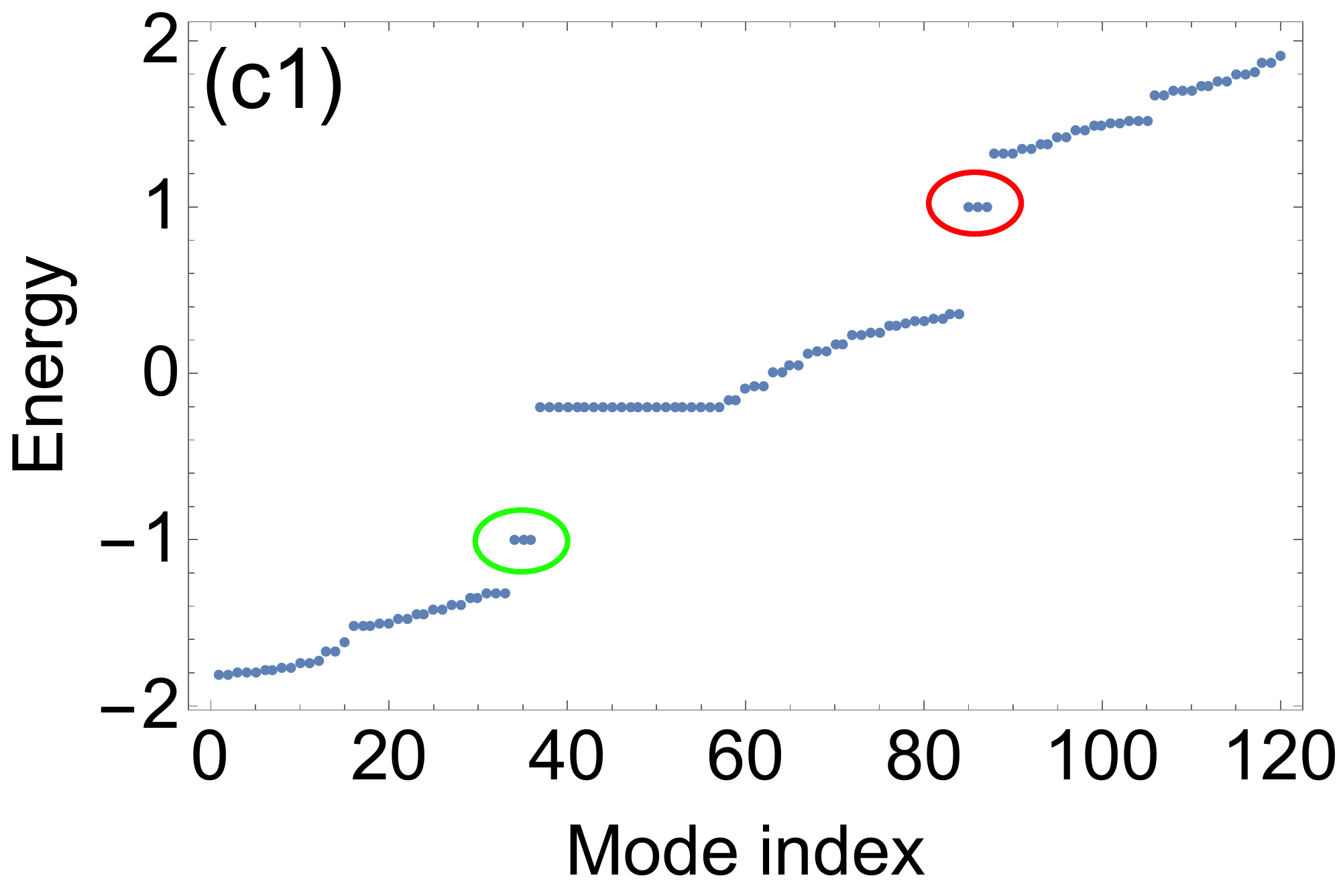}
        \end{minipage}
        \begin{minipage}[c]{0.35\hsize}
            \hspace{-15pt}
            \includegraphics[width=\hsize]{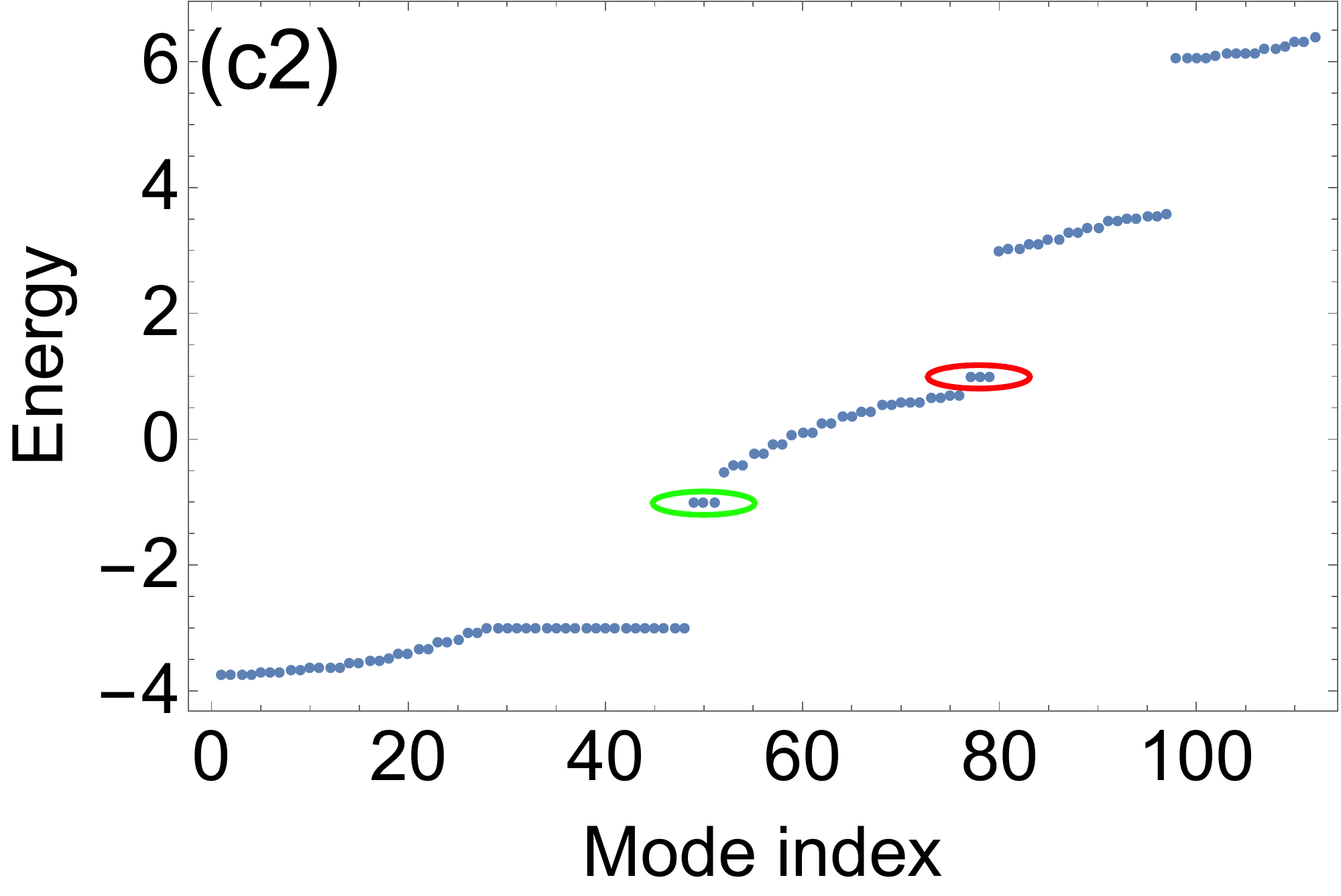}
        \end{minipage}
        \\
        \begin{minipage}[c]{0.22\hsize}
            \includegraphics[width=\hsize]{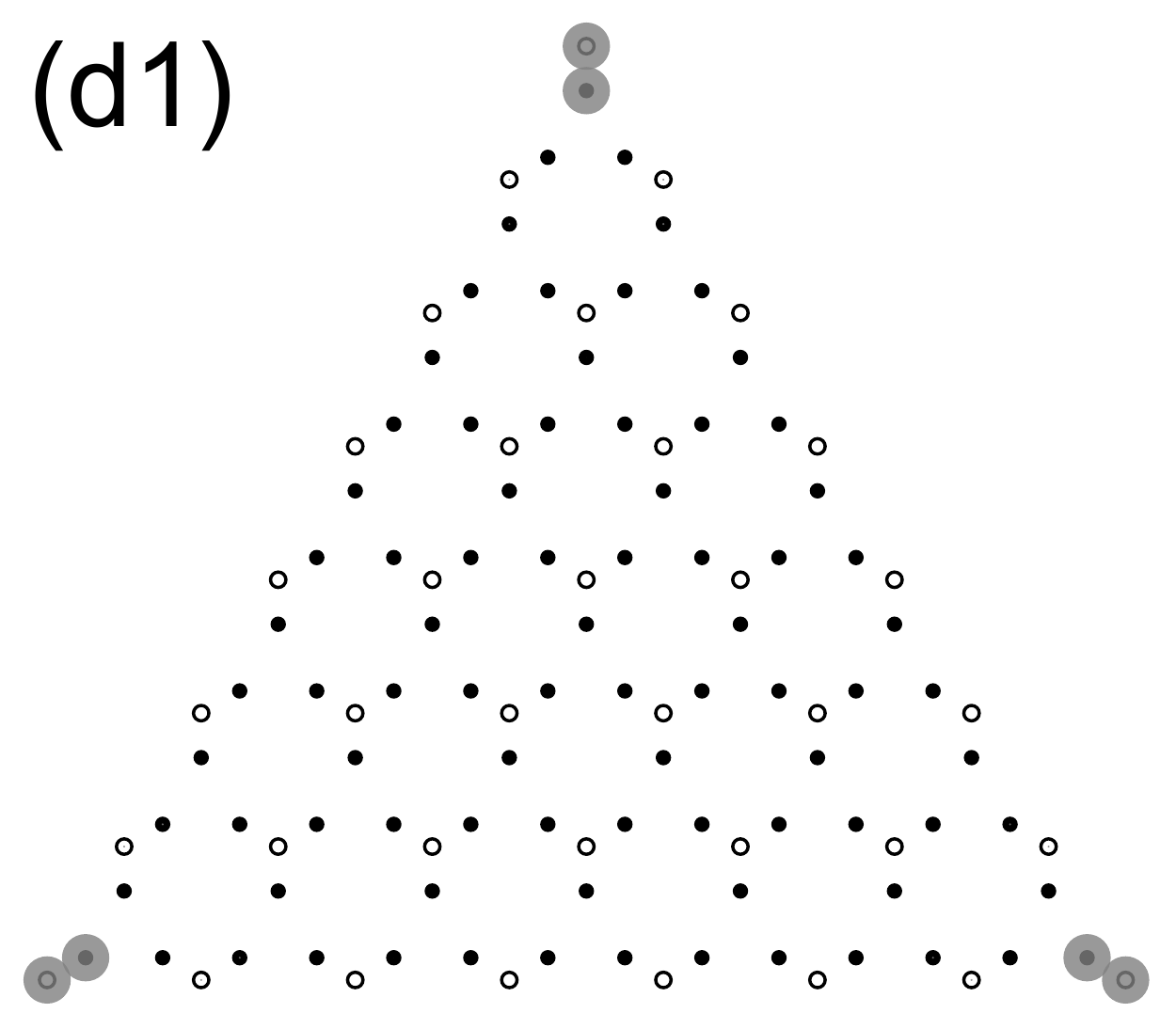}
        \end{minipage}
        %\hspace{10pt}
        \begin{minipage}[c]{0.22\hsize}
            \includegraphics[width=\hsize]{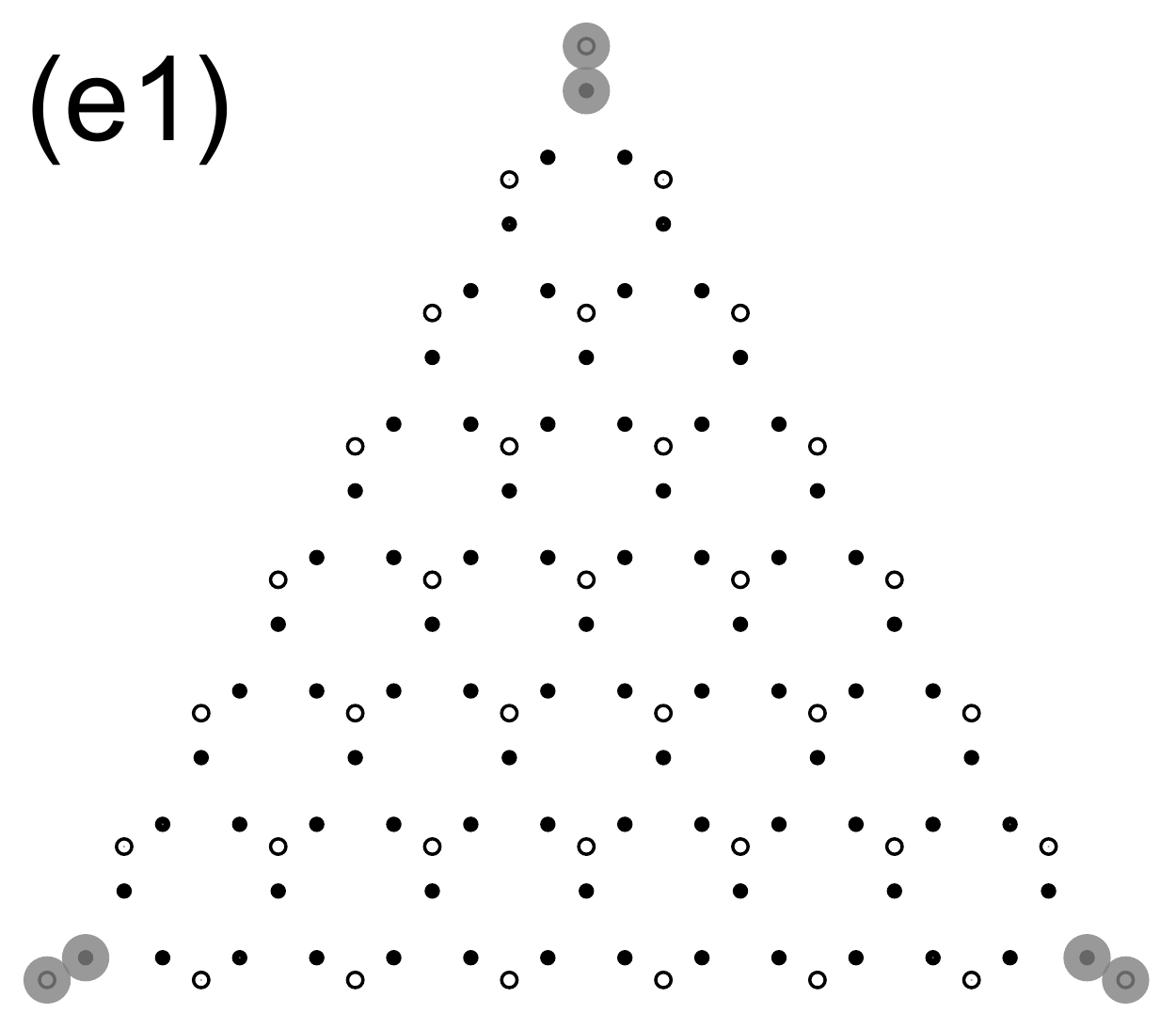}
        \end{minipage}
        \begin{minipage}[c]{0.22\hsize}
            \includegraphics[width=\hsize]{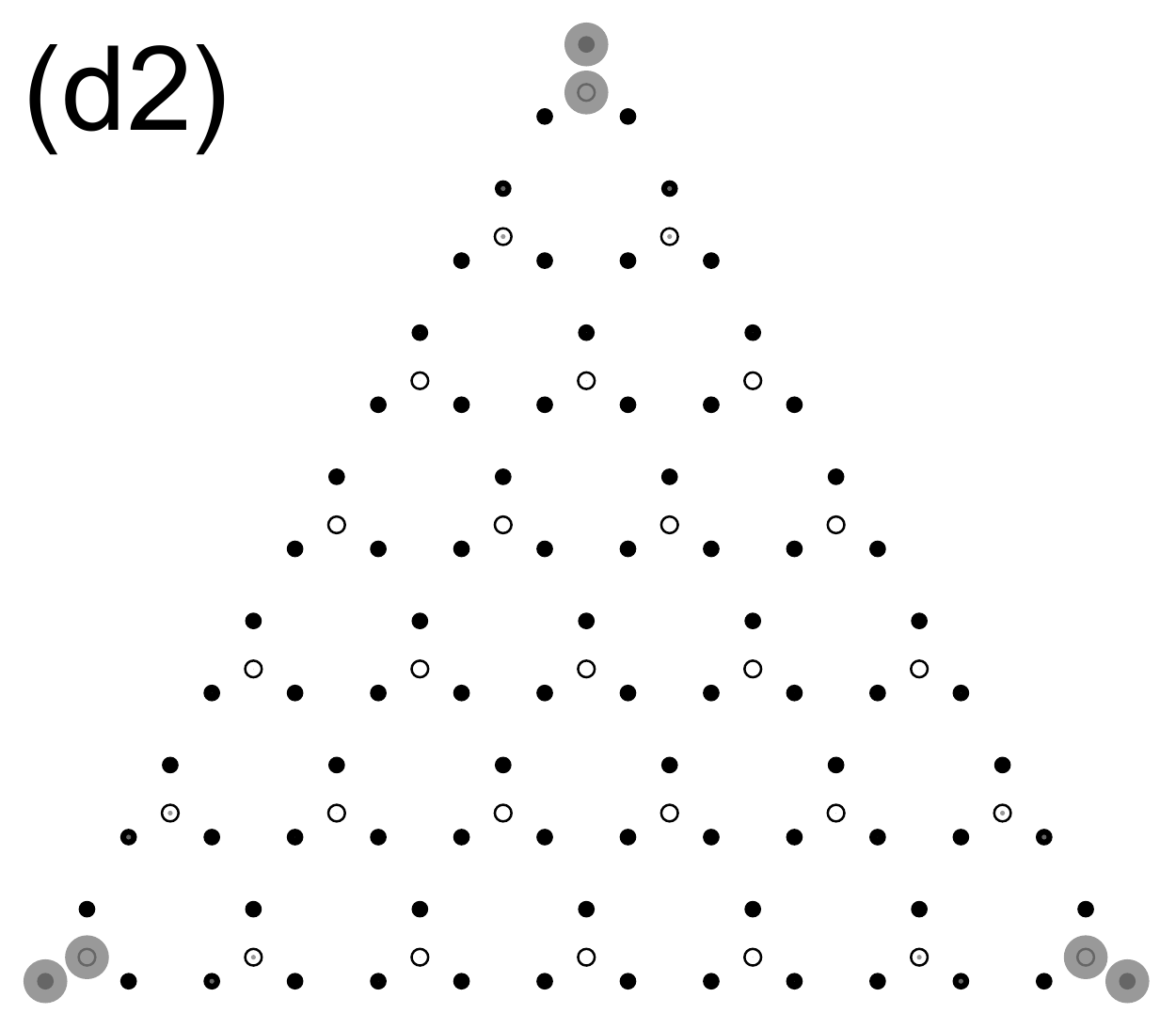}
        \end{minipage}
        %\hspace{10pt}
        \begin{minipage}[c]{0.22\hsize}
            \includegraphics[width=\hsize]{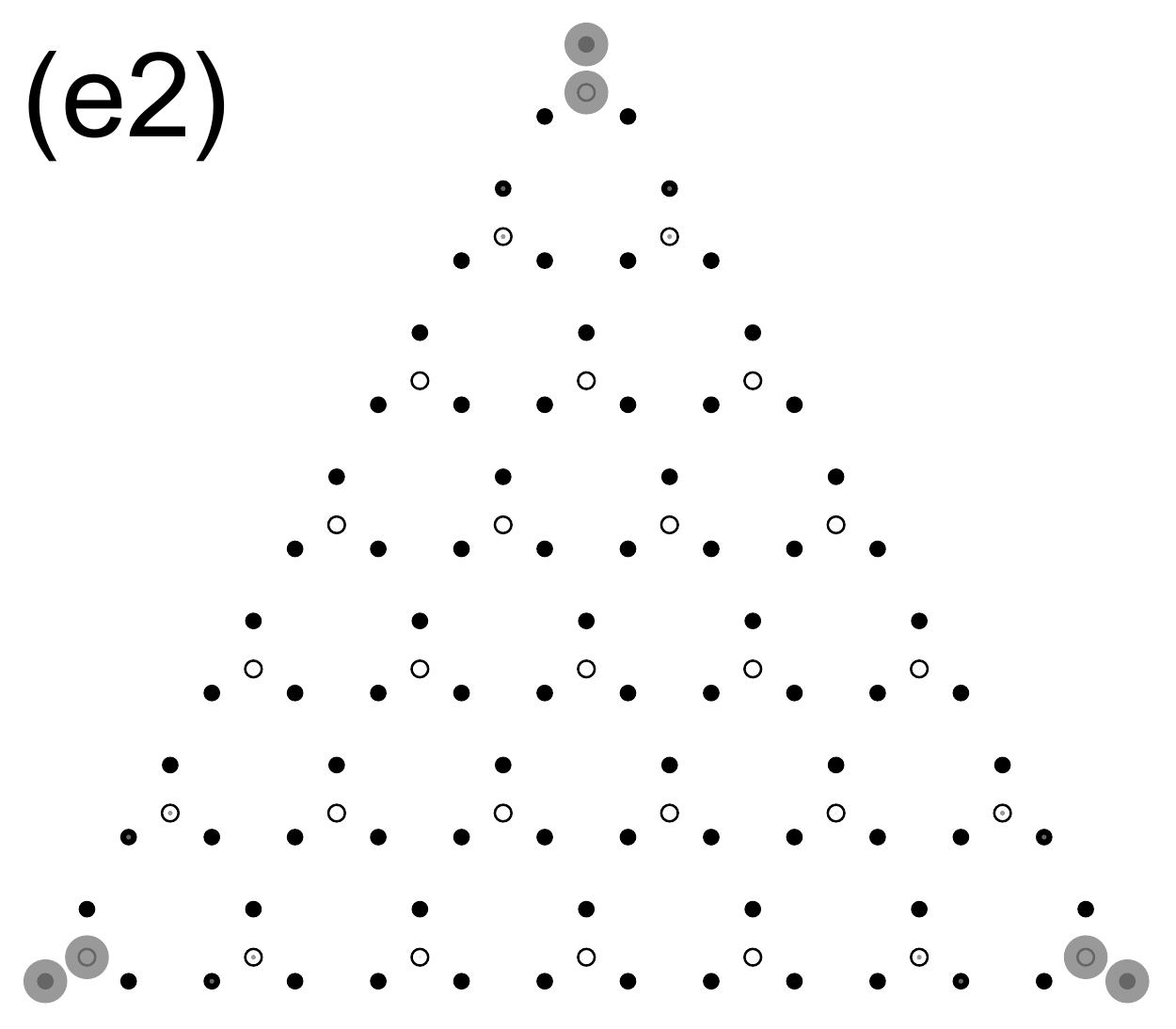}
        \end{minipage}
        \begin{minipage}[l]{0.25\hsize}
            \includegraphics[width=\hsize]{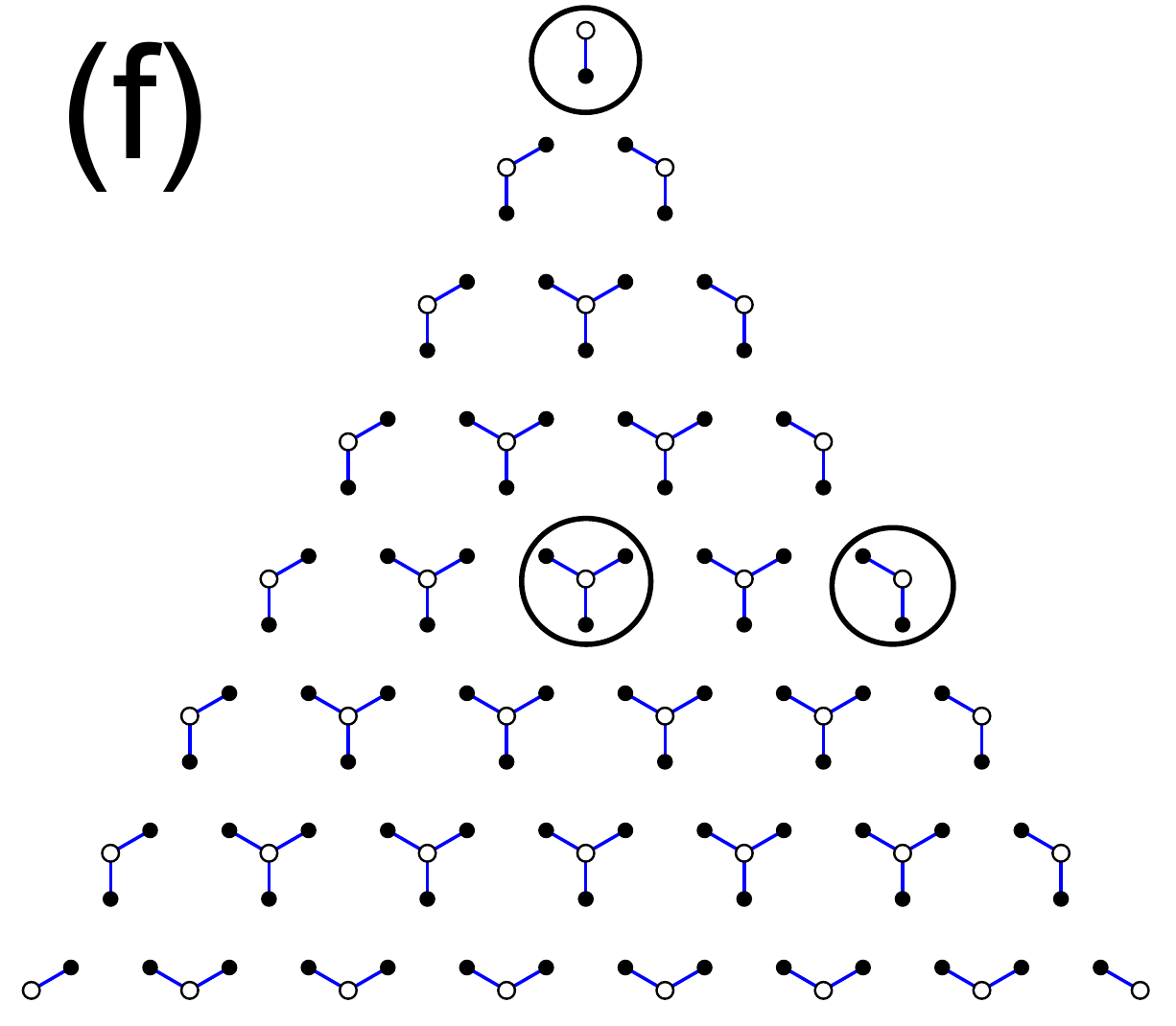}
        \end{minipage}
        \begin{minipage}[r]{0.34\hsize}
            \includegraphics[width=\hsize]{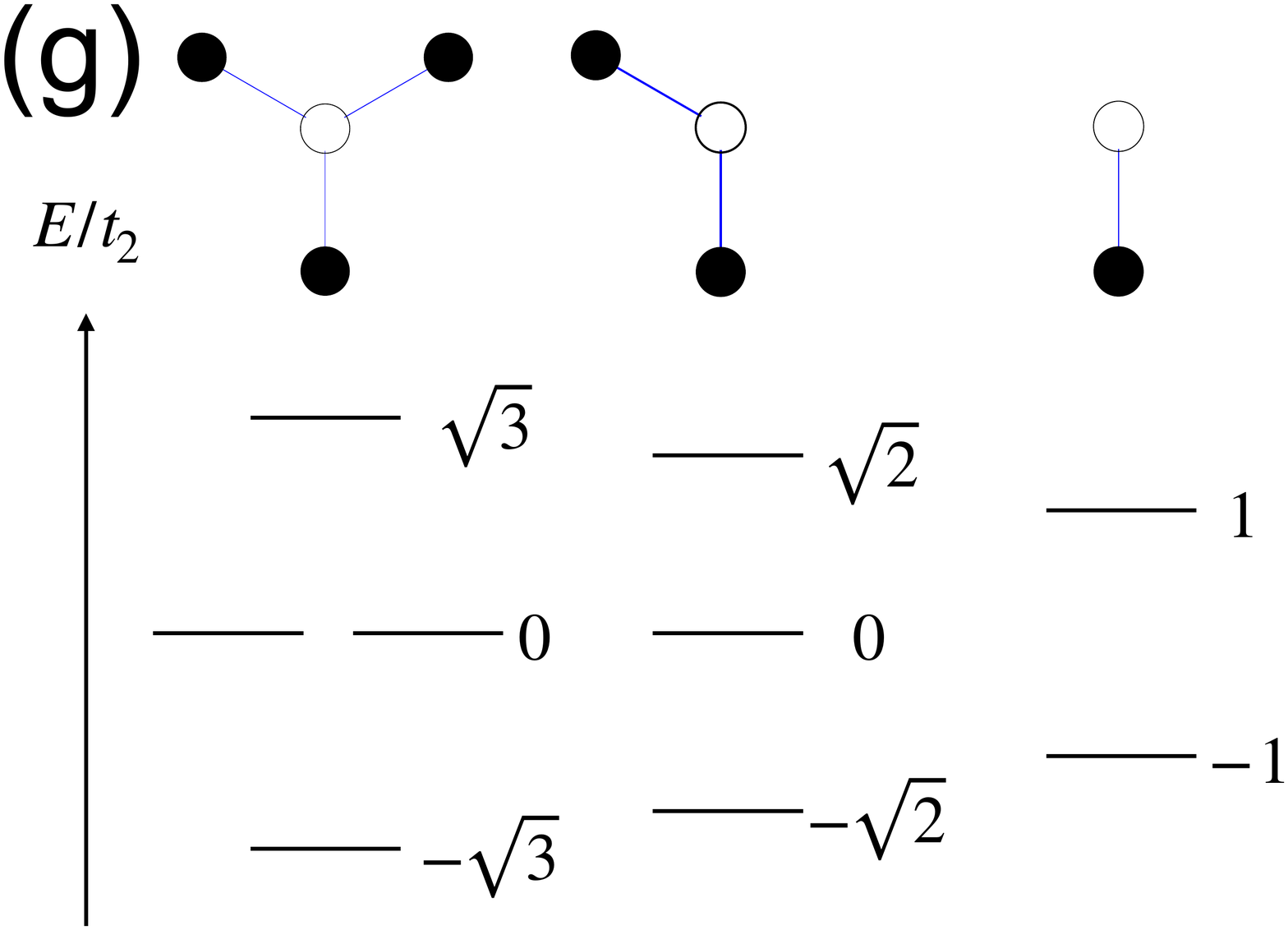}
        \end{minipage}\begin{minipage}[r]{0.25\hsize}
            \includegraphics[width=\hsize]{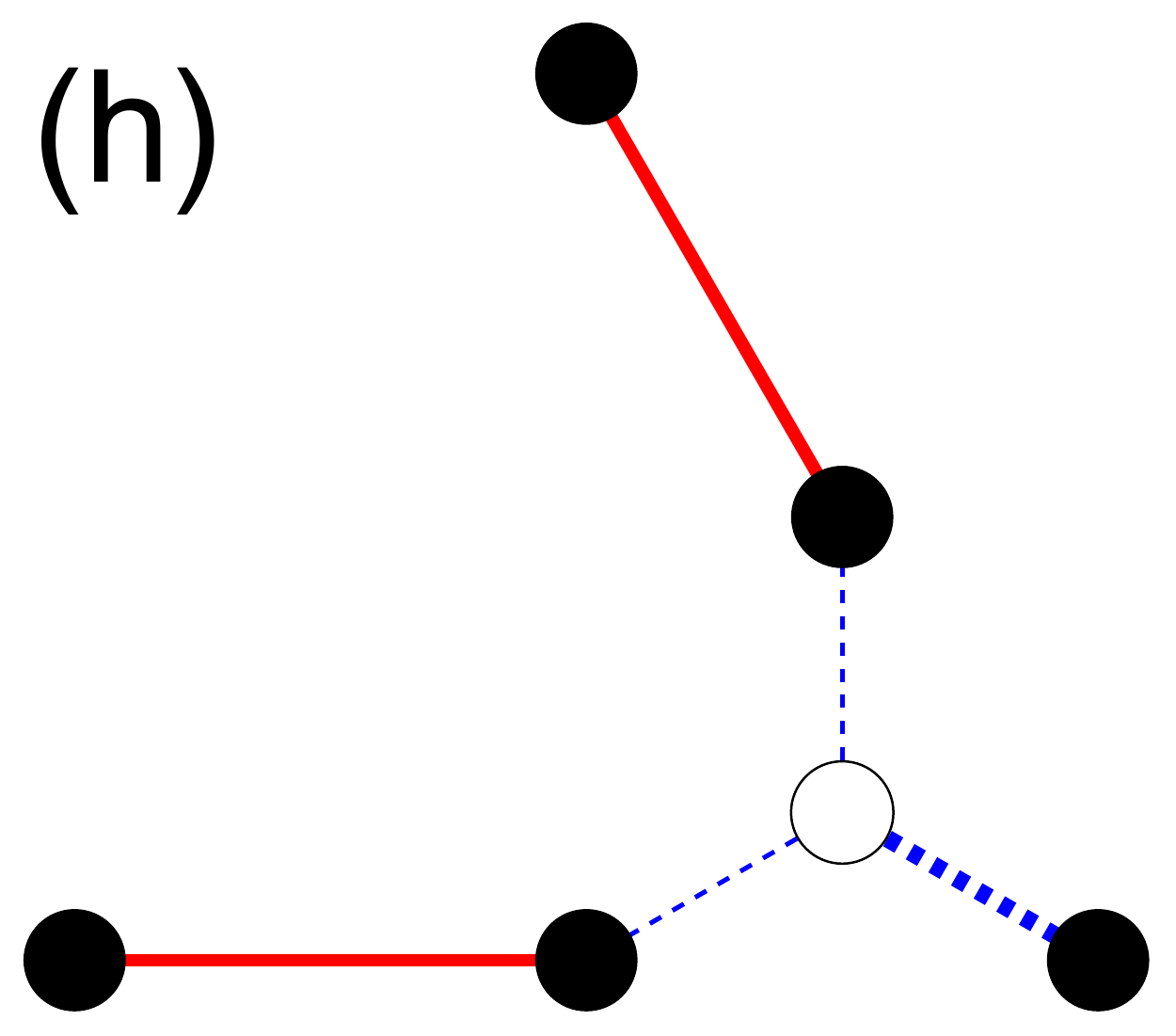}
        \end{minipage}
        \caption{(Color online)(a1-2) Finite systems under OBC, consisting of 84 black sites and 36 white sites(a1) and 84 black sites and 28 white sites(a2). 
        (b1-2) The energy spectrums. The horizontal axis is $t_1/t_2$.The  dashed line and the solid line correspond to the spectrum of periodic boundary conditions (PBC) and OBC respectively. The green and red lines correspond to the in-gap states with negative and positive energies, respectively.
        (c1) The energy spectrum of \textit{$S_M$} for  $t_1=0.2, t_2=1$. 
        (c2) The energy spectrum of \textit{$S_{dM}$} for  $t_1=3, t_2=1$. The in-gap corner states are encircled by green (lower) and red (upper) ellipses.
        The probability density distribution of
        (d1-2) the lower in-gap states and
        (e1-2) the upper in-gap states on \textit{$S_M$} and \textit{$S_{dM}$}. 
        The radii of gray circles represent the probability density.
        (f) The schematic picture of adiabatic limit of $t_1=0$. A tetramer, a trimer, and a dimer are encircled.
        (g) The energy levels of the tetramer, the trimer, and the dimer.
        (h) The schematic picture of a corner of \textit{$S_{dM}$} when $t_1 \gg t_2$.
        }
        \label{fig:martini_obc}
    \end{figure*}

    We first argue the higher-order topological phase of the martini model.
    We first examine the existence of the corner states, paying particular attention to the corner-shape dependence of the boundary modes. 
    To this end, we study two finite samples formed into triangle shape with different corner termination under the open boundary conditions (OBC), shown in Fig.~\ref{fig:martini_obc}(a1) and (a2).
    We call the finite sample shown in Fig.~\ref{fig:martini_obc}(a1) \textit{$S_M$} and that in \ref{fig:martini_obc}(a2) \textit{$S_{dM}$}.
    The energy spectrums of \textit{$S_M$} and \textit{$S_{dM}$} as a function of $t_1/t_2$ are shown in Fig.~\ref{fig:martini_obc}(b1) and (b2).
    In Fig.~\ref{fig:martini_obc}(b1), we see that there appear two in-gap states for $-1 < t_1/t_2 < 1/2$ (with positive energy) and for $-1/2 < t_1/t_2 < 1$ (with negative energy). 
    On the other hand, in Fig.~\ref{fig:martini_obc}(b2), we see that there appear two in-gap states for $ t_1/t_2 < -1, t_1/t_2 >2$ (with positive energy) and for $t_1/t_2 < -2, t_1/t_2 >1$ (with negative energy).
    As we have seen in Fig.~\ref{fig:martini_band},
    $t_1/t_2=-1$ and $t_1/t_2=1/\sqrt{2}$ ($t_1/t_2=-1/\sqrt{2}$ and $t_1/t_2=1$) are 
    the gap-closing points for the lower (upper) gap, indicating the topological phase transition at these points.
    In Fig.~\ref{fig:martini_obc}(c1) and (c2), we plot the energy spectrums of \textit{$S_M$} and \textit{$S_{dM}$} at $(t_1,t_2)=(0.2,1)$ and $(t_1,t_2)=(3,1)$. Green and red ellipses are the in-gap states with negative and positive energies, respectively. 
    We see that each in-gap state has three-fold degeneracy, which is because the sample hosts three corners.
    
    To further reveal the real-space properties of these in-gap states, we plot the real-space probability density distribution of these in-gap states in Fig.~\ref{fig:martini_obc}(d1) and (d2) for negative energy modes and Fig.~\ref{fig:martini_obc}(e1) and (e2) for positive energy modes.
    Note that we take the average over the three degenerate states.
    We see that the in-gap states have large amplitude at the corner and hence are corner states.
    \clearpage
    \clearpage
    It is worth noting that the corner states appear at finite energy.
    This is in contrast to the breathing kagome model where the corner state has zero energy\cite{Ezawa2018}.
    This can be understood by the following discussion of adiabatic limit\cite{Ezawa2018}.
    Namely, we focus on the case of $t_1=0$ which can be connected to HOTI phase without closing the band gap.
    When $t_1=0$, \textit{$S_M$} consists of the tetramers in the bulk, the trimers at the edges, and the three dimers at the corners as shown in Fig.~\ref{fig:martini_obc}(f), whereas \textit{$S_{dM}$} consists of the tetramers.
    The tetramers have the energy $\pm \sqrt{3}t_2$ and two-fold zero energy, the trimers have the energy $\pm \sqrt{2}t_2$ and zero energy, and the dimers have the energy $\pm t_2$ [Fig.~\ref{fig:martini_obc}(g)].
    As the energy levels of corner dimers are separated from those of tetramers and trimers, 
    these dimers serve as in-gap corner states. This holds even for finite $t_1$ as far as the band gap remains open, which accounts for the emergence of finite-energy corner states in \textit{$S_M$}.
    The appearance of the corner modes of \textit{$S_{dM}$} is understood as follows. When $t_1 \gg t_2$, at a corner of the sample,
    two sites are not connected to any of the red ($t_1$) bonds. 
    These two sites again form a dimer with the hopping $t_2$, resulting in the corner states with $E= \pm t_2$. See Fig.~\ref{fig:martini_obc}(h) for the schematic figure of the above argument.
    
\begin{figure}[!t]
        \centering
        \begin{minipage}[c]{0.49\hsize}
            \includegraphics[width=\hsize]{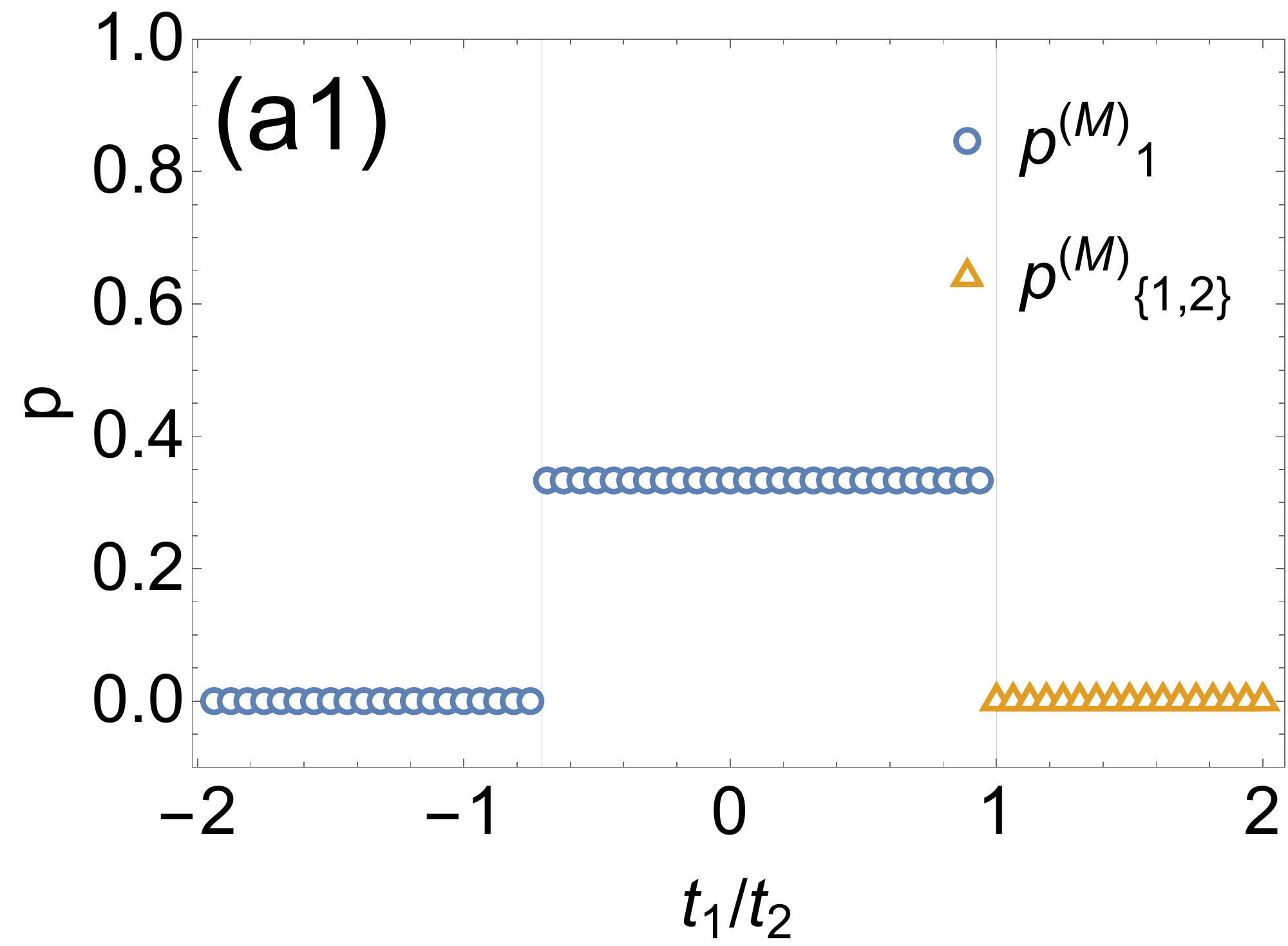}
        \end{minipage}
        \begin{minipage}[c]{0.49\hsize}
            \includegraphics[width=\hsize]{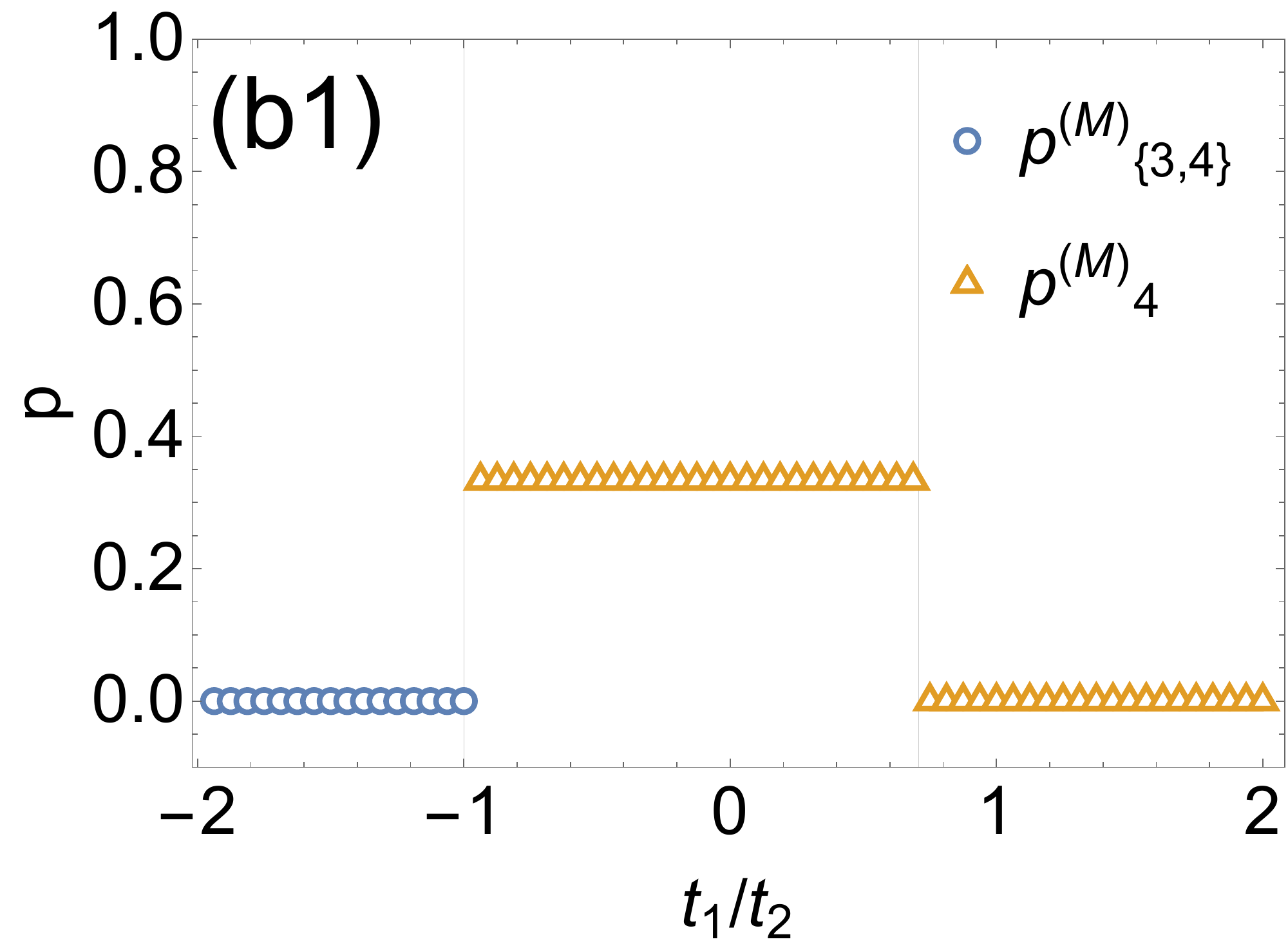}
        \end{minipage}
        \begin{minipage}[c]{0.49\hsize}
            \includegraphics[width=\hsize]{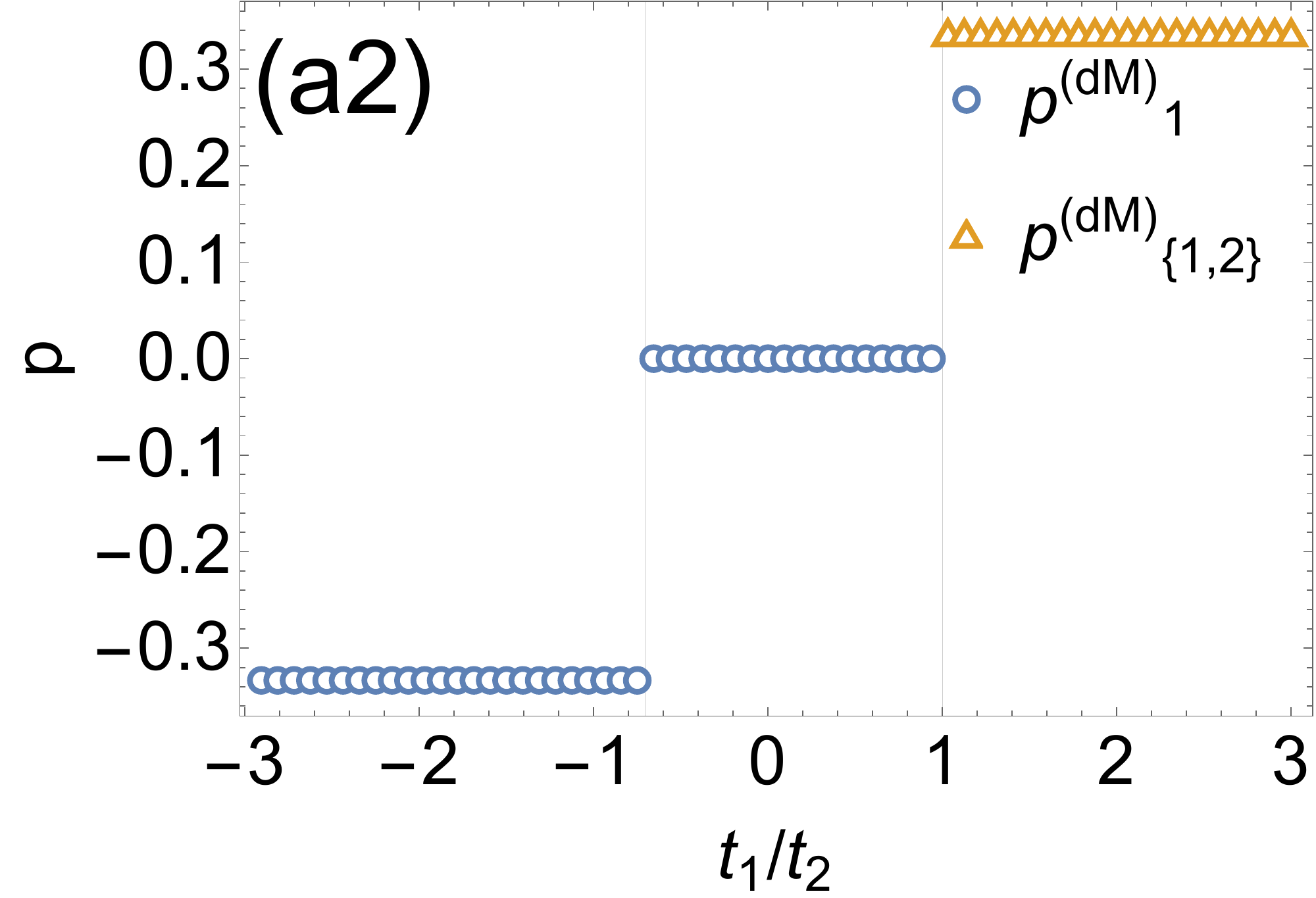}
        \end{minipage}
        \begin{minipage}[c]{0.49\hsize}
            \includegraphics[width=\hsize]{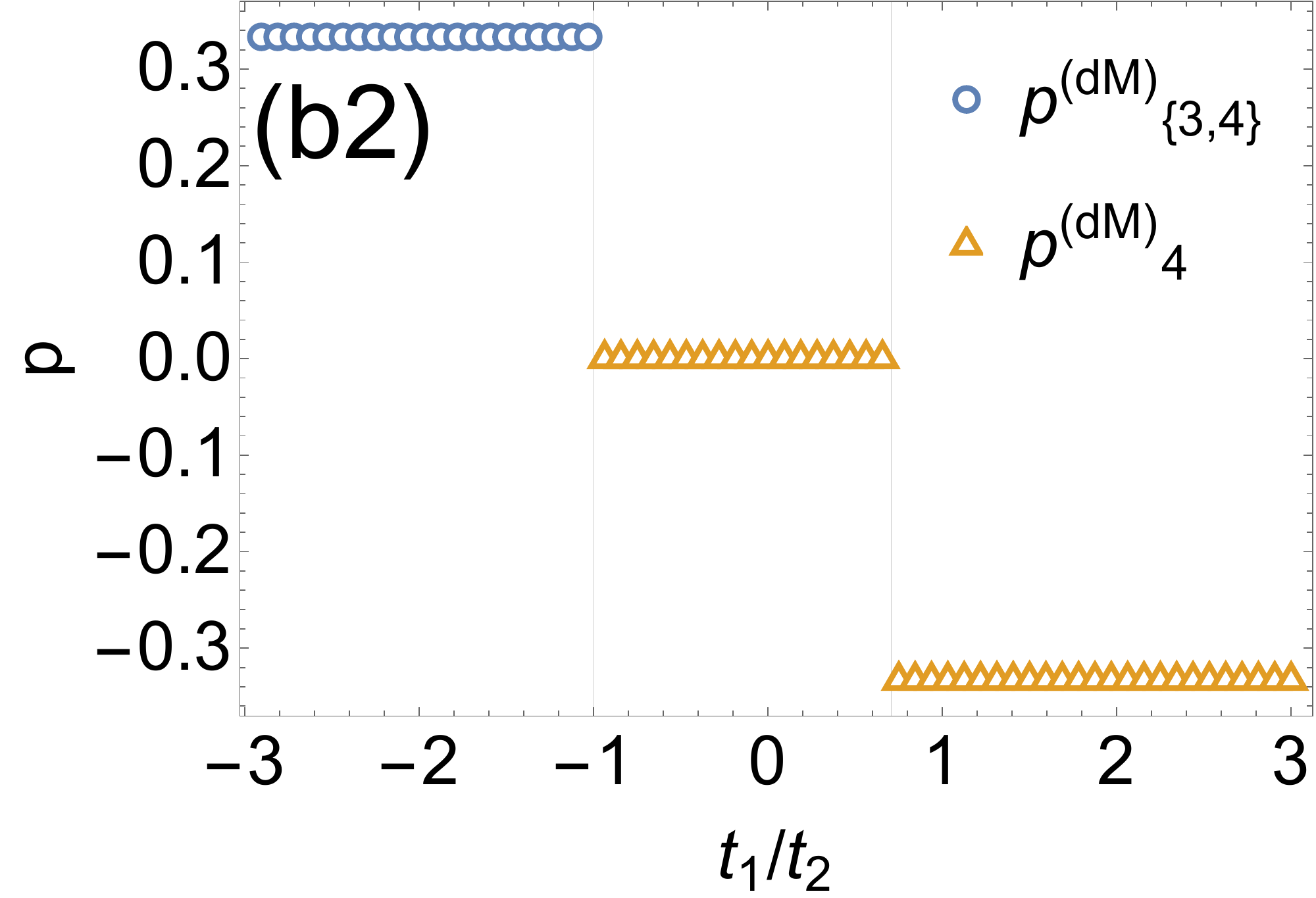}
        \end{minipage}
        \caption{(Color online)Numerical results of the polarization p of Eq. (\ref{Eq:polarization_sigle}) and (\ref{Eq:polarization_multiple}) as a function of $t_1/t_2$. 
        (a1-2) Blue circles are for the first band. Orange triangles are for the first and second bands. 
        (b1-2) Blue circles are for the third fourth band. Orange triangles are for the fourth band.}
        \label{fig:polarization_martini}
    \end{figure}

    Next, we discuss the topological origin of the in-gap corner states.
    Here, we use the $\mathbb{Z}_3$ topological invariant $p_n$, which is related to polarization\cite{Fang2012,Ezawa2018,Mizoguchi2020_sq}.
    For $C_3$-symmetric system, it takes the form:
    \begin{equation}
        \label{Eq:polarization_sigle}
        2\pi p_n = \arg{\xi_n({\rm{K}})} \ (\rm{mod} \ 2\pi),
    \end{equation}
    where
    \begin{equation}
        \label{Eq:xi_single}
        \xi_n(\bm{k}) = \bra{u_{\bm{k},n}}U_{\bm{k}}\ket{u_{\bm{k},n}}.
    \end{equation}
    Here $\ket{u_{{\bm{k}},n}}$ is an eigenstate of Hamiltonian $H_{\bm{k}}$ with band index $n$ and $U_{\bm{k}}$ is a $C_3$ operator that satisfies $H_{C_3 {\bm{k}}} = U_{{\bm{k}}} H_{{\bm{k}}} U^{\dagger}_{{\bm{k}}}$.   
    At K point, $H_{\bm{k}}$ and $U_{\bm{k}}$ are commutable since K point is a $C_3$-invariant momentum.
    This results in a $\mathbb{Z}_3$ quantization of $p_n$, $p_n = l/3$ with $l=0,1,2$.
    More generally, when the target bands have degeneracy points and the isolated band is not well-defined on such points, we can define the $p$ for a set of $N$ bands $\mathcal{N}=\{ N_1,N_2,\cdots,N_N \}$ as 
    \begin{equation}
        \label{Eq:polarization_multiple}
        2\pi p_{\mathcal{N}} = \arg{[\det{\Xi({\rm{K}})}]} \ (\rm{mod} \ 2\pi),
    \end{equation}
    where $\Xi_{i,j}(\bm{k})$ is the $N \times N$ matrix whose $i,j$ component is given as
    \begin{equation}
        \Xi_{i,j}(\bm{k}) = \bra{u_{\bm{k},i}} U_{\bm{k}} \ket{u_{\bm{k},j}} \ (i,j \in \mathcal{N}).
    \end{equation}
    To characterize the lower in-gap corner states, we numerically calculate $p_1$ for $-2 < t_1/t_2 < 1$ and $p_{\{1,2\}}$ for $1<t_1/t_2 <2$. 
    Similarly, to characterize the higher in-gap corner states, we numerically calculate $p_{\{3,4\}}$ for $-2 < t_1/t_2 < 1$ and $p_{4}$ for $1<t_1/t_2 <2$. 
    We plot the numerical result in Fig.~\ref{fig:polarization_martini}. In Fig.~\ref{fig:polarization_martini} (a1) and (b1), we use $U^{(\bm{M})}_{\bm{k}}$ as the $C_3$ operator, on the other hand, in Fig.~\ref{fig:polarization_martini}(a2) and (b2), we use $U^{(\bm{dM})}_{\bm{k}}$ as the $C_3$ operator.
    \begin{figure*}[!b]
        \centering
        \hspace{-15pt}
        \begin{minipage}[c]{0.45\hsize}
            \includegraphics[width=\hsize]{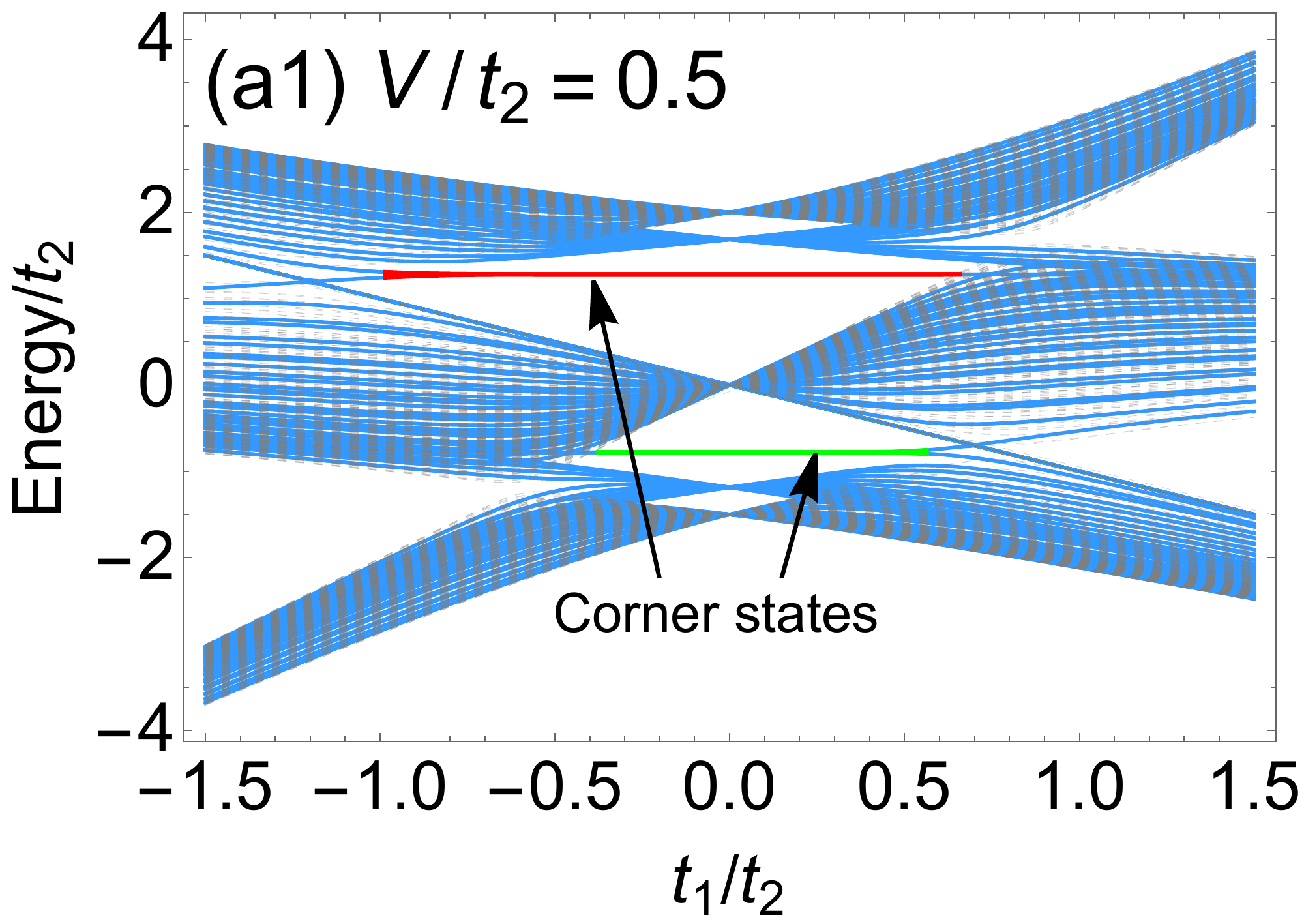}
        \end{minipage}
        \begin{minipage}[c]{0.44\hsize}
            \includegraphics[width=\hsize]{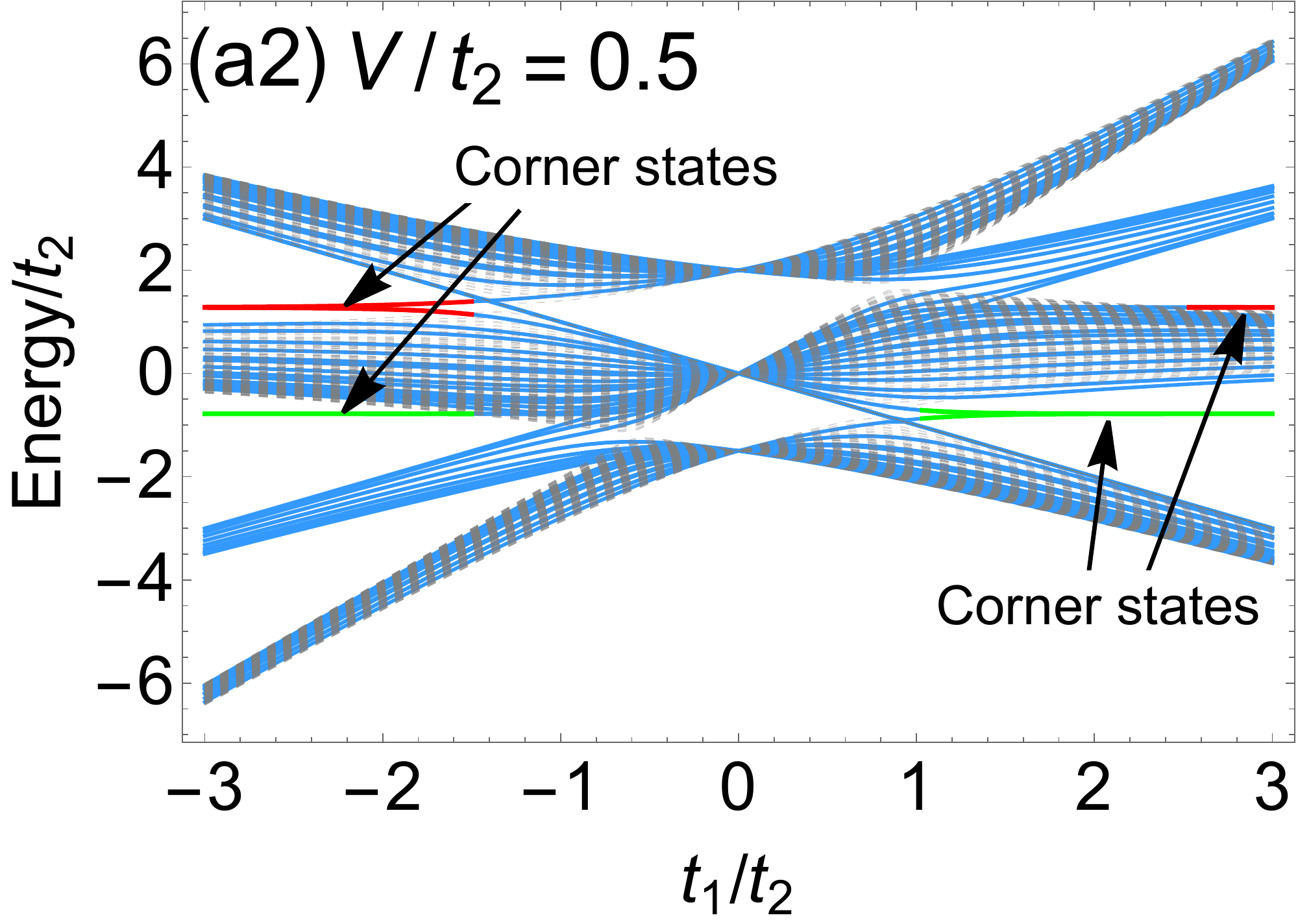}
        \end{minipage}
        \begin{minipage}[c]{0.2\hsize}
            \includegraphics[width=\hsize]{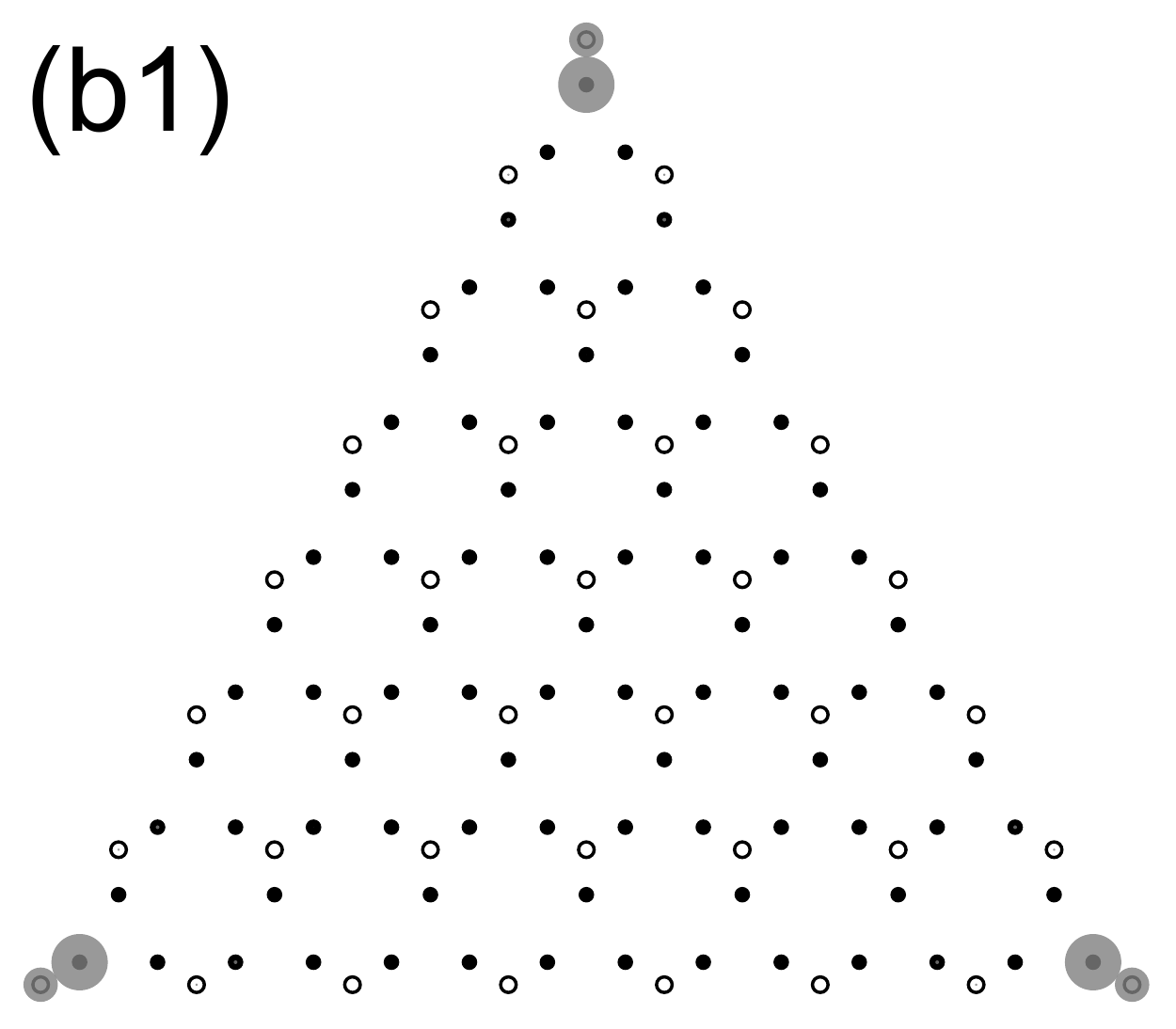}
        \end{minipage}
        \hspace{10pt}
        \begin{minipage}[c]{0.2\hsize}
            \includegraphics[width=\hsize]{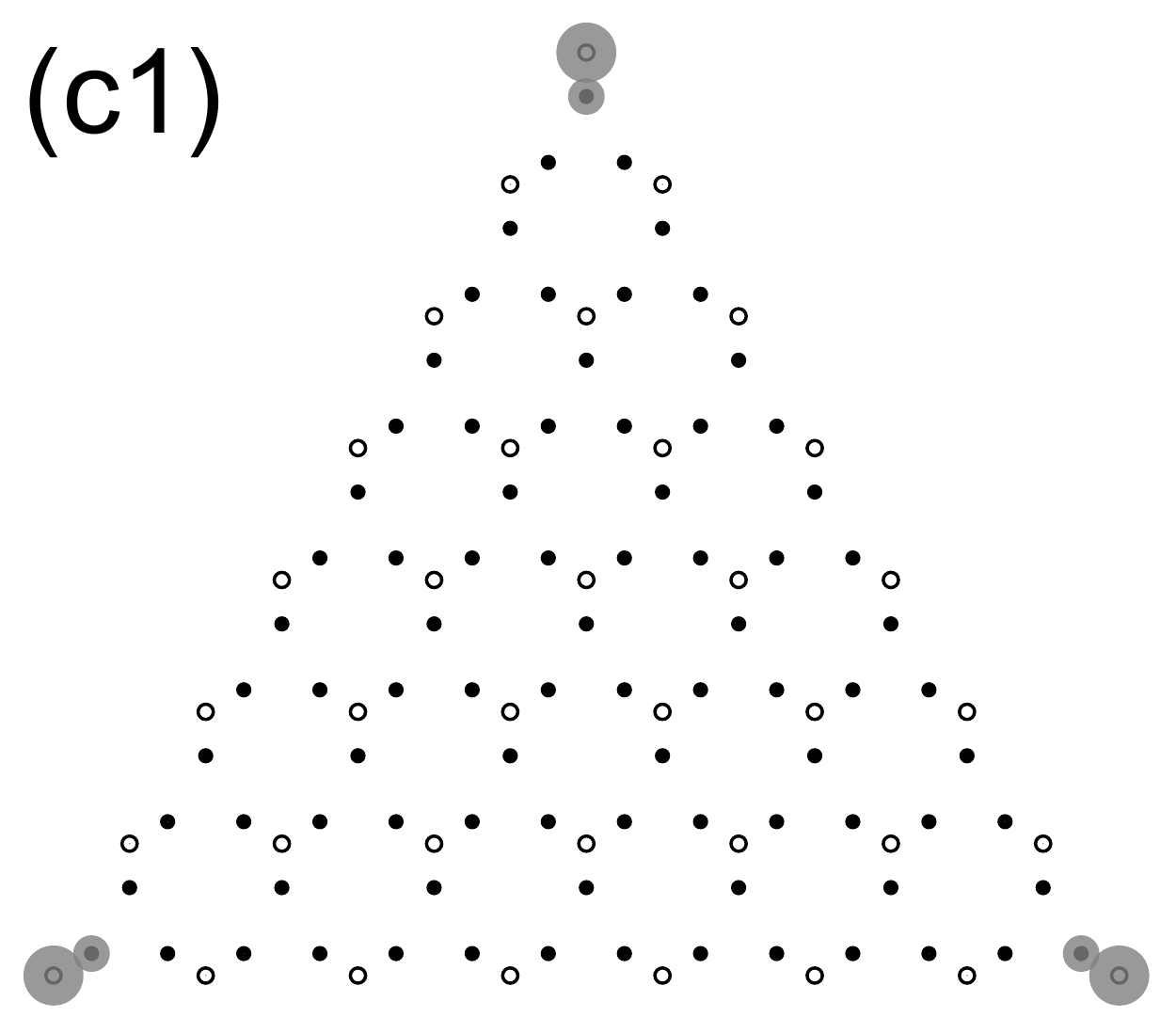}
        \end{minipage}
        \hspace{10pt}
        \begin{minipage}[c]{0.2\hsize}
            \includegraphics[width=\hsize]{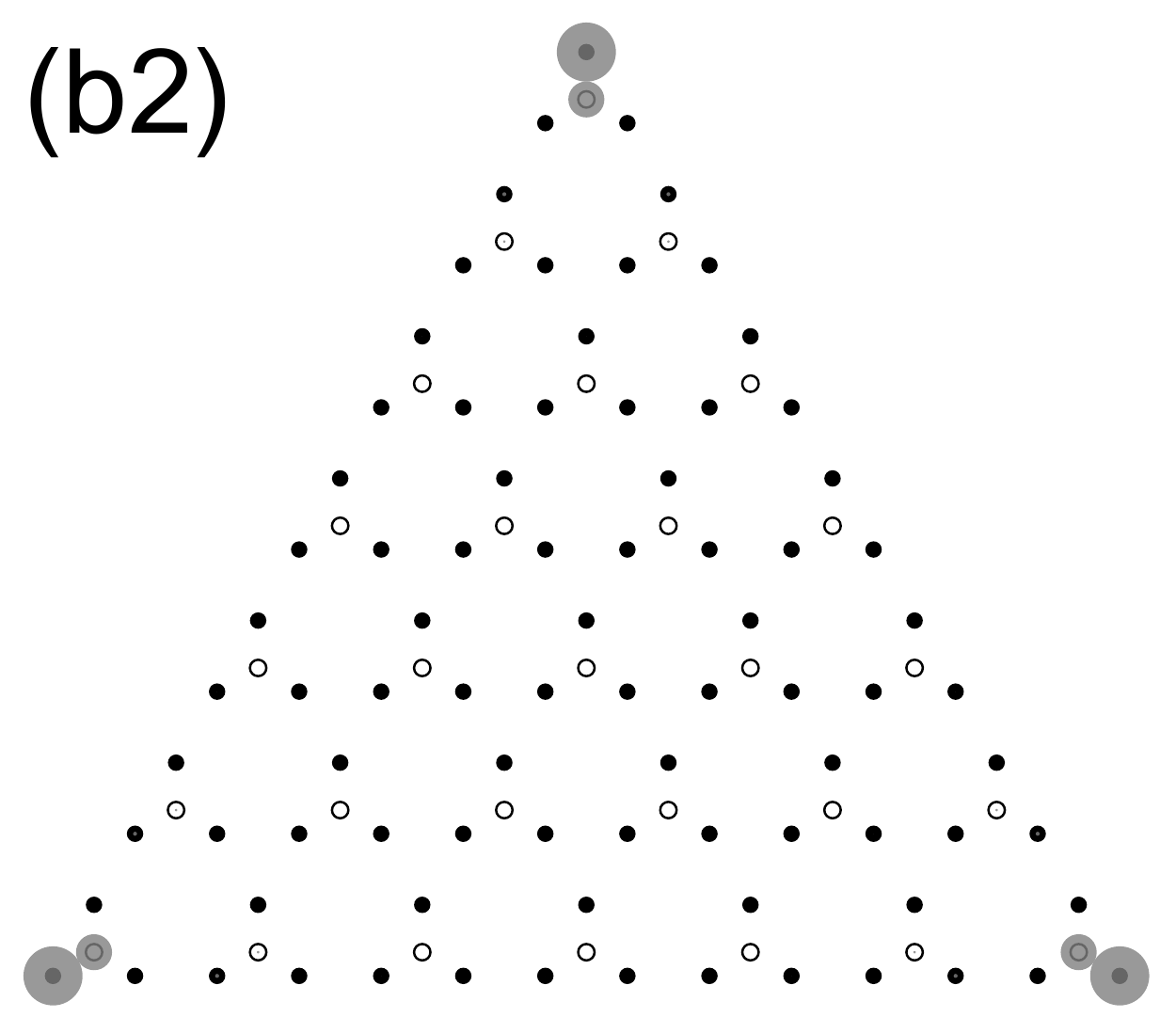}
        \end{minipage}
        \hspace{10pt}
        \begin{minipage}[c]{0.2\hsize}
            \includegraphics[width=\hsize]{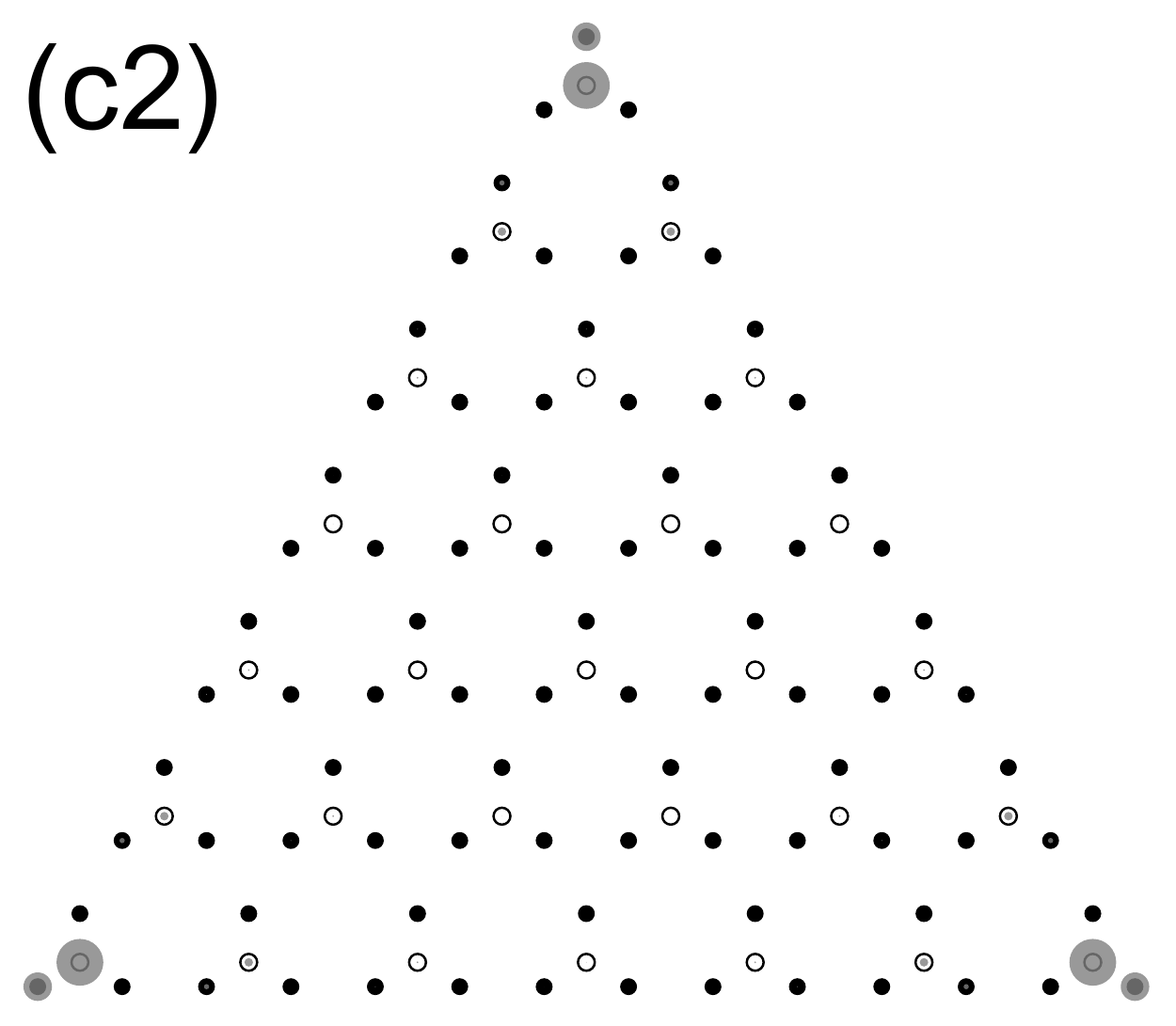}
        \end{minipage}
        \caption{(Color online)(a1-2) The energy spectrums of \textit{$S_M$} and \textit{$S_{dM}$} with $V=0.5$. The  dashed line and the solid line correspond to the spectrum of PBC and OBC respectively. The green and red lines correspond to the in-gap states  with negative and positive energies, respectively. The probability density distribution of (b1-2) lower in-gap states and (c1-2) upper in-gap states in (a1-2).}
        \label{fig:martini_obc_v05}
    \end{figure*}
    We see that $p$ is equal to $1/3$ for the parameters where the in-gap states appear, 
    namely, $-1/2 < t_1/t_2 <1$ ($t_1/t_2 <-2, t_1/t_2 >1$) for lower in-gap states and $-1 < t_1/t_2 < 1/2$ ($t_1/t_2 <-1, t_1/t_2 >2$) for higher in-gap states of \textit{$S_M$} (\textit{$S_{dM}$}).
    The value of $p$ jumps at $t_1/t_2 = -1/\sqrt{2}$ and $t_1/t_2 = 1$ for $p_1$ and $p_{\{1,2\}}$ and at $t_1/t_2 = -1$ and $t_1/t_2 = 1/\sqrt{2}$ for $p_4$ and $p_{\{3,4\}}$, 
    where the band-gap closes and topological transition occurs.
    These results establish the bulk-corner correspondence of the HOTI in the martini model.
    
    So far, we deal with the case of $V=0$, but we point out that the in-gap corner states appear even when $V \neq 0$. 
    Our numerical results regarding the corner states are shown in Fig.~\ref{fig:martini_obc_v05}.
    This result is essential for considering the square-root relation to the decorated honeycomb model, as we have shown in Eqs.~(\ref{eq:squaredHamiltonian_martini}) and (\ref{eq:squaredHamiltonian_dmartini}), as well as we discuss in the next subsection.
    
    \subsection{Decorated honeycomb lattice}
\begin{figure}[!tb]
        \centering
        \begin{minipage}[c]{0.6\hsize}
            \includegraphics[width=\hsize]{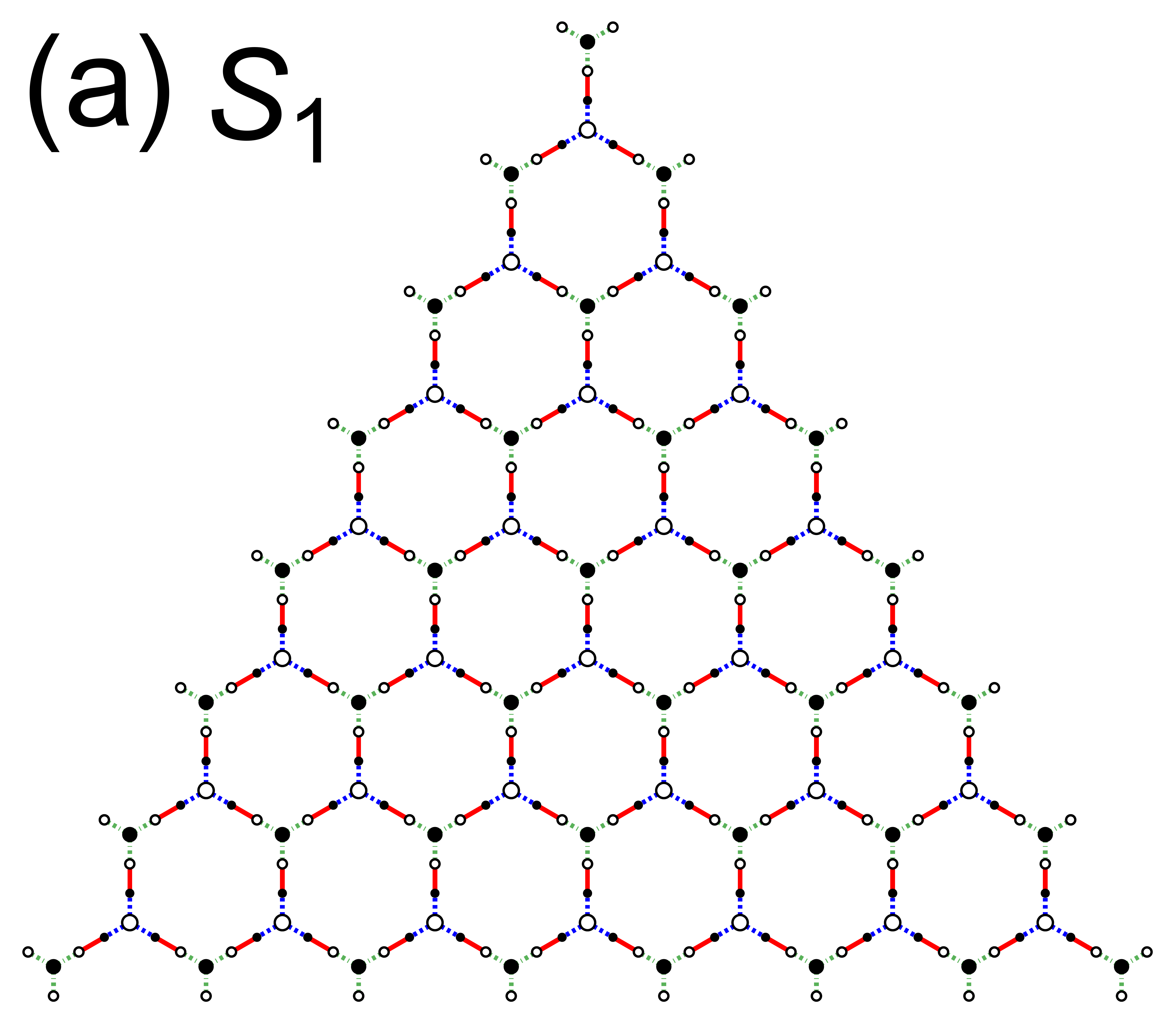}
        \end{minipage}\\
        \begin{minipage}[c]{0.7\hsize}
            \hspace{-15pt}
            \includegraphics[width=\hsize]{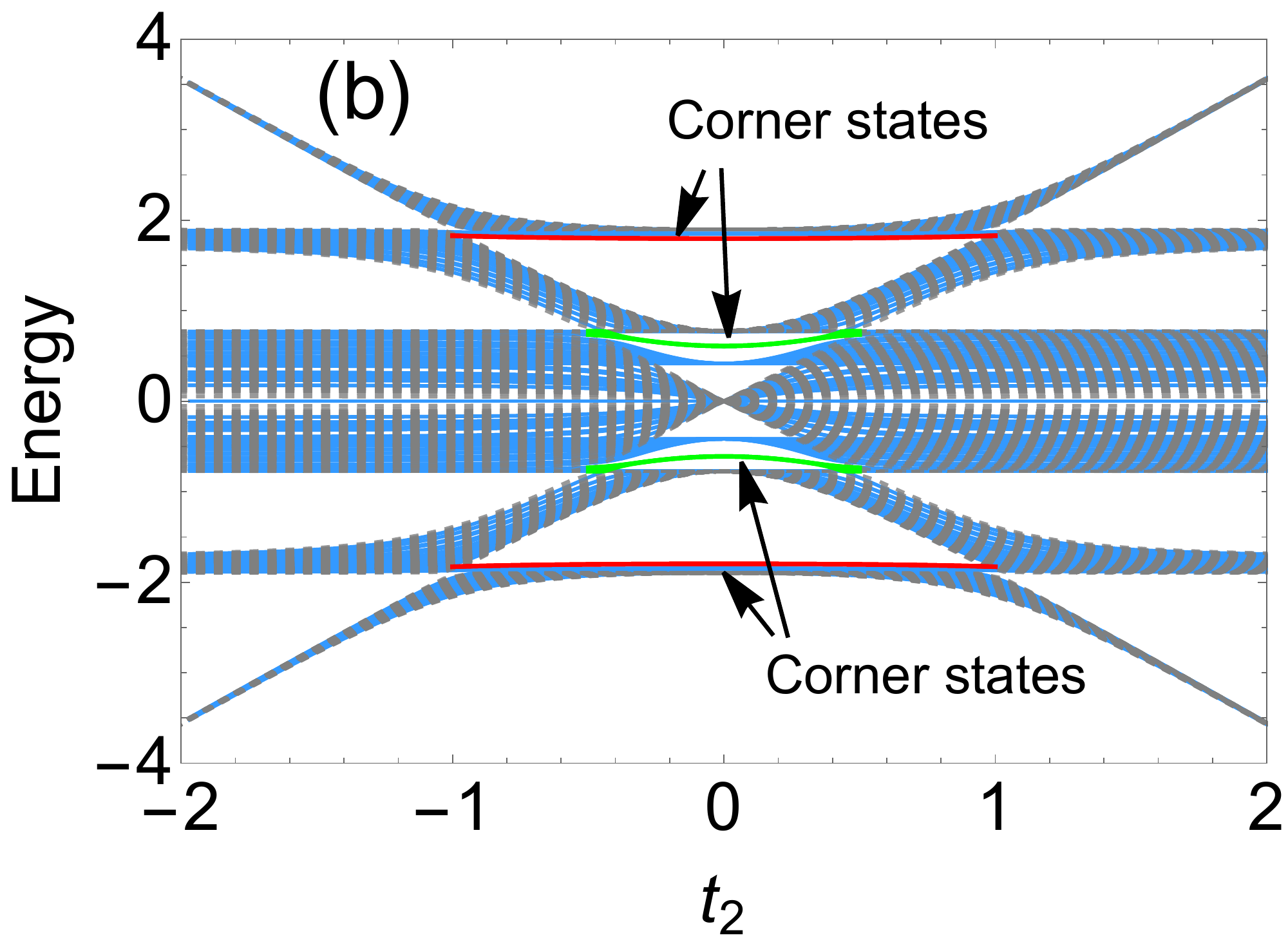}
        \end{minipage}
        \begin{minipage}[c]{0.7\hsize}
            \hspace{-15pt}
            \includegraphics[width=\hsize]{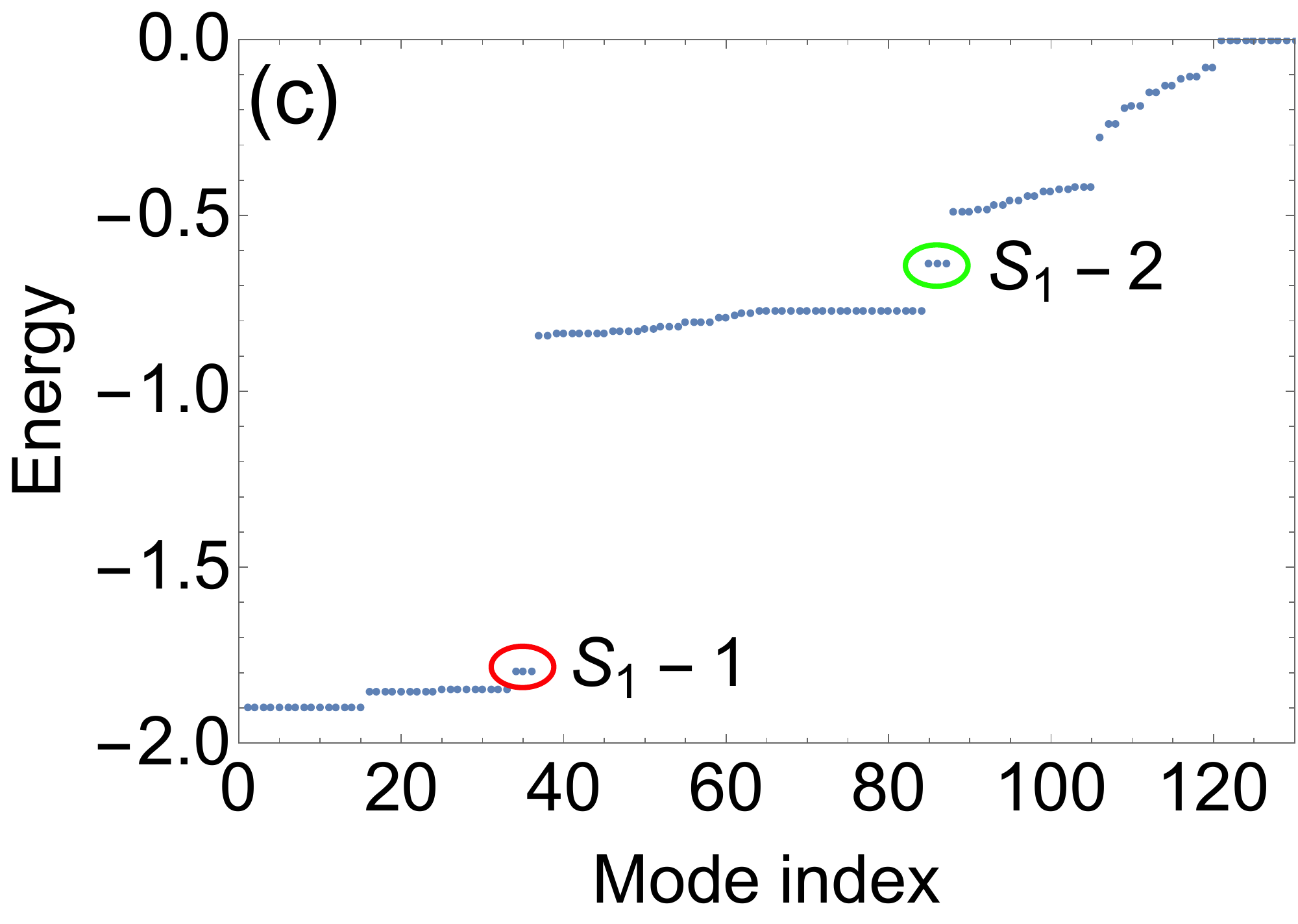}
        \end{minipage}\\
        \begin{minipage}[c]{0.45\hsize}
            \includegraphics[width=\hsize]{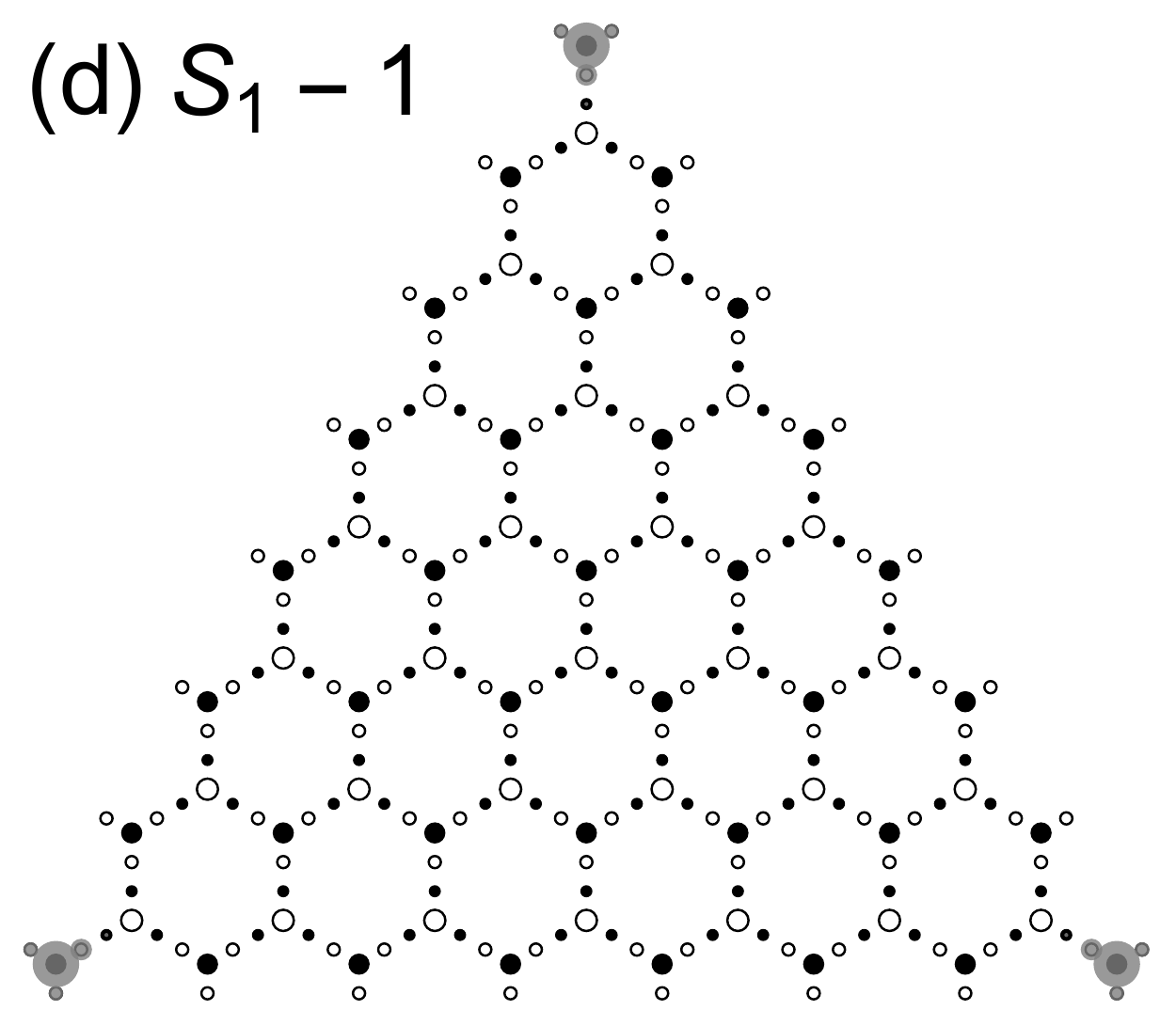}
        \end{minipage}
        \hspace{10pt}
        \begin{minipage}[c]{0.45\hsize}
            \includegraphics[width=\hsize]{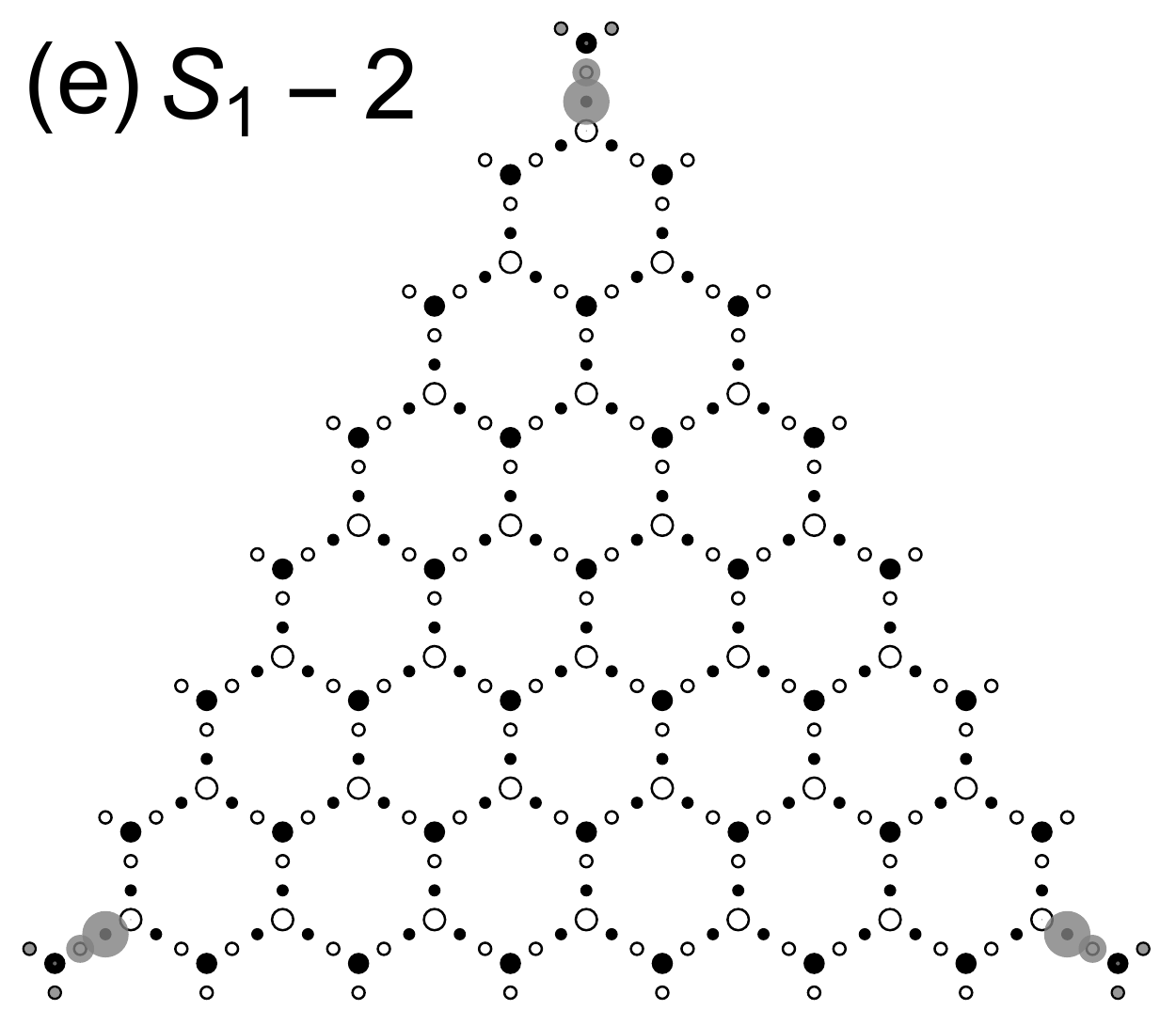}
        \end{minipage}
        \caption{(Color online)(a) Finite system under OBC, consisting of 120 black sites and 136 white sites of the decorated honeycomb model. 
        (b) The energy spectrum with $t_1=1,t_3=\sqrt{3/5}$. The horizontal axis is $t_2$. The  dashed line and the solid line correspond to the spectrum of PBC and OBC respectively. The green and red lines correspond to the in-gap states  with negative and positive energies, respectively. 
        (c) The energy spectrum of lower half energies for $t_1=1, t_2=0.2, t_3=\sqrt{3/5}$. The in-gap corner states are encircled by red and green ellipses.
        The probability density distribution of
        (d) the in-gap states encircled by the red ellipse (\textit{$S_1-1$}) and
        (e) the in-gap states encircled by the green ellipse (\textit{$S_1-2$}) in (c).
        The radii of gray circles represent the probability density.}
        \label{fig:honeycomb_obc_u}
    \end{figure}

\begin{figure}[!tb]
        \centering
        \begin{minipage}[c]{0.6\hsize}
            \includegraphics[width=\hsize]{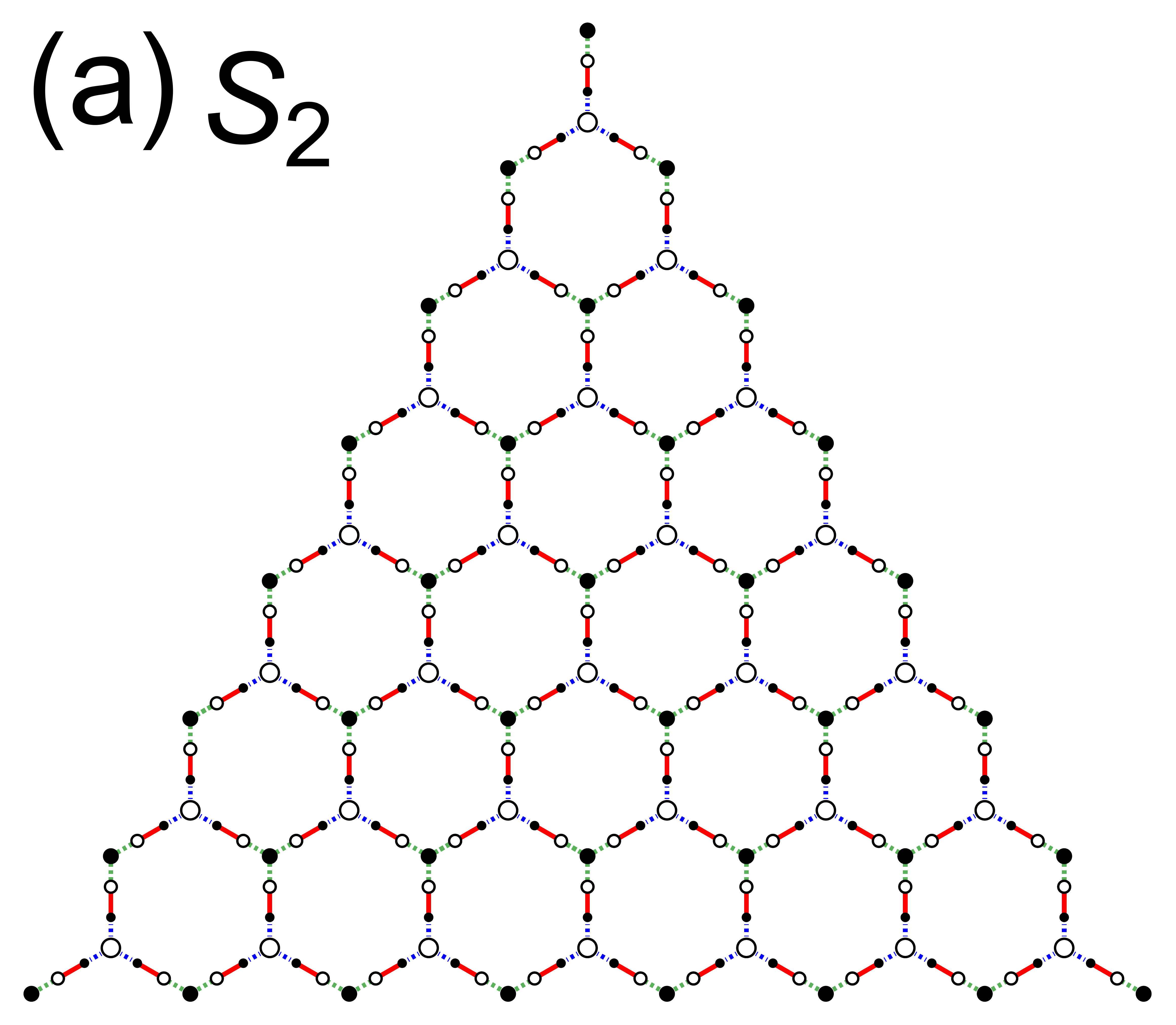}
        \end{minipage}\\
        \begin{minipage}[c]{0.7\hsize}
            \hspace{-15pt}
            \includegraphics[width=\hsize]{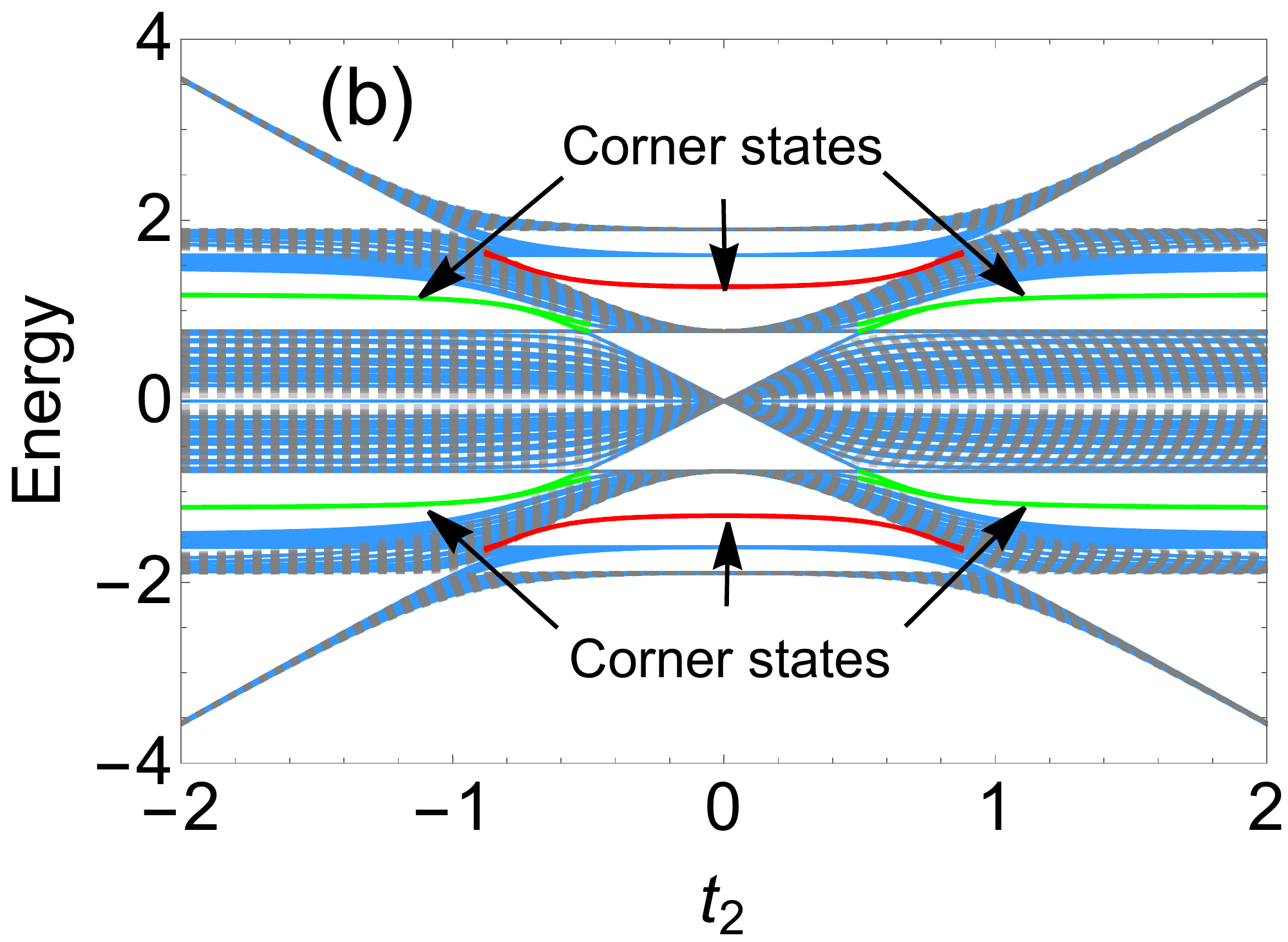}
        \end{minipage}\\
        \begin{minipage}[c]{0.7\hsize}
            \hspace{-15pt}
            \includegraphics[width=\hsize]{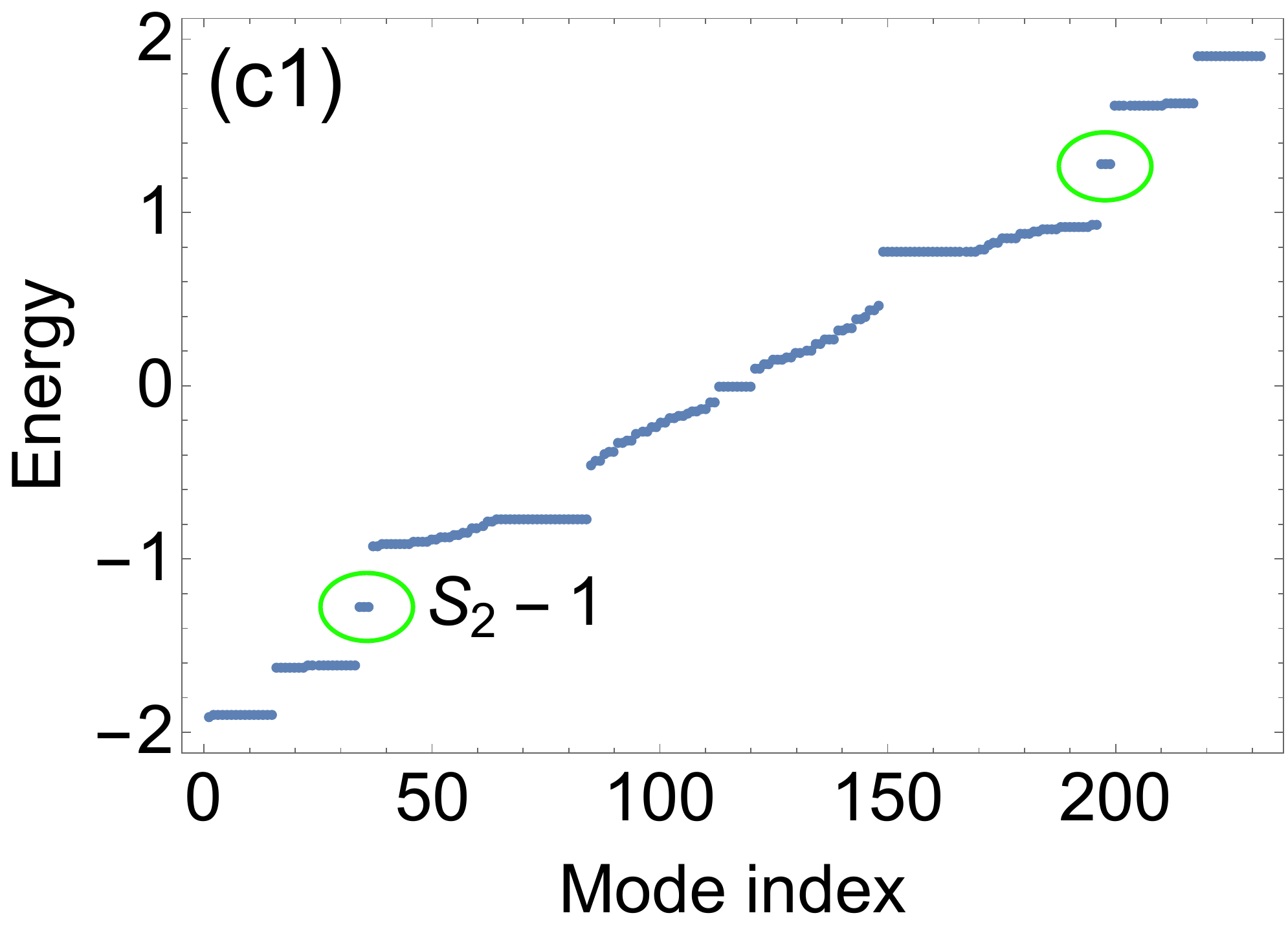}
        \end{minipage}
        \begin{minipage}[c]{0.7\hsize}
            \hspace{-15pt}
            \includegraphics[width=\hsize]{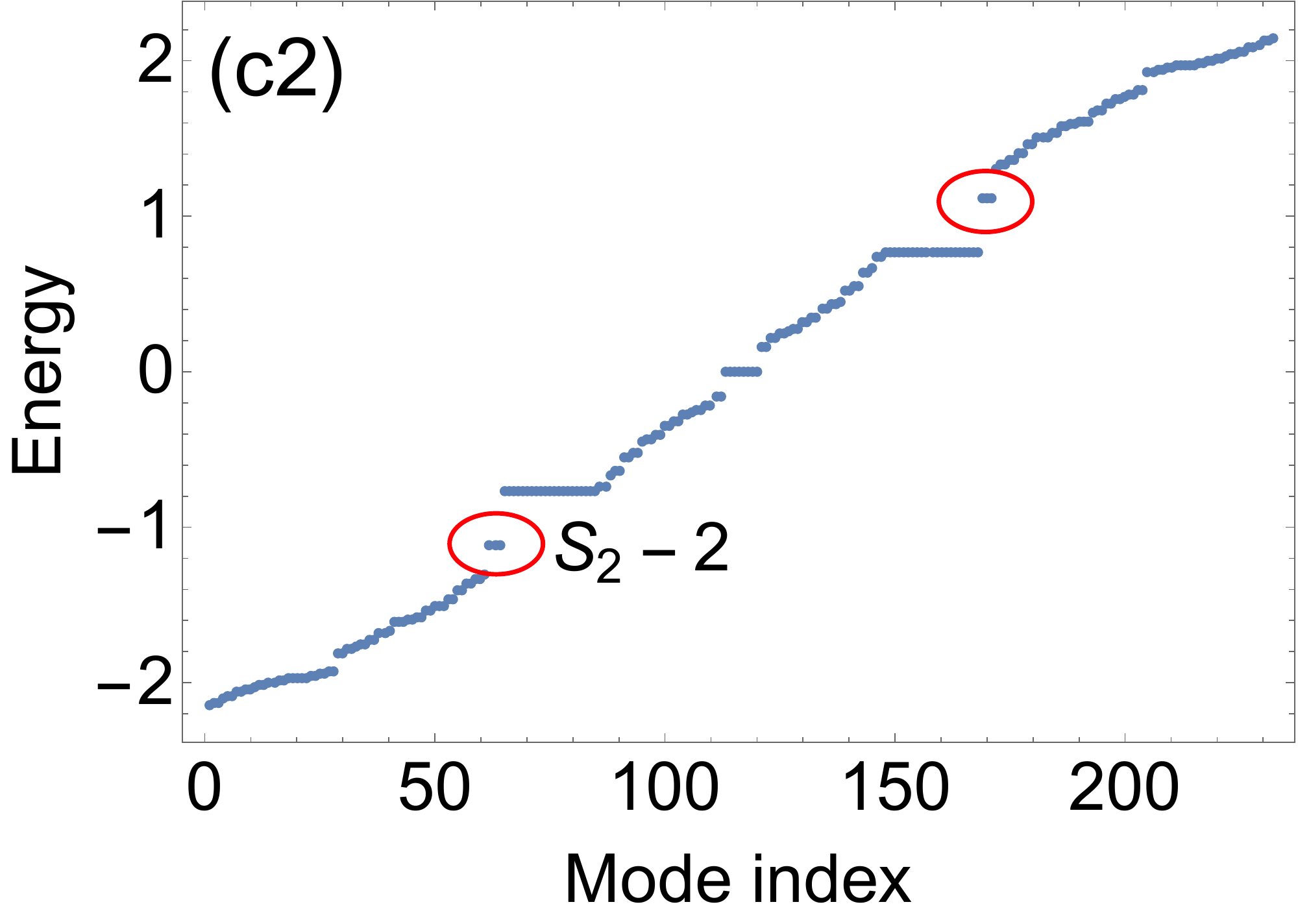}
        \end{minipage}\\
        \begin{minipage}[c]{0.4\hsize}
            \includegraphics[width=\hsize]{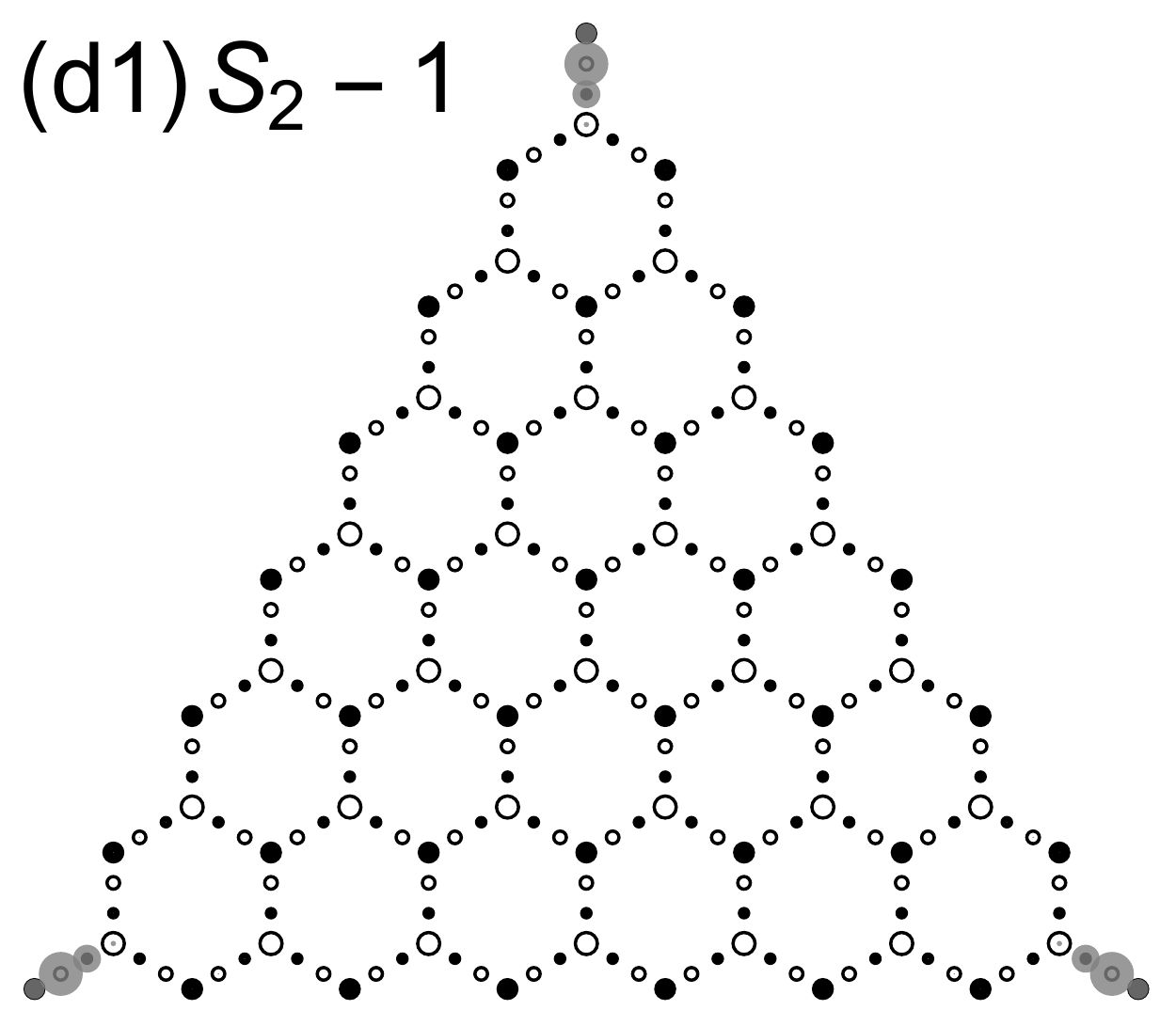}
        \end{minipage}
        \hspace{10pt}
        \begin{minipage}[c]{0.4\hsize}
            \includegraphics[width=\hsize]{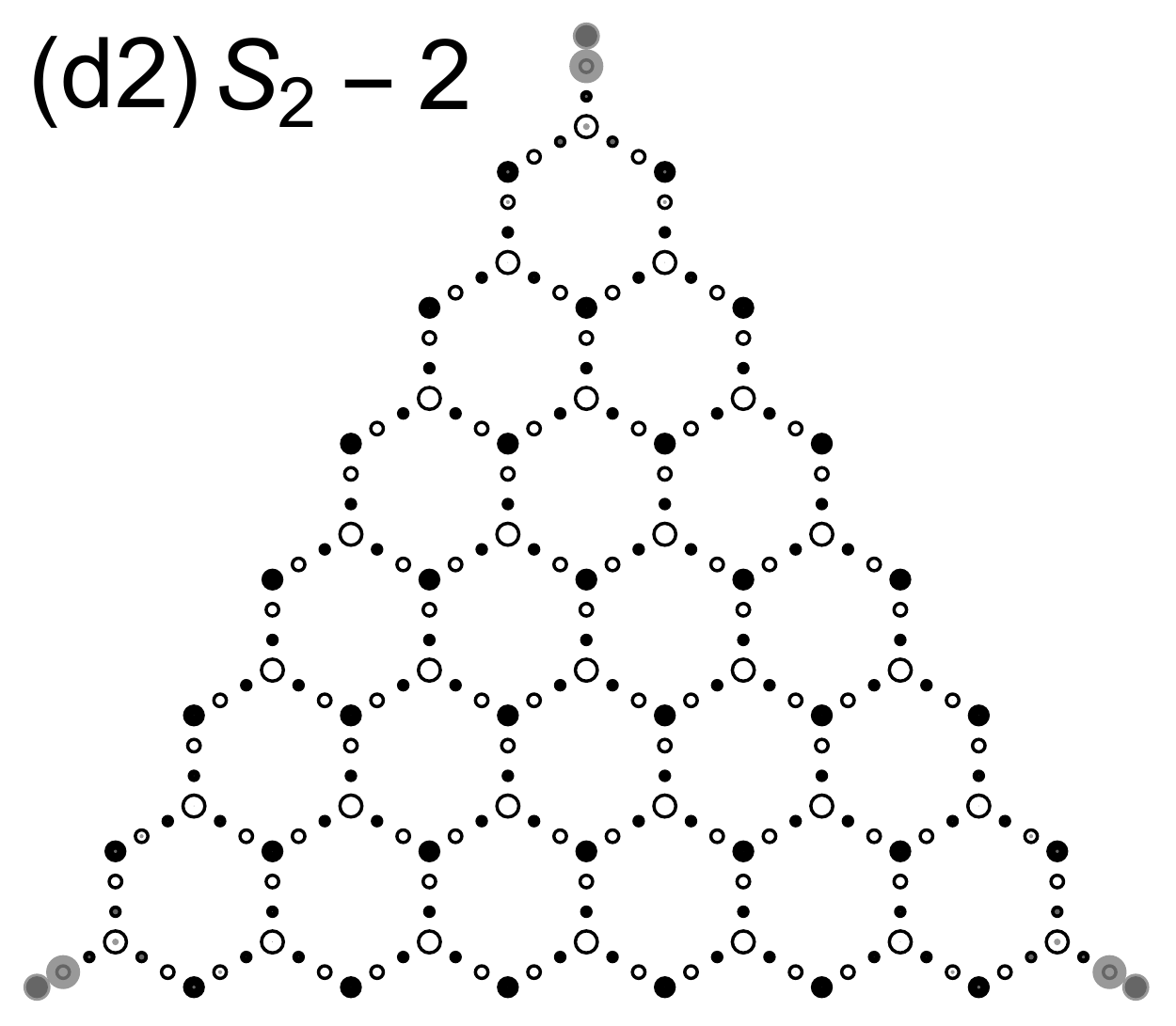}
        \end{minipage}
        \caption{(Color online)(a) Finite system under OBC, consisting of 120 black sites and 112 white sites of the decorated honeycomb model. 
        (b) The energy spectrum with $t_1=1,t_3=\sqrt{3/5}$. The horizontal axis is $t_2$. The  dashed line and the solid line correspond to the spectrum of PBC and OBC respectively. The green and red lines correspond to the in-gap states with negative and positive energies, respectively. 
        (c1) The energy spectrum for $t_1=1, t_2=0.3, t_3=\sqrt{3/5}$. The in-gap corner states are encircled by green ellipses.(c2) The energy spectrum for $t_1=1, t_2=1.0, t_3=\sqrt{3/5}$. The in-gap corner states are encircled by red ellipses.
        The probability density distribution of
        (d1) the lower in-gap states in (c1) and
        (d2) the lower in-gap states in (c2). 
        The radii of gray circles represent the probability density.}
        \label{fig:honeycomb_obc_d}
    \end{figure}

    Now we proceed to study the topological nature of the decorated honeycomb model.
    To this end, we study finite samples.
    We study two finite samples with different corner terminations[Fig.~\ref{fig:honeycomb_obc_u}(a) and \ref{fig:honeycomb_obc_d}(a)], to investigate how the corner states are inherited from the parents.
    In the following, we call the finite sample of Fig.~\ref{fig:honeycomb_obc_u}(a) [Fig.~\ref{fig:honeycomb_obc_d}(a)] \textit{$S_{\textit{1}}$} [\textit{$S_{\textit{2}}$}].

    Before going to the results, we remark on the reason why we adopt these two corner terminations. 
    To explain this, it is important to point out the parent-child relation under the OBC. 
    As is the case of the PBC, the squared Hamiltonian for the decorated honeycomb model is expressed as a direct sum of the upward martini (composed of black sites) and the downward martini (composed of white sites) models [Fig.~1(c)]. 
    However, unlike the PBC case, the sites on the edges and corners acquire additional on-site potentials after the squared operation, because those sites have fewer coordination numbers in the child lattice. 
    Then, the corner termination for \textit{$S_{\textit{1}}$} (\textit{$S_{\textit{2}}$}) is chosen such that the no black (white) site on the edges and corners has a coordination-number imbalance while letting such imbalance for the white (black) sites. 
    In other words, the corner termination is chosen such that one of the two parent Hamiltonians becomes free from the un-uniformity at the boundaries. 
    Consequently, the ``natural" parent for a given corner termination is the other block that does not have un-uniformity. 
    To be concrete, for \textit{$S_{\textit{1}}$} (\textit{$S_{\textit{2}}$}), the upward martini model of $S_M$ (the upward martini model of $S_{dM}$) is the natural parent. 
    
    Based on the above insight, we argue the corner modes and the topological characterization.
    We first argue the results in Fig.~\ref{fig:honeycomb_obc_u}, i.e., those for \textit{$S_1$}.
    We see that there appear two types of in-gap states represented by the green line and red line for $|t_2|<1/2$ and $|t_2|<1$.
    As shown in Fig.~\ref{fig:band_dhoney}, $|t_2|=1$ and $|t_2|=1/2$  are the gap-closing points at $\Gamma$ point and K point, indicating the topological phase transitions occur at these points.
    In Fig.~\ref{fig:honeycomb_obc_u} (c), we show the energy spectrum for $(t_1,t_2,t_3)=(1,0.2,\sqrt{3/5})$.
    Green and red ellipses indicate the lower and upper in-gap states. 
    We see that each in-gap state has three-fold degeneracy, which is again because the sample hosts three corners.
    We plot the real-space probability density distribution of the in-gap states encircled by a green ellipse in Fig.~\ref{fig:honeycomb_obc_u} (d).
    We also plot that of the in-gap states encircled by a red ellipse in Fig.~\ref{fig:honeycomb_obc_u} (e).
    We again take the average over the three degenerate states.
    We see that, in both Figs.~\ref{fig:honeycomb_obc_u}(d) and \ref{fig:honeycomb_obc_u}(e), the in-gap states have large amplitude at the corner and hence are corner states.
    Meanwhile, there is a difference between Figs.~\ref{fig:honeycomb_obc_u}(d) and \ref{fig:honeycomb_obc_u}(e) with respect to sublattice dependence.
    Namely, the corner state of the bottom left corner of Fig.~\ref{fig:honeycomb_obc_u}(d) has a large amplitude at sublattice (iii), 
    on the other hand, that of the bottom left corner of Fig.~\ref{fig:honeycomb_obc_u}(e) has a large amplitude at sublattice (iii) and sublattice (I).
    We call the in-gap states shown in Fig.~\ref{fig:honeycomb_obc_u}(d) [Fig.~\ref{fig:honeycomb_obc_u}(e)] \textit{$S_{\textit{1}}$-1} [\textit{$S_{\textit{1}}$-2}].
    \begin{figure}[!t]
        \centering
        \begin{minipage}[c]{0.49\hsize}
            \includegraphics[width=\hsize]{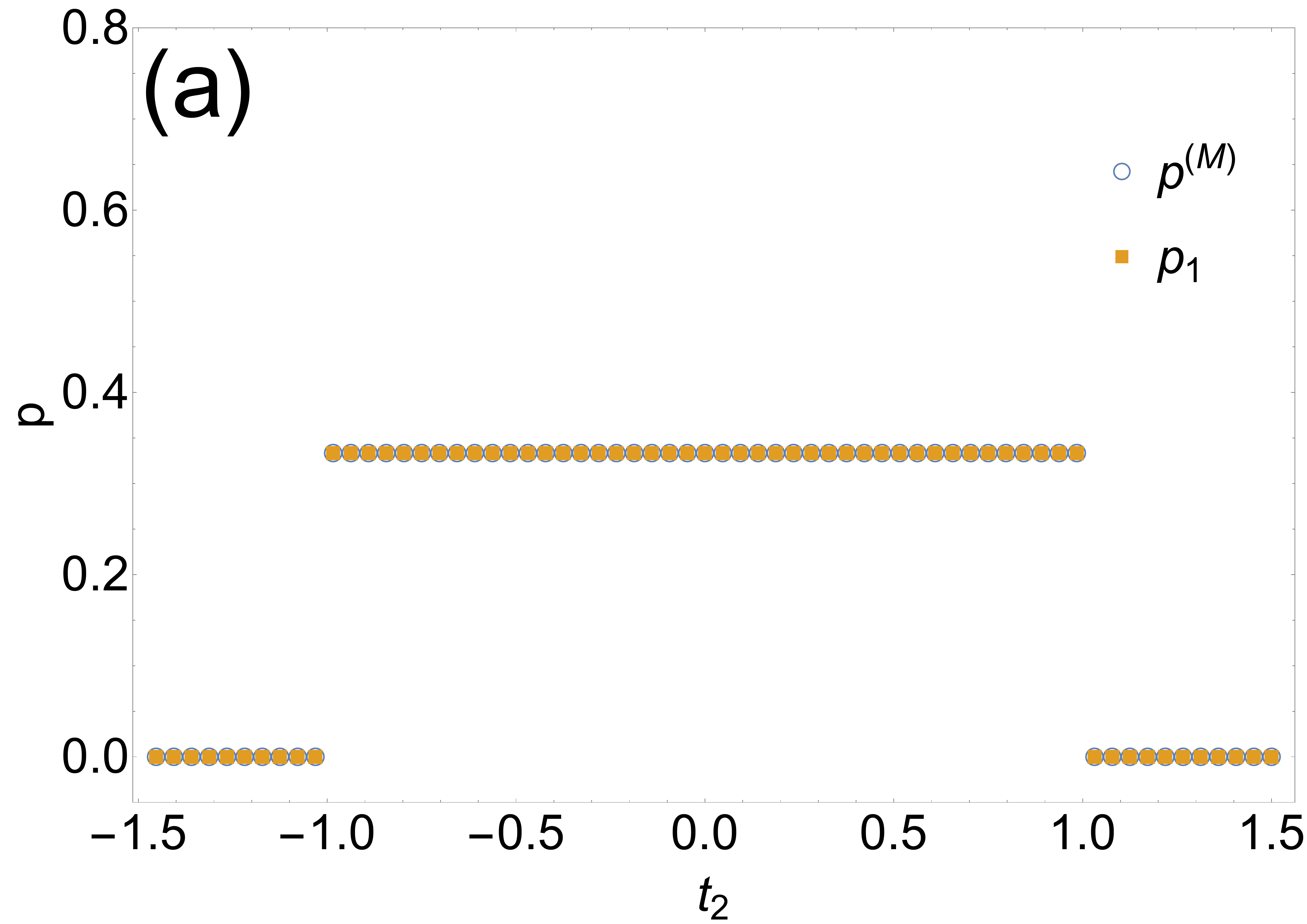}
        \end{minipage}
        \begin{minipage}[c]{0.49\hsize}
            \includegraphics[width=\hsize]{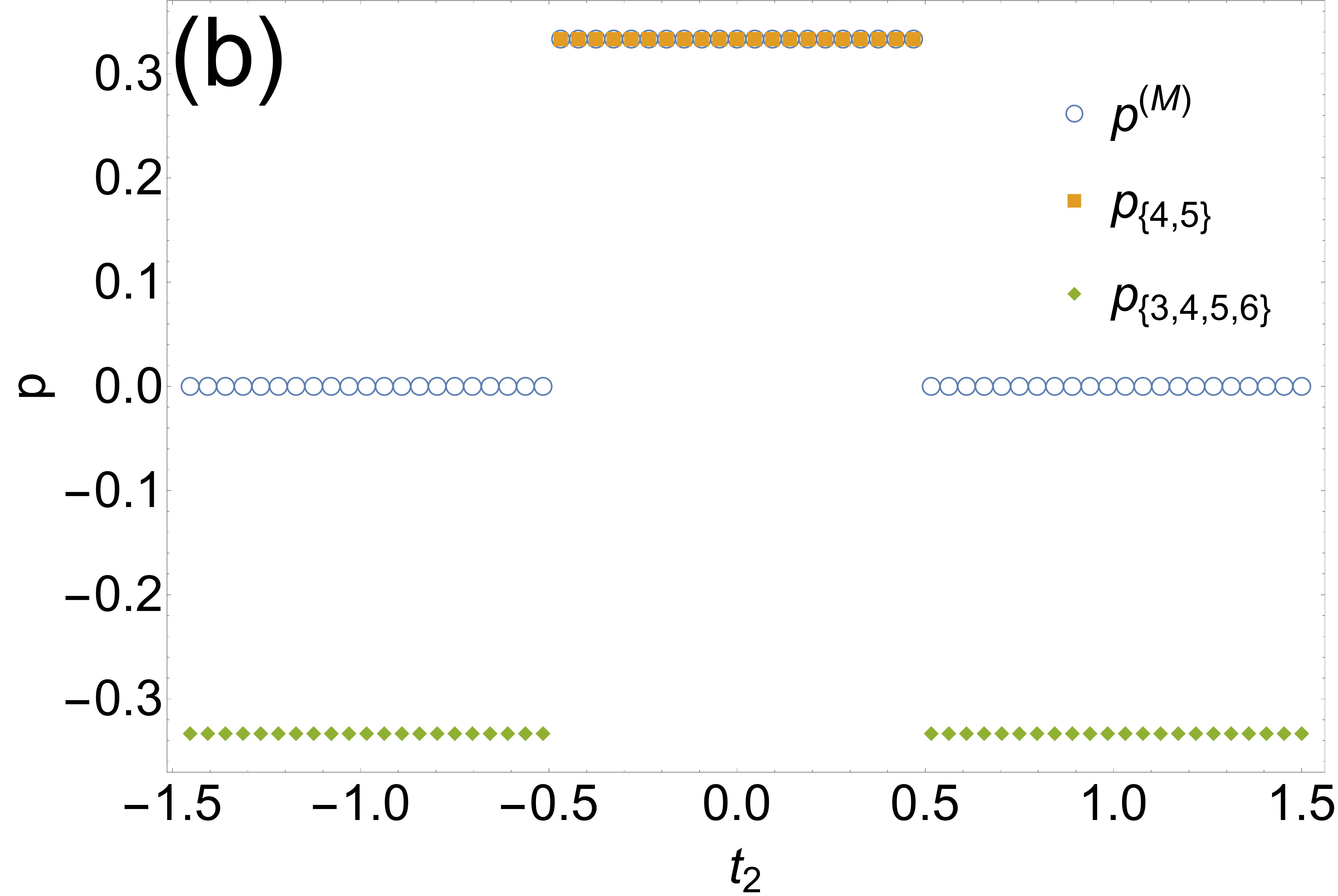}
        \end{minipage}
        \begin{minipage}[c]{0.49\hsize}
            \includegraphics[width=\hsize]{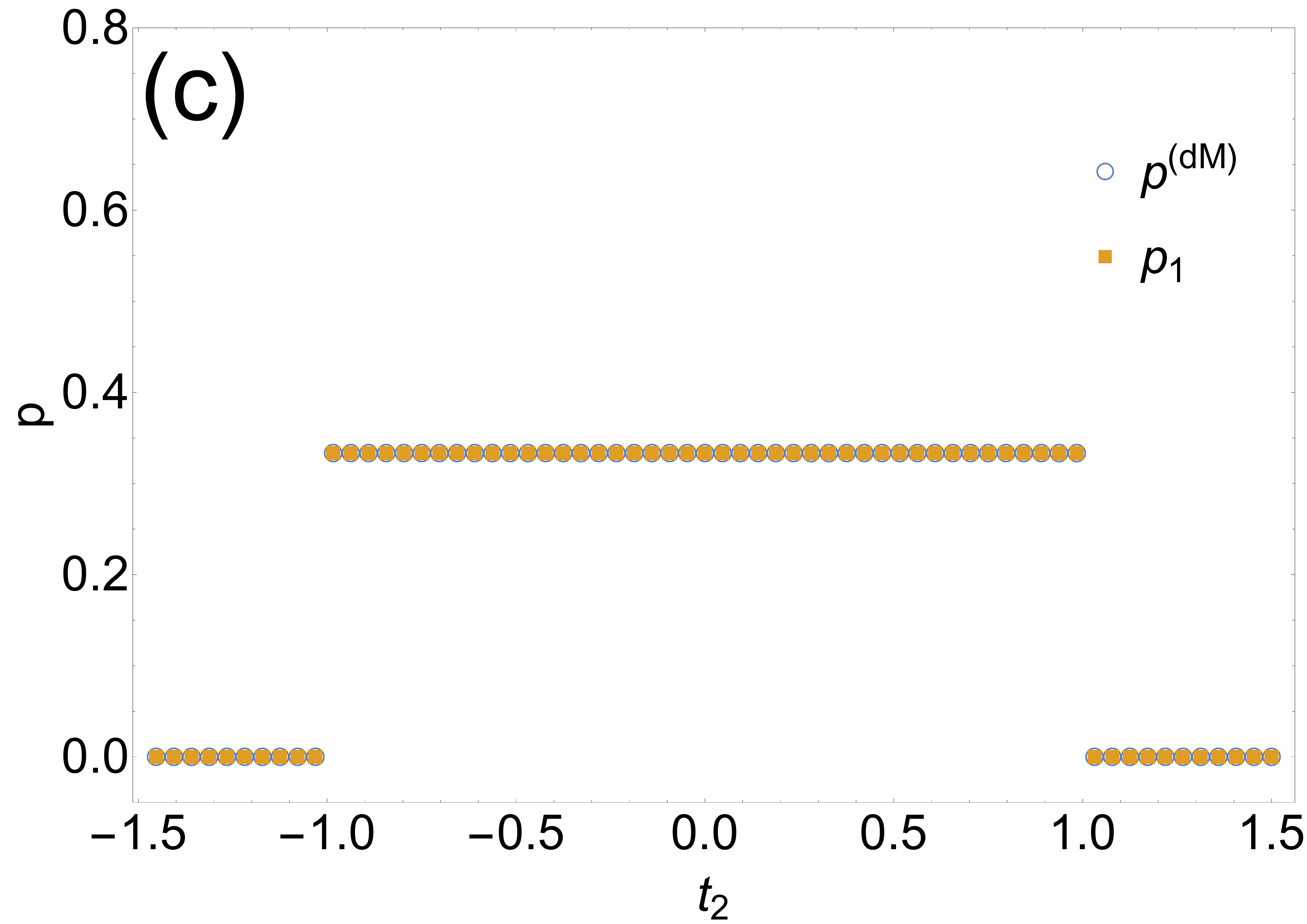}
        \end{minipage}
        \begin{minipage}[c]{0.49\hsize}
            \includegraphics[width=\hsize]{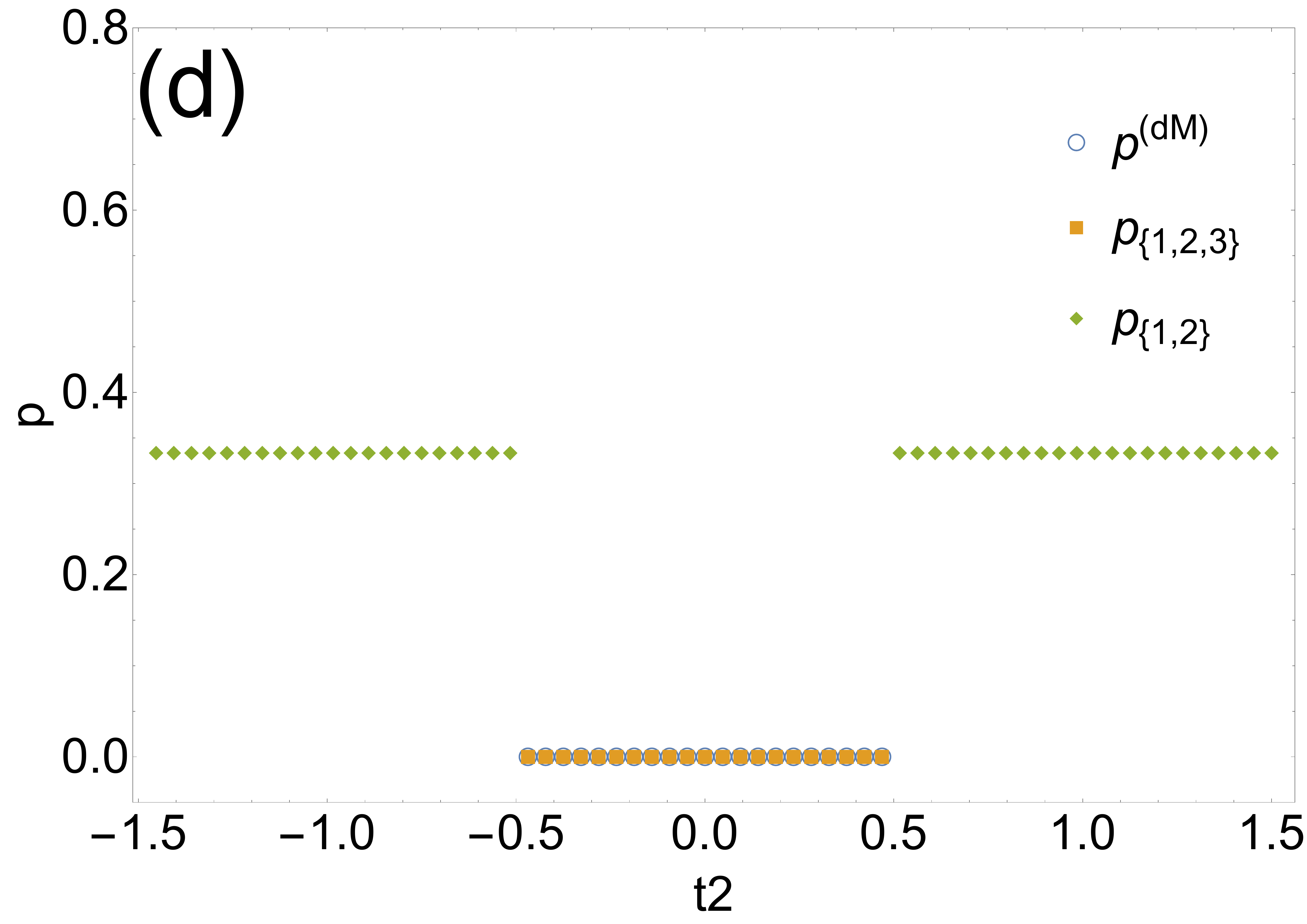}
        \end{minipage}
        \caption{(Color online)Numerical results of the polarization $p$ of Eq. (\ref{Eq:polarization_sigle}) and (\ref{Eq:polarization_multiple}) as a function of $t_2$. 
        The blue circles in these pictures are for the corresponding parent Hamiltonian.
        (a) The $p$ characterize \textit{$S_{\textit{1}}$-1}. Orange squares are for the first band of the decorated honeycomb model. (b) The $p$ characterize \textit{$S_{\textit{1}}$-2}. Orange squares are for the fourth and fifth bands of the decorated honeycomb model. Green rhombuses are for the third, fourth, fifth, and sixth bands of the decorated honeycomb model.
        (c) The $p$ characterize \textit{$S_{\textit{2}}$-1}. Orange squares are for the first band of the decorated honeycomb model. (d) The $p$ characterize \textit{$S_{\textit{2}}$-2}. Orange squares are for the first, second, and third bands of the decorated honeycomb model. Green rhombuses are for the first and second bands of the decorated honeycomb model.
        \label{fig:polarization_honeycom}}
    \end{figure}
    Similarly, we also see the corner states for the sample \textit{$S_{\textit{2}}$} , as shown in Fig.~8. 
    Interestingly, in contrast to \textit{$S_{\textit{1}}$}, we see that the corner states, the red corner states (\textit{$S_{\textit{2}}$-1}) and the green corner states (\textit{$S_{\textit{2}}$-2}) do not coexist in the same parameter region. 
    Rather, \textit{$S_{\textit{2}}$-1} appears for $|t_2| \lesssim 1$ while \textit{$S_{\textit{2}}$-2} appears for $|t_2| \gtrsim 1$ [Fig.~8(b)], which can be accounted for by the corresponding result for the parent Hamiltonian.
    
    We discuss the topological origin of the in-gap corner states.
    Here, we use $p$ of Eqs. (\ref{Eq:polarization_sigle}) and (\ref{Eq:polarization_multiple}) as a topological invariant.
    We characterize the in-gap states \textit{$S_{\textit{1}}$-1} and \textit{$S_{\textit{1}}$-2} by the polarization of the decorated honeycomb model and that of the upward martini model.
    To be concrete, to characterize the in-gap state \textit{$S_{\textit{1}}$-1}, we numerically calculate $p_1$ and corresponding $p$ of upward martini model $p^{(\bm{M})}$.
    To characterize the in-gap state \textit{$S_{\textit{1}}$-2}, we also numerically calculated $p_{\{4,5\}}$ for $|t_2|<1/2, p_{\{3,4,5,6\}}$ for $|t_2|>1/2$ and corresponding $p$ of the upward martini model $p^{(\bm{M})}$.
    Similarly, we characterize the in-gap states \textit{$S_{\textit{2}}$-1} and \textit{$S_{\textit{2}}$-2} by the polarization of the decorated honeycomb model and that of the downward martini model.
    To be concrete, to characterize the in-gap state \textit{$S_{\textit{2}}$-1}, we numerically calculate $p_1$ and corresponding $p$ of the downward martini model $p^{(\bm{dM})}$.
    To characterize the in-gap state \textit{$S_{\textit{2}}$-2}, we also numerically calculated $p_{\{1,2\}}$ for $|t_2|<1/2, p_{\{1,2,3\}}$ for $|t_2|>1/2$ and corresponding $p$ of of downward martini model $p^{(\bm{dM})}$.
    We plot the numerical result in Fig.~\ref{fig:polarization_honeycom}.
    We see that the $p$ is equal to $1/3$ for the parameters where the in-gap states appear, namely,
    $|t_2| <1$ for \textit{$S_{\textit{1}}$-1} and \textit{$S_{\textit{2}}$-1}, and $|t_2|>1/2$ for \textit{$S_{\textit{1}}$-2} and \textit{$S_{\textit{2}}$-2}.
    The value of $p$ jumps at $|t_2| = 1$ for $p_1$ and $|t_2| = 1/2$ for $p_{\{1,2\}}$, $p_{\{1,2,3\}}$,  $p_{\{4,5\}}$,  $p_{\{3,4,5,6\}}$, where the band-gap closes and topological transition occurs.
    the $p$s for the parents and the children are identical, except for the region $|t_2| > 1/2$ in Fig.~9(b). 
    In Appendix, we explain the origin of this behavior. 
    We emphasize that this does not mean the breakdown of the inheritance of topology from the parent to the child. 
    
    The results in Fig.~\ref{fig:polarization_honeycom} establish the bulk-corner correspondence of the HOTI in the decorated honeycomb model.
    The results also indicate that the emergence of the corner states on the decorated honeycomb model is inherited from the martini model, which means the realization of square-root HOTI on the decorated honeycomb model.
    \section{Summary}
    \label{Sec:summary}
    In this paper,we have proposed a concrete example of the HOTI and its square-root.
    Namely, we have shown that the conventional HOTI is realized in the martini lattice model, and the square-root HOTI is realized in the decorated honeycomb lattice with two sites on each edge of the hexagon.
    The emergence of the corner states and their topological protection by the bulk $\mathbb{Z}_3$ invariant have been confirmed.
    
    Remarkably, we find that the corner states in the decorated honeycomb model are sensitive to the corner termination. 
    From the viewpoint of the square-root topology, this can be accounted for by the fact that the un-uniformity around the boundary appears in one of the blocks of the squared Hamiltonian: Consequently, the ``natural'' parent for a given  corner termination is the other block which do not have un-uniformity.
    
    On the experimental realization, we believe that both the martini lattice and the decorated honeycomb lattice are feasible in artificial topological systems, such as photonic crystals, phononic crystals, and electric circuits.
    In particular, for the decorated honeycomb model, the energy of corner state can be switched by varying $t_2$ [Fig.~\ref{fig:honeycomb_obc_d}(b)], which might serve as intriguing property for a corner-mode-based-engineering.

    \section*{Acknowledgement}
    This work is supported by JSPS KAKENHI, Grands No. JP17H06138, No. JP20K14371 (T.M.) and JST-CREST JPMJCR19T1(Y.H.).
	
    \renewcommand{\thetable}{\Alph{section}.\arabic{table}}
    \renewcommand{\thefigure}{\Alph{section}.\arabic{figure}}
    \setcounter{figure}{0}
    \setcounter{section}{1}
    \section*{Appendix: Relation of the polarization between parents and child}
    We point out the generic relation of the polarization between the parents and the child. 
    Note that the case of $E^{(\bm{M})}_{{\bm{k}},i} = 0$ is an exception, which occurs at K point with $i = 1$. 
    
    We first argue the non-zero energy modes, in particular, for the positive-energy modes of the decorated honeycomb lattice model. 
    A similar argument holds for the negative energy sector. 
    The positive-energy eigenvector of Eq.~(\ref{Eq:eigstate_honey}) and the block diagonal form of the $C_3$ operator of Eq.~(13) lead to
    \begin{align}
    \label{eq:eigenEq_c3_2}
            U^{(\bm{DH})}_{\rm{K}}\ket{u_{{\rm{K}},i+4}} &= \frac{1}{\sqrt{2 E_{{\rm{K}},i}^{(\bm{M})}}}\left(
                \begin{array}{c}
                    \sqrt{E_{{\rm{K}},i}^{(\bm{M})}} U^{(\bm{M})}_{\rm{K}} \ket{u_{{\rm{K}},i}^{(\bm{M})}}   \\
                    U^{(\bm{dM})}_{\rm{K}}\ket{u_{{\rm{K}},i}^{(\bm{dM})}} 
                \end{array}
                \right) \nonumber \\
                &= \frac{1}{\sqrt{2 E_{{\rm{K}},i}^{(\bm{M})}}}\left(
                \begin{array}{c}
                    \xi^{(\bm{M})}_i({\rm{K}}) \sqrt{E_{{\rm{K}},i}^{(\bm{M})}} \ket{u_{{\rm{K}},i}^{(\bm{M})}}   \\
                    \xi^{(\bm{dM})}_i({\rm{K}}) \ket{u_{{\rm{K}},i}^{(\bm{dM})}} 
                \end{array}
                \right), \tag{A.1}
    \end{align}
    where $\xi_{i}^{(\bm{M})}({\rm{K}})$ and $\xi_{i}^{({\bm{dM}})}({\rm{K}})$ are the eigenvalues of $U^{(\bm{M})}_{{\rm{K}}}$ and $U^{{\bm{dM}}}_{{\rm{K}}}$ for $\ket{u^{({\bm{M}})}_{{\rm{K}}, i+4}}$ and $\ket{u^{({\bm{dM})}}_{{\rm{K}}, i+4}}$, respectively. 
    Since $\ket{u_{{\rm{K}},i+4}}$ is the eigenstate of $U_{{\rm{K}}}^{(\bm{DH})}$, the eigenvalue $\xi_{i+4}({\rm{K}})$ has to be identical to $\xi_{i}^{({\bm{M}})}({\rm{K}})$ and $\xi_{i}^{({\bm{dM}})}({\rm{K}})$. 
    Consequently, for $p$, the relation
    \begin{align}
        \label{eq:inv_child_and_parents}
        p_n = p^{(\bm{M})} = p^{(\bm{dM})}, \tag{A.2}
    \end{align}
    holds. Indeed, we see that this relation holds in Fig.~\ref{fig:polarization_honeycom}, except for Fig.~\ref{fig:polarization_honeycom}(b). 
    
\begin{figure}[H]
        \centering
        \begin{minipage}[c]{\hsize}
            \includegraphics[width=\hsize]{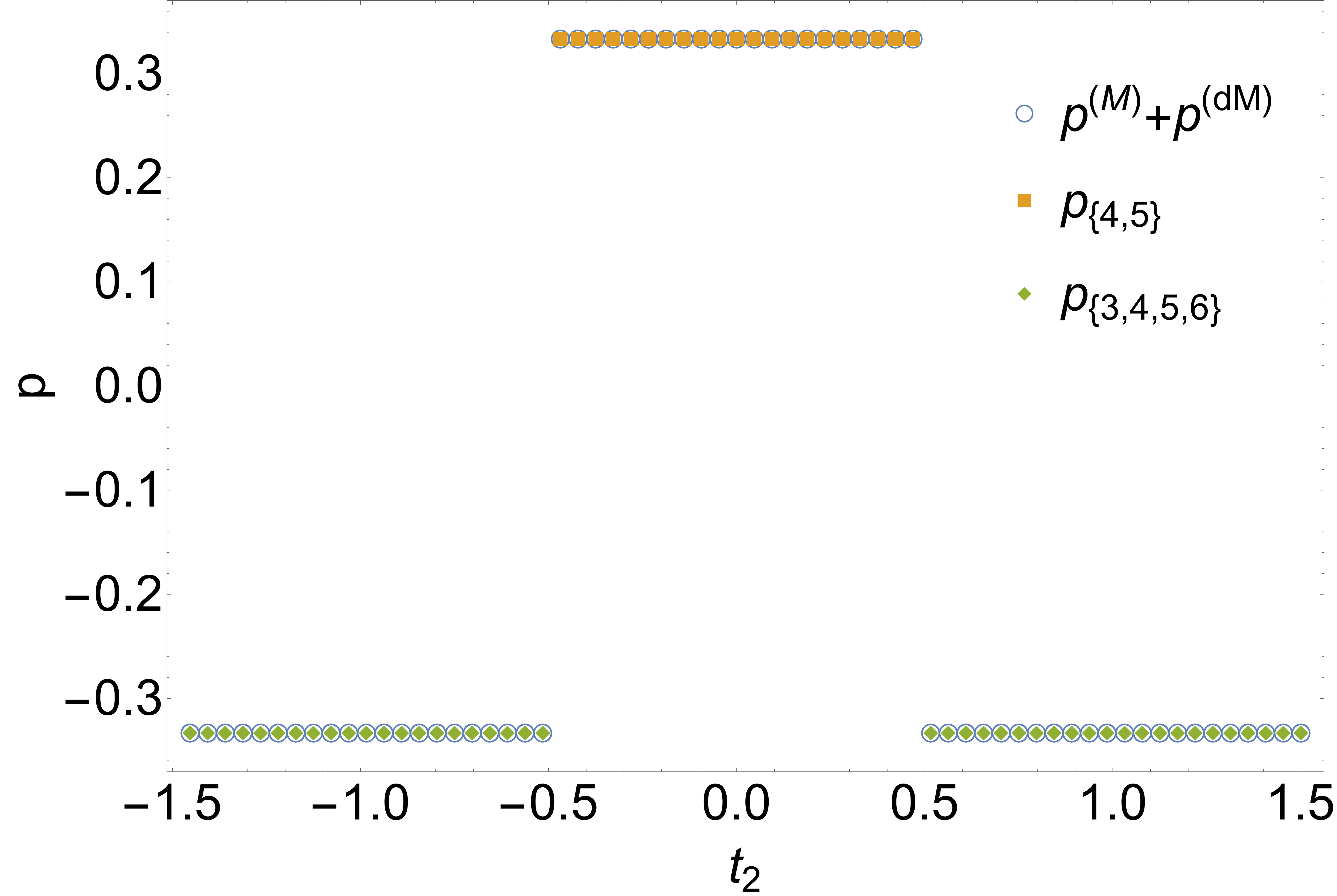}
        \end{minipage}
        \caption{(Color online)Numerical results of the polarization $p$ of Eq. (\ref{Eq:polarization_sigle}) and (\ref{Eq:polarization_multiple}) as a function of $t_2$. 
        Blue circles are for the corresponding parent Hamiltonian.
        Orange squares are for the fourth and fifth bands of the decorated honeycomb lattice model. Green rhombuses are for the third, fourth, fifth, and sixth bands of the decorated honeycomb lattice model.
        \label{fig:polarization_honeycom_dmartini}}
    \end{figure}

    In fact, in Fig.~\ref{fig:polarization_honeycom}(b), we consider the polarization including the case of $E^{({\bm{M}})}_{{\rm{K}},i} = 0$, corresponding to the fourth and the fifth bands for $H_{{\rm{K}}}^{({\bm{DH}})}$. 
    As we have mentioned in Sect. 2.3, the zero-energy eigenstate of $H^{({\bm{DH}})}_{{\rm{K}}}$ can be chosen as Eq.~(\ref{eq:zero-energy_eigenstate}). %[upward martiniのみに値を持つ波動関数とdownward martiniのみに値を持つ波動関数をSect. 2.3に記載しておく]。
    Both of these two states are the eigenstates of $U_{{\rm{K}}}^{({\bm{DH}})}$, but their eigenvalues are, respectively, $\xi_{1}^{({\bm{M}})}({\rm{K}})$, and $\xi_{1}^{({\bm{dM}})}({\rm{K}})$, which are not necessarily identical in this case. 
    Then, unlike the case of finite-energy states, we can argue the parent-child correspondence is only for a set of these two bands. 
    Specifically, $p_{\{4,5\}}$ satisfies
    \begin{align}
        \label{eq:polarization_sum}
        p_{\{4,5\}} = p^{({\bm{M}})}_1 + p^{({\bm{dM}})}_1. \tag{A.3}
    \end{align}
    In Fig.~\ref{fig:polarization_honeycom_dmartini}, we plot the $t_2$ dependence of $p^{({\bm{dM}})},p^{({\bm{M}})}$, and $p_{\{4,5\}}$ /  $p_{\{3,4,5,6\}}$corresponding to the choice of bands of Fig.~\ref{fig:polarization_honeycom}(b). 
    We see that the relation of Eq.~(\ref{eq:polarization_sum}) indeed holds. 
    
    \bibliographystyle{jpsj}
    \bibliography{70633}

\begin{thebibliography}{10}

\bibitem{Haldane1988}
F.~D.~M. Haldane: Phys. Rev. Lett. {\bfseries 61} (1988) 2015.

\bibitem{Kane2005}
C.~L. Kane and E.~J. Mele: Phys. Rev. Lett. {\bfseries 95} (2005) 146802.

\bibitem{Kane2005_2}
C.~L. Kane and E.~J. Mele: Phys. Rev. Lett. {\bfseries 95} (2005) 226801.

\bibitem{Bernevig2006}
B.~A. Bernevig, T.~L. Hughes, and S.-C. Zhang: Science {\bfseries 314} (2006)
  1757.

\bibitem{Hasan2010}
M.~Z. Hasan and C.~L. Kane: Rev. Mod. Phys. {\bfseries 82} (2010) 3045.

\bibitem{Qi2011}
X.-L. Qi and S.-C. Zhang: Rev. Mod. Phys. {\bfseries 83} (2011) 1057.

\bibitem{Hatsugai1993}
Y.~Hatsugai: Phys. Rev. Lett. {\bfseries 71} (1993) 3697.

\bibitem{Hatsugai1993_2}
Y.~Hatsugai: Phys. Rev. B {\bfseries 48} (1993) 11851.

\bibitem{Benalcazar2017}
W.~A. Benalcazar, B.~A. Bernevig, and T.~L. Hughes: Science {\bfseries 357}
  (2017) 61.

\bibitem{Benalcazar2017_2}
W.~A. Benalcazar, B.~A. Bernevig, and T.~L. Hughes: Phys. Rev. B {\bfseries 96}
  (2017) 245115.

\bibitem{Hashimoto-Wu-Kimura2017}
K.~Hashimoto, X.~Wu, and T.~Kimura: Phys. Rev. B {\bfseries 95} (2017) 165443.

\bibitem{Schindler2018}
F.~Schindler, A.~M. Cook, M.~G. Vergniory, Z.~Wang, S.~S.~P. Parkin, B.~A.
  Bernevig, and T.~Neupert: Science Advances {\bfseries 4} (2018) eaat0346.

\bibitem{Ezawa2018}
M.~Ezawa: Phys. Rev. Lett. {\bfseries 120} (2018) 026801.

\bibitem{Hayashi2018}
S.~Hayashi: Communications in Mathematical Physics {\bfseries 364} (2018) 343.

\bibitem{Araki2019}
H.~Araki, T.~Mizoguchi, and Y.~Hatsugai: Phys. Rev. B {\bfseries 99} (2019)
  085406.

\bibitem{Takane2019}
Y.~Takane: Journal of the Physical Society of Japan {\bfseries 88} (2019)
  094712.

\bibitem{Watanabe2020}
H.~Watanabe and S.~Ono: Phys. Rev. B {\bfseries 102} (2020) 165120.

\bibitem{Takahashi2021}
R.~Takahashi, T.~Zhang, and S.~Murakami: Phys. Rev. B {\bfseries 103} (2021)
  205123.

\bibitem{Arkinstall2017}
J.~Arkinstall, M.~H. Teimourpour, L.~Feng, R.~El-Ganainy, and H.~Schomerus:
  Phys. Rev. B {\bfseries 95} (2017) 165109.

\bibitem{Dirac1928}
P.~A.~M. Dirac and R.~H. Fowler: Proceedings of the Royal Society of London.
  Series A, Containing Papers of a Mathematical and Physical Character
  {\bfseries 117} (1928) 610.

\bibitem{Kane2014}
C.~L. Kane and T.~C. Lubensky: Nature Physics {\bfseries 10} (2014) 39.

\bibitem{Attig2017}
J.~Attig and S.~Trebst: Phys. Rev. B {\bfseries 96} (2017) 085145.

\bibitem{Attig2019}
J.~Attig, K.~Roychowdhury, M.~J. Lawler, and S.~Trebst: Phys. Rev. Research
  {\bfseries 1} (2019) 032047.

\bibitem{Naumis2021}
G.~G. Naumis, L.~A. Navarro-Labastida, E.~Aguilar-M\'endez, and
  A.~Espinosa-Champo: Phys. Rev. B {\bfseries 103} (2021) 245418.

\bibitem{Navarro-Labastida2022}
L.~A. Navarro-Labastida, A.~Espinosa-Champo, E.~Aguilar-Mendez, and G.~G.
  Naumis: Phys. Rev. B {\bfseries 105} (2022) 115434.

\bibitem{Kremer2020}
M.~Kremer, I.~Petrides, E.~Meyer, M.~Heinrich, O.~Zilberberg, and A.~Szameit:
  Nature Communications {\bfseries 11} (2020) 907.

\bibitem{Song2020}
L.~Song, H.~Yang, Y.~Cao, and P.~Yan: Nano Letters {\bfseries 20} (2020) 7566.

\bibitem{Yan2020}
M.~Yan, X.~Huang, L.~Luo, J.~Lu, W.~Deng, and Z.~Liu: Phys. Rev. B {\bfseries
  102} (2020) 180102.

\bibitem{Wu2021}
H.~Wu, G.~Wei, Z.~Liu, and J.-J. Xiao: Opt. Lett. {\bfseries 46} (2021) 4256.

\bibitem{Yan2021}
W.~Yan, D.~Song, S.~Xia, J.~Xie, L.~Tang, J.~Xu, and Z.~Chen: ACS Photonics
  {\bfseries 8} (2021) 3308.

\bibitem{Kang2021}
J.~Kang, T.~Liu, M.~Yan, D.~Yang, X.~Huang, R.~Wei, J.~Qui, G.~Dong, Z.~Yang,
  and F.~Nori: arXiv:2109.00879 (2021).

\bibitem{Geng2021}
Z.-G. Geng, Y.-G. Peng, H.~Lv, Z.~Xiong, Z.~Chen, and X.-F. Zhu: Journal of
  Physics: Condensed Matter {\bfseries 34} (2021) 104001.

\bibitem{Mizoguchi2020_sq}
T. Mizoguchi, Y. Kuno, and Y. Hatsugai: Phys. Rev. A \textbf{102} (2020) 033527
  [Erratum \textbf{104} (2021) 029906].

\bibitem{Ezawa2020}
M.~Ezawa: Phys. Rev. Research {\bfseries 2} (2020) 033397.

\bibitem{Mizoguchi2021}
T.~Mizoguchi, T.~Yoshida, and Y.~Hatsugai: Phys. Rev. B {\bfseries 103} (2021)
  045136.

\bibitem{Marques2021}
A.~M. Marques, L.~Madail, and R.~G. Dias: Phys. Rev. B {\bfseries 103} (2021)
  235425.

\bibitem{Dias2021}
R.~G. Dias and A.~M. Marques: Phys. Rev. B {\bfseries 103} (2021) 245112.

\bibitem{Yoshida2021}
T.~Yoshida, T.~Mizoguchi, Y.~Kuno, and Y.~Hatsugai: Phys. Rev. B {\bfseries
  103} (2021) 235130.

\bibitem{Marques2021_2}
A.~M. Marques and R.~G. Dias: Phys. Rev. B {\bfseries 104} (2021) 165410.

\bibitem{Song2022}
L.~Song, H.~Yang, Y.~Cao, and P.~Yan: Nature Communications {\bfseries 13}
  (2022) 5601.

\bibitem{Cheng2022}
W.~Cheng, X.~Zhang, M.-H. Lu, and Y.-F. Chen: Phys. Rev. B {\bfseries 105}
  (2022) 094103.

\bibitem{Zhang2022}
R.-L. Zhang, Q.-P. Wu, M.-R. Liu, X.-B. Xiao, and Z.-F. Liu: Annalen der Physik
  {\bfseries n/a} 2100497.

\bibitem{Roychowdhury2022}
K.~Roychowdhury, J.~Attig, S.~Trebst, and M.~J. Lawler: arXiv:2207.09475
  (2022).

\bibitem{PhysRevB.106.L060305}
R.~W. Bomantara: Phys. Rev. B {\bfseries 106} (2022) L060305.

\bibitem{PhysRevResearch.4.033109}
W.~Deng, T.~Chen, and X.~Zhang: Phys. Rev. Research {\bfseries 4} (2022)
  033109.

\bibitem{10.21468/SciPostPhys.13.2.015}
L.~Zhou, R.~W. Bomantara, and S.~Wu: SciPost Phys. {\bfseries 13} (2022) 015.

\bibitem{Miyahara2005}
S.~Miyahara, K.~Kubo, H.~Ono, Y.~Shimomura, and N.~Furukawa: Journal of the
  Physical Society of Japan {\bfseries 74} (2005) 1918.

\bibitem{Scullard2006}
C.~R. Scullard: Phys. Rev. E {\bfseries 73} (2006) 016107.

\bibitem{McClarty2020}
P.~A. McClarty, M.~Haque, A.~Sen, and J.~Richter: Phys. Rev. B {\bfseries 102}
  (2020) 224303.

\bibitem{Barreteau2017}
C.~Barreteau, F.~Ducastelle, and T.~Mallah: Journal of Physics: Condensed
  Matter {\bfseries 29} (2017) 465302.

\bibitem{Mizoguchi2021_FB}
T.~Mizoguchi, H.~Katsura, I.~Maruyama, and Y.~Hatsugai: Phys. Rev. B {\bfseries
  104} (2021) 035155.

\bibitem{Kubo2006}
K.~Kubo, C.~Hotta, S.~Miyahara, and N.~Furukawa: Physica B: Condensed Matter
  {\bfseries 378-380} (2006) 273.
\newblock Proceedings of the International Conference on Strongly Correlated
  Electron Systems.

\bibitem{Baughman1987}
R.~H. Baughman, H.~Eckhardt, and M.~Kertesz: The Journal of Chemical Physics
  {\bfseries 87} (1987) 6687.

\bibitem{Longuinhos2014}
R.~Longuinhos, E.~A. Moujaes, S.~S. Alexandre, and R.~W. Nunes: Chemistry of
  Materials {\bfseries 26} (2014) 3701.

\bibitem{Li2015}
Z.~Li, M.~Smeu, A.~Rives, V.~Maraval, R.~Chauvin, M.~A. Ratner, and E.~Borguet:
  Nature Communications {\bfseries 6} (2015) 6321.

\bibitem{Lee2020}
J.~M. Lee, C.~Geng, J.~W. Park, M.~Oshikawa, S.-S. Lee, H.~W. Yeom, and G.~Y.
  Cho: Phys. Rev. Lett. {\bfseries 124} (2020) 137002.

\bibitem{Fang2012}
C.~Fang, M.~J. Gilbert, and B.~A. Bernevig: Phys. Rev. B {\bfseries 86} (2012)
  115112.

\end{thebibliography}
\end{document}